\documentclass[table,dvipsnames]{aa}
\usepackage{epic,eepic}
\usepackage{arydshln}
\usepackage{amstext}
\usepackage{amsmath,bbold,mathrsfs,bm,mathtools}
\usepackage{graphicx}
\usepackage{tabularx}
\usepackage{textcomp}
\usepackage{txfonts}
\usepackage{bm}
\usepackage{time}
\usepackage[labelformat=simple]{subcaption}

\usepackage{capt-of}

\usepackage{color}
\usepackage{url}
\usepackage{upgreek}

\usepackage{natbib,twoopt}
\bibpunct{(}{)}{,}{a}{}{,}

\makeatletter
\newcommandtwoopt{\citeads}[3][][]{\href{http://adsabs.harvard.edu/abs/#3}%
{\def\hyper@linkstart##1##2{}%
\let\hyper@linkend\@empty\citealp[#1][#2]{#3}}}
\newcommandtwoopt{\citepads}[3][][]{\href{http://adsabs.harvard.edu/abs/#3}%
{\def\hyper@linkstart##1##2{}%
\let\hyper@linkend\@empty\citep[#1][#2]{#3}}}
\newcommandtwoopt{\citetads}[3][][]{\href{http://adsabs.harvard.edu/abs/#3}%
{\def\hyper@linkstart##1##2{}%
\let\hyper@linkend\@empty\citet[#1][#2]{#3}}}
\newcommandtwoopt{\citeyearads}[3][][]%
{\href{http://adsabs.harvard.edu/abs/#3}
{\def\hyper@linkstart##1##2{}%
\let\hyper@linkend\@empty\citeyear[#1][#2]{#3}}}
\newcommandtwoopt{\citeauthorads}[3][][]%
{\href{http://adsabs.harvard.edu/abs/#3}
{\def\hyper@linkstart##1##2{}%
\let\hyper@linkend\@empty\citeauthor[#1][#2]{#3}}}
\makeatother

\usepackage{epic}
\usepackage{fancyvrb}
\usepackage{rotating}
\usepackage{float,afterpage}

\usepackage{multirow,multicol}
\usepackage{array}
\usepackage{balance}
\usepackage[table,dvipsnames]{xcolor}

\usepackage{hyperref} %
\usepackage{overpic}

\usepackage{xifthen}

\usepackage{soulutf8}

\def\Fourier{\mathfrak{F}}

\def\Mao{{M_\text{AO}}}
\def\Mmfbd{{M_\text{MFBD}}}
\def\Msim{{M_\text{sim}}}

\def\abs#1{\left\lvert\, #1 \,\right\rvert}

\makeatletter
\def\url@leostyle{%
\@ifundefined{selectfont}{\def\UrlFont{\sf}}{\def\UrlFont{\tiny\ttfamily}}}
\makeatother
\def\imwidth{32.2mm}

\title{Multi-frame blind deconvolution and phase diversity with statistical
inclusion of uncorrected high-order modes}

\titlerunning{MFBD and PD with statistical inclusion of uncorrected
high-order modes}

\author{Mats~G.~L{\"o}fdahl and Tomas Hillberg}
\authorrunning{M.G.~L{\"o}fdahl \& T. Hillberg}

\institute{Institute for Solar Physics, Dept. of Astronomy, Stockholm
University, Albanova University Center, 106\,91 Stockholm, Sweden
}

\date{Compiled \now\ on \today.}

\abstract%
{Images collected with ground-based telescopes suffer blurring and
distortions from turbulence in the Earth’s atmosphere. Adaptive
optics (AO) can only partially compensate for these effects. Neither
multi-frame blind deconvolution (MFBD) methods nor speckle
techniques perfectly restore AO-compensated images to the correct
power spectrum and contrast. MFBD methods can only estimate and
compensate for a finite number of low-order aberrations, leaving a
tail of uncorrected high-order modes. Restoration of AO-corrected
data with speckle interferometry depends on calibrations of the AO
corrections together with assumptions regarding the height
distribution of atmospheric turbulence.}%
{We seek to develop an improvement to MFBD image restoration that
combines the use of turbulence statistics to account for high-order
modes in speckle interferometry with the ability of MFBD methods to
sense low-order modes that can be partially corrected by AO and/or
include fixed or slowly changing instrumental aberrations.}%
{We modify the MFBD image-formation model by supplementing the fitted
low-order wavefront aberrations with tails of random high-order
aberrations. These tails follow Kolmogorov statistics scaled to
estimated or measured values of Fried’s parameter, $r_0$, that
characterize the strength of the seeing at the moment of data
collection. We refer to this as statistical diversity (SD). We test
the implementation of MFBD with SD with noise-free synthetic data,
simulating many different values of $r_0$ and numbers of modes
corrected with AO.} %
{Statistical diversity improves the contrasts and power spectra of
restored images, both in accuracy and in consistency with varying
$r_0$, without penalty in processing time. Together with focus
diversity (FD, or traditional phase diversity), the results are
almost perfect. SD also reduces errors in the fitted wavefront
parameters. MFBD with SD and FD seems to be resistant to errors of
several percent in the assumed $r_0$ values.} %
{The addition of SD to MFBD methods shows great promise for improving
contrasts and power spectra in restored images. Further studies with
real data are merited.} %


\begin{document}

\maketitle

\noindent

\section{Introduction}
\label{sec:introduction}

Astronomical images from ground-based telescopes are degraded by
random wavefront aberrations caused by turbulence in the Earth's
atmosphere. For high-resolution solar observations, a combination of
adaptive optics (AO) and image restoration has been very successful in
delivering image data collected in good seeing with resolution near
the diffraction limit of the telescope. However, it has proved harder
to restore images to the correct power spectrum and contrast.

\citet{gonsalves79wavefront} introduced phase diversity (PD) as a way
to constrain the problem of retrieving the wavefront phase from image
plane data. \citet{paxman92joint} derived the model fitting rigorously
for different noise models and extended the formalism to multiple
diversities. PD is based on images recorded with a (preferably known)
difference in the wavefront phase, most often a parabola corresponding
to a slight defocus of one or more cameras. \citet{paxman92phase}
combined PD with multi-frame blind deconvolution \citep[MFBD;
][]{schulz93multi-frame} and referred to this as phase-diverse speckle
interferometry (PDS). MFBD
methods (including PD and PDS) in the present context are based on fitting a model of the
space-invariant image formation process to the data, where both the
wavefront aberrations over the pupil and (at least formally) the image
pixel intensities are the fitted parameters.
\citetads{1994A&AS..107..243L} and \citetads{1994SPIE.2302..268S}
independently developed PDS for high-resolution solar images. These
developments form the core of the methods
\citepads{2002SPIE.4792..146L,2005SoPh..228..191V,2021A&A...653A..68L}
used during two decades for restoration of image data from the Swedish
1m Solar Telescope \citepads[SST;][]{2003SPIE.4853..341S}. The
space-invariant model is only valid within an isoplanatic patch with
a size on the order of only a few arcseconds. Because of this, the
restoration is usually performed in 4--5\arcsec{} squared subfields
where the assumption of space invariance is approximately true. This
size is a trade-off between the smaller isoplanatic patch and the need
for subfields of sufficient size to contain the necessary fine structure. A
restored version of a larger field of view (FOV) is formed by
mosaicking the restored subfields.

Solar PD- and MFBD-based image restoration can restore good data to
diffraction-limited resolution. This is very useful for science based
on the spatial shapes of features in the photosphere and tracking
motions. Examples of this include motions and shapes of magnetic
elements \citep[e.g.][]{berger98measurements_motion,berger04solar},
formation of micropores \citep{berger02observation_aas}, spatial
structure and magnetic field in sunspot penumbrae
(\citeauthor{scharmer02dark} \citeyear{scharmer02dark},
\citeauthor{langhans05inclination} \citeyear{langhans05inclination},
\citeauthorads{2008ApJ...689L..69S} \citeyearads{2008ApJ...689L..69S},
\citeauthor{spruit10striation} \citeyear{spruit10striation}), and
faculae as hot-wall granulation \citep{carlsson04observational}.
During the last decade, SST observations have been based on the CRISP
\citepads{2008ApJ...689L..69S} and CHROMIS \citepads{2017psio.confE..85S}
spectro(polari)meters, where narrowband (NB) data are collected in
short bursts of approximately ten frames per wavelength tuning and polarization
state, and synchronized wideband (WB) data are collected throughout the
line scan. Such data are usually processed with multi-object MFBD
\citepads[MOMFBD;][]{2005SoPh..228..191V,2002SPIE.4792..146L}, where
the WB data constrain the aberrations identified for the individual NB
states and restored images are calculated for all NB states as well as
for the WB. Physical quantities at multiple heights in the photosphere
and chromosphere are then fit by inverting models of the solar
atmosphere with codes such as those of \citetads{2015A&A...577A...7S},
\citet{delacruz19stic}, and \citet{delacruz19method}. This makes
better control over stray light even more important. Examples of such
science are studies by \citetads{2019A&A...627A.101V},
\citetads{2019ApJ...870...88E}, \citetads{2021A&A...648A..54R}, and
\citetads{2022A&A...664A...8M}.

However, the contrasts and power spectra of AO-corrected observations
that are processed with MFBD-based methods have not been reliably
restored. Comparisons based on synthetic data, with objects from
magnetohydrodynamic (MHD) simulations and aberrations following
Kolmogorov statistics, reveal that observed data were lacking in
contrast. \citet{scharmer10high-order} found that high-order wavefront
aberrations uncorrected by AO and image restoration was a
significant cause. \citet{lofdahl12sources} investigated conventional
large-angle stray light in the instrumentation and found that it was a
minor cause after an adjustment of the position of the AO wavefront
sensor (WFS). \citet{lofdahl16off-disk} measured light scattered by
the atmosphere with very wide point spread functions (PSFs) and
found that this was also not a main reason for the low contrast.
Recently, \citetads{2019A&A...626A..55S} reported on the optical
performance of the SST after an upgrade of the AO system and some
other optics. The conclusion from these investigations, as well as
from analysis by \citet[][supporting material]{scharmer11detection}
using the intensity in sunspot umbrae as a constraint, is that the
remaining stray light comes from a PSF that is not wider than a few
arcseconds and that high-order modes not corrected by AO or image
restoration are one of the major causes of the missing contrast. The
resulting PSF wings are a well-known effect known in night-time
astronomy as a diffuse halo around stellar objects.

\citet{scharmer10high-order} proposed a method for compensating for
the uncorrected high-order modes with a post-restoration
deconvolution. The corrective PSF was calculated from random
high-order wavefronts following the correct statistics. This
compensation worked in the ideal case, where all Karhunen--Lo\`eve
\citep[KL; e.g., ][]{roddier90atmospheric} modes up to a
certain cutoff were perfectly corrected by AO and/or image
restoration. For such data, the contrast after compensation for the
high-order modes is consistent for varying $r_0$. However, there is no
perfect cutoff in the modes corrected by MFBD. MFBD partially
``compensates'' for the effect of uncorrected high-order modes by
introducing errors in the modes included in the MFBD model. Therefore,
the post-restoration deconvolution led to an over-compensation of the
contrast after MFBD processing with the number of wavefront parameters
typically used.

\citetads{2021A&A...653A..17S} built on the work by
\citet{scharmer10high-order} by introducing the efficiencies with
which the AO corrects the individual KL modes. Saranathan et al.
simulated wavefronts with Kolmogorov statistics with the $r_0$
reported by the AO WFS, degraded synthetic images from MHD simulations
based on the corresponding transfer functions, and then used MFBD on
both the real data and the synthetic data. The ratios of the standard
deviations of the mode amplitudes in the two restorations yield the
efficiencies of the AO for all modes. The corresponding contributions
are subtracted from the synthetic wavefronts, allowing a straylight
PSF to be constructed with the efficiencies taken into account. As in
the method by \citet{scharmer10high-order}, data can be deconvolved
with this PSF post-restoration. One drawback with this method is that
the parallel MFBD processing has to be done with all KL modes that are
involved in the AO correction; in the SST case, this means at least
the $\sim$230 KL modes used for modeling the AO modes (they included
250 modes). Further assumptions are that the MFBD restoration is
perfect up to some number of modes and that the MFBD processing
with 250 modes works similarly well for the real data and for the
synthetic data. \citetads{2021A&A...653A..17S} report on the
processing of a single dataset collected at 630~nm while $r_0$ on
average was 13~cm, increasing the restored contrast from 9\% to
12.3\%, closer to the contrast expected from MHD simulations but still
2.5--3 percentage points lower. With only a single dataset, the
consistency of the restored contrast for datasets collected in
different seeing conditions remains unclear. Also, the reported
contrast is in the center of the FOV where the correction works best.
For consistent correction outside the center of the FOV, the
time-consuming calculation of the efficiencies would have to be run at
several distances from the WFS FOV.

\citetads{1996ApJ...466.1087P} restored 470~nm solar granulation data
from the Swedish Vacuum Solar Telescope \citep{scharmer85concepts}
with PDS as well as speckle interferometry (simply referred to as
Speckle hereafter), and compared the results. While the restored
images were virtually identical in appearance and resolution, the
Speckle restorations had significantly more power and RMS contrast
(12.6\% for Speckle vs. 11.0\% for PDS).
\citetads{1996ApJ...466.1087P} found that a plausible explanation was
that the lower contrast for PDS was caused by the low-order
parameterization of the wavefronts, with only 15 Zernike polynomials
\citep{noll76zernike}.

Methods based on Speckle \citep{labeyrie70attainment} differ from MFBD
methods in that they are not based on a necessarily finite
parameterization of the wavefront; they independently reconstruct the
Fourier phases (which encode the locations of features and gradients
in an image) and the Fourier amplitude (i.e., the relative weights of
large and small intensity variations in the image) of the restored
object. The Fourier phases are reconstructed by use of a process that
is independent of the statistical properties of the wavefront. Here,
we are more interested in the Fourier amplitudes, the square of which
is the power spectrum. For this, the statistics of atmospheric
turbulence are used to infer the appropriate speckle transfer
functions (STFs), which are corrections to the Fourier amplitudes (and
thereby to the power spectra and contrasts). Without AO correction,
these STFs depend temporally only on Fried's parameter, $r_0$
\citep{fried66optical}, which was measured with the spectral ratio
method \citep{luhe84estimating}. The spectral ratio is the ratio of
the square of a long-exposure transfer function and an average of the
squares of an ensemble of short-exposure transfer functions. Assuming
the object does not change, this can be calculated from a set of
images that statistically sample the random turbulence in the
atmosphere, usually about 100 exposures recorded during a suffiently
long time interval such that statistical averages can be considered
relevant. The spatial frequency at which the spectral ratio tends to
zero is a measure of $r_0$. Like MFBD methods, Speckle is based on a
space-invariant image-formation model, applied in subfields of larger
FOVs. \citetads{1993A&A...268..374V} developed this method for solar
observations into the Kiepenheuer Institut Speckle Imaging Package
(KISIP).

There are several other Speckle codes that are or have been used for
restoration of solar images. We do not attempt a full account here,
but mention a few developments that lead up to the Speckle
restoration used for DKIST data \citepads{2020SPIE11452E..1XB}, as we
believe this should be state of the art in 2022.

The implementation of AO on major high-resolution solar telescopes
\citepads[see][for an account of the history]{2011LRSP....8....2R}
posed a challenge for Speckle. AO compensation has no impact on the
reconstruction of the Fourier phases, and so restoration to diffraction-limited resolution did not require any changes
\citepads[e.g.,][]{2005SoPh..227..217D}. However, the AO correction
changes the statistics of the wavefront aberrations and thereby the
shapes of the spectral ratios as well as the STFs. The problem is
two-fold: (1) At the lock point of the AO WFS, these quantities depend
not only on $r_0$, but also on the efficiencies with which the AO can
correct aberrations at different spatial scales. This means $r_0$
cannot be measured from the spectral ratios without accounting for the
AO correction and the STFs cannot be calculated from $r_0$ alone. (2)
Due to anisoplanatism, the AO correction efficiencies decline with the
field angle (the distance from the WFS lock point), by varying degree
depending on the spatial scales of the wavefront aberrations. (At
large field angles, the wavefront can actually be degraded by the AO.)
This means the STFs depend on $r_0$, the efficiencies, and the field
angle.

\citetads{1997JOSAA..14.1949M} developed the formalism for how Zernike
polynomial components of atmospheric wavefronts decorrelate with
distance in the focal plane for particular atmospheric turbulence
profiles. Using this and an assumption of a two-layer atmosphere with
one phase screen at the pupil and one at 10~km altitude,
\citetads{2007ApOpt..46.8015W} modified the spectral ratio method to
support AO correction. These authors measured the efficiencies of the AO
correction at the lock point as decomposed into Zernike polynomials.
With this input, they could calculate the corresponding spectral ratio
and STF models as a function of $r_0$, AO efficiency, and field angle,
building a library for lookup in the reconstruction process. Using a
least squares fit to find the spectral ratio model that best fits the
data, both $r_0$ and the AO correction efficiency are found for a
particular data set. A metric indicating how well the predicted
decorrelation of the Zernike modes matches the data is computed as a
side product. The best-matched spectral-ratio model immediately
provides the STF to be applied to the Fourier amplitude.
\citetads{2008A&A...488..375W} implemented this approach in KISIP.

Publications that report on the contrasts in AO-corrected data
collected in varying conditions and  restored with this
implementation appear to be scarce. However,
\citetads{2007ApJ...668..586U} restored 115~min of G-band (430~nm)
image data from the Dunn Solar Telescope. They measured $r_0$ values
with a distribution centered on 10.2~cm with most values between 8 and
15~cm during the collection interval. The restored granulation
contrasts ranged from 7\% to 14\% of the average intensity, correlating
with the raw data contrasts as well as with the measured $r_0$ (their
Figs.~2 and~3). Their Fig.~1 (left half) shows a restored image. The
image quality shows strong variations over the FOV, with the contrast
in the upper left corner being much lower than in the center and lower left.

Further developments include measuring $r_0$ with the AO WFS instead
of the spectral ratio \citepads{2010ApOpt..49.1818W}, which reduces the
number of parameters that have to be measured with the spectral ratio
method. \citetads{2017A&A...607A..83P} showed that with a good
measurement of $r_0$, the transfer functions at the AO lock point can
be reliably calculated. The estimation of the transfer function at
other field angles is less certain because the distribution of the
turbulent layers in the atmosphere along the line of sight can change
from day to day and even on a timescale of hours due to changes in
elevation, and also differs between observatory sites. Monitoring this
and maintaining a database of STFs for all possible turbulence
profiles would be a major undertaking. How well the two-layer model
with height calibration represents realistic scenarios is an open
question, but this approach seems to work well most of the time (F.
Wöger, private communication).

Deep learning with neural networks (NNs) has seen great progress in
many kinds of image processing in recent years, with MFBD of
high-resolution solar image data being one of them.
\citet{asensio_ramos18real-time} demonstrated an extremely fast
algorithm based on this latter technique. After being trained with
output from the MOMFBD code (traditional MFBD, with or without PD),
the MFBD NN is able to provide similar results several orders of
magnitude faster. Like traditional MFBD methods, it handles
anisoplanatism by restoring subfields that are subsequently mosaicked
to form the restored version of a large FOV. Carlos Diaz Baso recently
designed a version of this
NN\footnote{https://github.com/ISP-SST/mfbdNN}, which is now installed
at the SST, providing an almost real-time view of the target area
during observations with a resolution that is similar to the science
data delivered by the SSTRED pipeline \citepads{2021A&A...653A..68L};
it can also be used for making high-quality quick-look movies.

\citetads{2021A&A...652A..50W} similarly trained their NN with
Speckle-restored images. After training, it could output images of
similar quality much faster and with the additional feature that it
does not assume spatially invariant blurring, and so has the potential
to handle anisoplanatism better.

Incorporating a physical model of the image formation, including
wavefront phases as sums of KL modes, \citetads{2021A&A...646A.100A}
demonstrated a promising development path to NN MFBD for scientific
use. This implementation is trained with only raw data, and does not require
already restored data for this step. With the training outlined by these latter authors, their NN produced restored images of inferior quality compared to
traditional MFBD restorations based on the same data sets. This was
apparently due to estimation of the wavefronts only succeeding for a
few of the modes. However, as the authors point out, the main purpose
of the paper was to provide a framework for this kind of
processing and they point out several ways to improve the quality of the
restorations in the future, including using more diverse data for
training as well as adding support for PD data.

A relevant point here is that the method of
\citetads{2021A&A...646A.100A} is based on the same image-formation
model as traditional MFBD, including a finite parameterization of the
wavefront phase. Therefore, when some of the issues with their demo dataset
are fixed, their NN MFBD method should have the same weakness when it comes to restoring
the data to correct and consistent power.

In the present paper, we propose a statistical compensation for the
high-order modes that are not included in the finite parameterization
of the wavefront phases in MFBD processing. In the methods developed
by \citet{scharmer10high-order} and \citetads{2021A&A...653A..17S},
this compensation was applied after the MFBD image restoration. Here,
the compensation is instead included in the MFBD image-formation model
itself.

To a good approximation, the fitted low-order modes are enough to
model the core of the PSF, successfully restoring the structures and
resolution of the real scene. The details in the less-well-determined
high-order modes are of less importance, but they combine to produce
extended wings of the PSFs, which affect the contrast. If any
AO-corrected modes are included in the MFBD-fitted modes, the
unaltered high-order components therefore only need to be corrected in
a statistical sense, using a model of the atmospheric turbulence. This
is similar to Speckle, except that Speckle uses the statistics for the
entire wavefronts.

To implement this high-order wavefront compensation, we use the PD
concept in a novel way. We introduce random but fixed contributions
from the uncorrected high-order wavefront modes in a statistical
sense as early as in the MFBD-based image restoration step. A fully
developed method is outside the scope of this paper; here we provide a
proof of concept with synthetic data. We ignore complications such as
noise, anisoplanatism, or modeling of a real AO system, but
demonstrate that SD has the potential to improve estimation of the
PSFs and make the restored contrast consistently more correct, without
significantly increasing the computing time. We also investigate the
method's sensitivity to a few different kinds of errors.

We revisit the theory of MFBD and PD in Sect.~\ref{sec:theory} as well
as the method of \citet{scharmer10high-order} and then go on to
develop the proposed method. In Sect.~\ref{sec:synthetic-data} we
describe our synthetic data, simulating the effects of atmospheric
turbulence with varying strength as well as AO correction of varying
numbers of modes. We demonstrate some aspects of the proposed method
in Sect.~\ref{sec:processing} and discuss the results in
Sect.~\ref{sec:discussion}. We end with some conclusions in
Sect.~\ref{sec:conclusions}.

\section{Theory}
\label{sec:theory}

\subsection{Image formation}

\label{sec:image-formation}

We use a space-invariant image-formation model, including additive
Gaussian white noise. The optical system is characterized by a
generalized pupil function, which can be written for a wavefront with
number $j\in\{1,\ldots,J\}$ (for real data, usually equivalent to a
time coordinate) and a diversity component $\theta_{jk}$ as
\begin{equation}
\label{eq:pupil}
P_{jk}(x_\text{p}) = A(x_\text{p}) \cdot \exp\bigl\{i\bigl(\phi_{j}(x_\text{p})+\theta_{jk}(x_\text{p})\bigr)\bigl\},
\end{equation}
where $x_\text{p}$ is the 2D spatial coordinate in the pupil plane
(which we now drop for brevity in the notation), $A$ is a binary
function that encodes the geometrical shape and size of the pupil, and
$\phi_j$ are the instantaneous phase variations of a wavefront
emanating from the object as it reaches the pupil.

The phase of an instantaneous wavefront with index $j$ can be expanded
in suitable basis functions, $\{\psi_m\}$,  or modes as
\begin{equation}
\label{eq:1}
\phi_j = \sum_{m=1}^\infty\alpha_{jm}\psi_m.
\end{equation}
We use KL modes as the basis functions $\{\psi_m\}$, disregarding the
inconsequential piston mode. This choice corresponds to MOMFBD
processing of real SST data \citepads{2021A&A...653A..68L} and
approximately to the control modes of the SST AO
(\citeauthor{scharmer03adaptive} \citeyear{scharmer03adaptive},
\citeyearads{2019A&A...626A..55S}) together with the correlation
tracker. The KL coefficients follow normal distributions,
\begin{equation}
\label{eq:6}
\alpha_{jm} \sim \mathcal{N}\left(0,\Biggl(\frac{D}{r_0}\Biggr)^{5/6} \sigma_m\right) ,
\end{equation}
where $D$ is the telescope aperture diameter and $\sigma_m^2$ are the
atmospheric variances of the KL modes.
With correction of $M$ modes by use of AO and/or MFBD, it is
useful to write this as a part with correction, $\phi_{jM}$, plus a
tail of higher order uncorrected modes, $\tau_{jM}$,
\begin{equation}
\label{eq:5}
\phi_j  = \sum_{m=1}^M\alpha_{jm}\psi_m + \!\!\! \sum_{m=M+1}^\infty\alpha_{jm}\psi_m = \phi_{jM} + \tau_{jM}.
\end{equation}

\begin{figure}[!tbh]
\def\tilewidth{0.2\linewidth}
\centering
\def\tabrow#1#2#3#4#5{\hspace{-2.5mm}\raisebox{7mm}{#1} &
\includegraphics[angle=90,width=\tilewidth]{wfs_#5_0873} &
\includegraphics[angle=90,width=\tilewidth]{tails_#5_0873} &
\raisebox{7mm}{#2} &
\raisebox{7mm}{#3} &
\raisebox{7mm}{#4}}
\begin{tabular}{cc@{\,}cccc}
&& &  & \hbox to 0 pt{\hss Strehl ratio of $\tau_{jM}(r_0)$\hss} & \\
\cline{4-6}
\noalign{\vspace{1mm}}
$M$ & $\phi_{jM}$ & $\tau_{jM}$ & 6 cm  & 10 cm & 15 cm \\[1mm]
\hline
\noalign{\vspace{1mm}}
\tabrow{1002}{0.941}{0.974}{0.987}{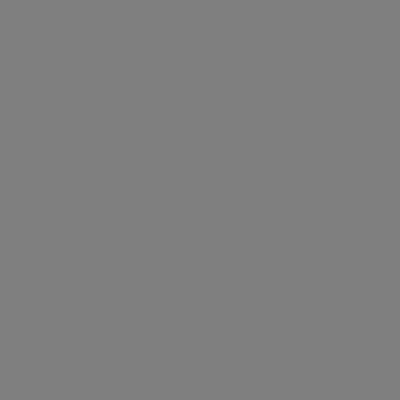}\\[-1mm]
\tabrow{102}{0.607}{0.808}{0.897}{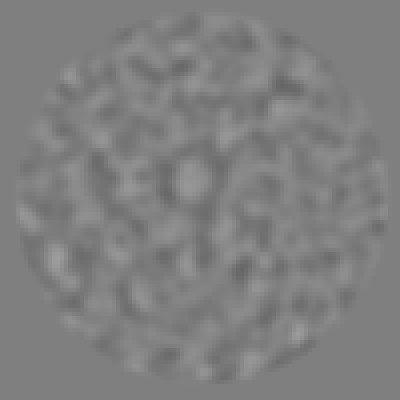}\\[-1mm]
\tabrow{82}{0.543}{0.771}{0.876}{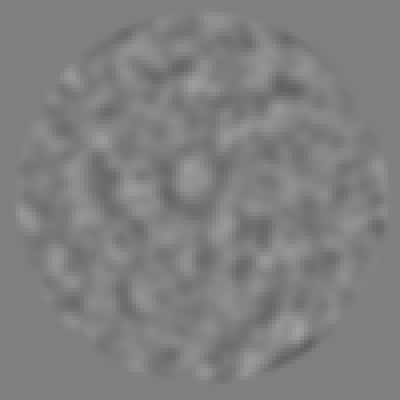}\\[-1mm]
\tabrow{62}{0.454}{0.714}{0.843}{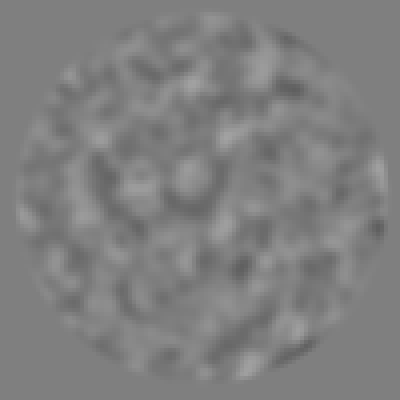}\\[-1mm]
\tabrow{52}{0.396}{0.673}{0.818}{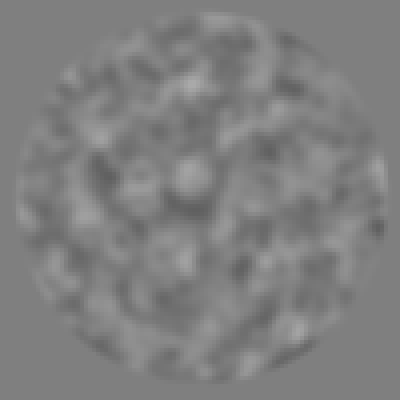}\\[-1mm]
\tabrow{42}{0.324}{0.618}{0.783}{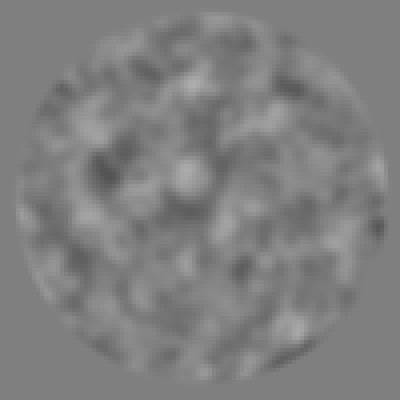}\\[-1mm]
\tabrow{32}{0.235}{0.539}{0.730}{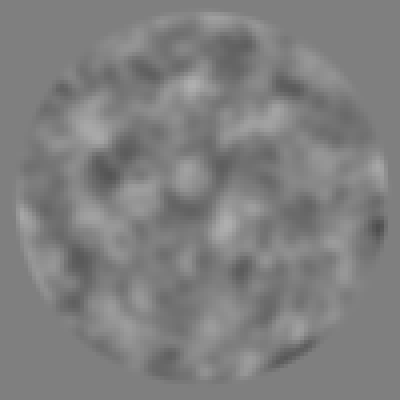}\\[-1mm]
\tabrow{22}{0.129}{0.418}{0.641}{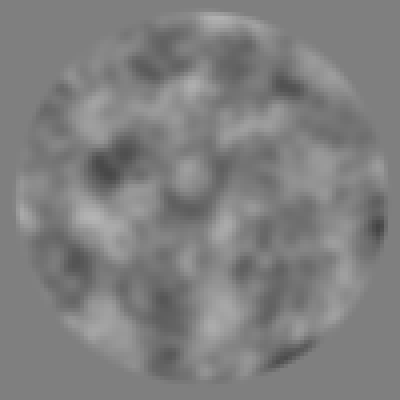}\\
\hline
\end{tabular}
\caption{Example of an instantaneous wavefront phase $\phi_j$ split into
$\phi_{jM}$, a sum of the first $M$ modes, and $\tau_{jM}$, the
residual tail of higher-order modes; see Eq.~(\ref{eq:5}). The
Strehl ratios are calculated for $r_0=6$, 10, and 15~cm
(corresponding to 1\farcs7, 1\farcs0, and 0\farcs5 seeing,
respectively) with Eqs.~(\ref{eq:14}) and~(\ref{eq:15}), assuming
a 97cm aperture and perfect compensation of $\phi_{jM}$. The
wavefront images are calculated with an upper limit of $M=1002$; see  $\Msim$ in Sect.~\ref{sec:wavefronts} for comparison.}
\label{fig:wf-and-tails}
\end{figure}

We illustrate the splitting of a sample random realization of $\phi_j$
for a few values of $M$ in Fig.~\ref{fig:wf-and-tails}. As $M$ grows,
$\tau_{jM}$ becomes flatter and loses all of its larger features. For
the Strehl ratios, we use the wavefront variance after optimal
modal correction with $M$ modes, which can be written as
\begin{equation}
\sigma_\text{corr}^2(M) = 0.3 (D/r_0)^{5/3} M^{-0.92}
\label{eq:14}
\end{equation}
\citep[][based on a figure by
\citetads{1998PASP..110..837R}]{scharmer10high-order}, where $D$ is
the telescope diameter (97~cm), and the expression for the Strehl
ratio $S_M$ as a function of the variance,
\begin{equation}
S_M = \exp(-\sigma_\text{corr}^2(M)).
\label{eq:15}
\end{equation}
We note that as $M$ includes the tip and tilt modes, which do not affect
the contrast of individual exposures, correction of $M$ modes
corresponds to correction of $M-2$ modes with curvature.

With a diversity term
\begin{equation}
\theta_{jk} = a_k Z_4,
\label{eq:12}
\end{equation}
we can represent PD in the form of defocus (Zernike polynomial~4) by
$a_k$ radians RMS. For PD with two cameras, we often use $a_1=0$ and
$a_2$ selected so that $a_2 Z_4$ has a peak-to-valley (PTV) of
approximately one wave. For reasons that are made apparent in
Sect.~\ref{sec:tail-diversity}, we hereafter refer to PD with
defocus as ``focus diversity'' (FD), using PD as a more general term.

The optical transfer function (OTF), can be expressed as the
auto-correlation of the pupil function,
\begin{equation}
\label{eq:3}
S_{jk} = \Fourier\left\{ \abs{\Fourier^{-1}\{P_{jk}\}}^2 \right\},
\end{equation}
where $\Fourier\{\cdot\}$ denotes the Fourier transform and
$\Fourier^{-1}\{\cdot\}$ its inverse. The isoplanatic formation of an
image, $d_{jk}$, is the convolution of an object $f$ and the PSF
$s_{jk}=\Fourier^{-1}\{S_{\!jk}\}$. In the Fourier domain, we can write
this as
\begin{equation}
D_{jk}(u) = F(u) \cdot S_{\!jk}(u) + N_{jk}(u),
\label{eq:image-formation}
\end{equation}
where $N_{jk}$ is an additive noise term with Gaussian statistics, and
$u$ is the 2D spatial frequency coordinate. We also drop this
coordinate for brevity (as well as the spatial coordinate $x$ in the
focal plane).

\subsection{Multi-frame blind deconvolution}
\label{sec:mfbd}

For a derivation of the MFBD algorithm used, along with PD, we refer
the reader to \citetads{2002SPIE.4792..146L} or
\citetads{2005SoPh..228..191V}. Information of various forms about the
circumstances under which the data were collected, such as multiple
cameras, multiple objects, and multiple PDs, can be brought to bear on
the model-fitting problem by specifying linear equality constraints
and using them to reduce the number of free parameters. Here we give a
brief account, with the notation needed to understand PD and the novel
way in which we use the concept and for the interpretation of the
results.

By fitting the image formation model to the data, we can estimate
low-order representations of the wavefronts, $\hat\phi_{j\Mmfbd}$,
given by Eq.~(\ref{eq:5}) and estimates of the $\Mmfbd$ first
wavefront coefficients. We use the circumflex, $\hat\cdot$, to
indicate quantities that are estimated (or based on other estimated
quantities) by minimizing a maximum-likelihood metric that is valid
for additive Gaussian noise,
\begin{equation}
\label{eq:2}
L = \sum_{j,k} \sum_{u} \abs{ \hat D_{jk}(\hat\phi_{j\Mmfbd}, \theta_{jk}) - D_{jk} }^2,
\end{equation}
where sums over $j$ and $k$ should be understood as running over
whatever combinations of those indices exist in the dataset. The
estimated data frame can be written as
\begin{equation}
\label{eq:4}
\hat{D}_{jk}(\hat{\phi}_{j\Mmfbd}, \theta_{jk}) = \hat{F} \hat{S}_{\!jk}(\hat{\phi}_{j\Mmfbd}, \theta_{jk}),
\end{equation}
through Eqs.~(\ref{eq:pupil}) and (\ref{eq:3}). The pixel intensities
of the object $f$ are formally unknowns in the model but the additive
Gaussian noise assumption allows us to write the estimate of the
deconvolved object as
\begin{equation}
\label{eq:F}
\hat F = H \frac{\sum_{j,k} D_{jk} \hat S^*_{\!jk}(\hat \phi_{j\Mmfbd}, \theta_{jk})}
{\sum_{j,k}\abs{\hat S_{\!jk}(\hat \phi_{j\Mmfbd}, \theta_{jk})}^2},
\end{equation}
where the superscript asterisk denotes complex conjugation and $H$ is
a low-pass filter; for example, the optimal low-pass filter of
\citetads{1994A&AS..107..243L}.

Equation~(\ref{eq:F}) can then be plugged into Eq.~(\ref{eq:4}) to form an
expression for $\hat D_{jk}$ without explicit reference to the unknown
object $F$. This eliminates $\hat F$ also in the error metric of
Eq.~(\ref{eq:2}) and makes $\{\alpha_{jm}\}$ the only explicit model
parameters, which reduces the size of the problem significantly.

We note that in Eq.~(\ref{eq:2}) we introduce a mismatch between the
observational data and the image formation model in that the
atmospheric wavefront aberrations of the real data, $D_{jk}$, are a
sum of an infinite number of modes, $\phi_j$, while that of
$\hat D_{jk}$ is only a sum of $\Mmfbd$ modes, $\hat \phi_{j\Mmfbd}$.
The difference is the tail of uncorrected modes, $\tau_{j\Mmfbd}$.

\begin{figure}[!htb]
\centering
\includegraphics[bb=132 577 318 669,width=0.7\linewidth,clip]{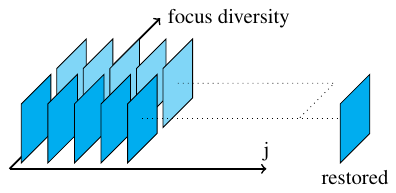}
\caption{Schematic representation of constraints and diversities for
a dataset with five atmospheric wavefronts and FD.}
\label{fig:constraints_without_tails}
\end{figure}

A dataset for MFBD with FD (also called PDS) is illustrated in
Fig.~\ref{fig:constraints_without_tails}. The fitting of the model to
the data is constrained by the fact that all images depict the same
object, and also by the pairwise identical wavefront $\phi_j$ with a
fixed FD addition, $\theta_{jk}$. MFBD without FD is illustrated by
the same figure but without the lighter blue frames.

\subsection{Post-restoration correction}
\label{sec:post-rest-corr}

\citet{scharmer10high-order} proposed a method to compensate for stray
light resulting from $\tau_{j\Mmfbd}$. Their approach was to use the
low-order correction with MFBD and then compensate the restored image
for the average effect of hundreds of versions of
$\tilde\tau_{j\Mmfbd}$, where we use a tilde, $\tilde\cdot$, to
indicate that these are random realizations of $\tau_{\Mmfbd}$, based
on KL coefficients according to Eq.~(\ref{eq:6}). The $r_0$ needed in
Eq.~(\ref{eq:6}) was measured with a WFS.
\citet{scharmer10high-order} wrote the compensated object estimate in the Fourier domain as
\begin{equation}
\label{eq:23}
F \approx \hat F / \hat S,
\end{equation}
where the division with $\hat S$ is a deconvolution with a transfer
function corresponding to the effect of the high-order modes. This
transfer function can be written as
\begin{equation}
\hat S
= \frac{\sum_{j,k} \langle S_{\!jk} \rangle {\hat S}_{\!jk}^*  }
{\sum_{j,k} {\abs{{\hat S}_{\!jk}}^2}},
\label{eq:22}
\end{equation}
where ${\hat S}_{\!jk}$ is calculated from the estimated
$\hat\phi_{\Mmfbd}$ while $S_{\!jk}$ is calculated from
$\hat\phi_{\Mmfbd}$ plus the random $\tilde\tau_{\Mmfbd}$. The angular
brackets, $\langle\cdot\rangle$, indicate an ensemble average over
many independent realizations of $\tau_{j\Mmfbd}$. The transfer
function $\hat S$ removes the effect of the MFBD
restoration and replaces it with a correction for both the
MFBD-estimated low-order wavefront component and an average high-order
component based on statistics.

The main problem with this method is that MFBD correction does not
lead to a clean separation between the included low-order modes and
the uncorrected high-order modes. Instead, because of the model
mismatch mentioned above, the algorithm introduces errors in the
$\Mmfbd$ included modes to try to compensate the OTFs for the effects
of the missing high-order modes.

\subsection{Statistical diversity}
\label{sec:tail-diversity}

\citet{scharmer10high-order} showed that the exact realizations of the
high-order components of the random wavefronts do not change the
degraded images significantly, as long as they follow the correct
statistics. As the MFBD error metric in Eq.~(\ref{eq:2}) involves the
differences between real images and synthetic images based on the
estimated wavefront, this insensitivity makes it hard for MFBD to
correctly identify the high-order aberrations.

We now introduce a statistical compensation for $\tau_{j\Mmfbd}$ in
the MFBD model. For each estimated wavefront in a dataset, we add
$\tilde\tau_{j\Mmfbd}$ as a statistical diversity (SD) wavefront
contribution, where we again use a tilde to indicate that it is based
on random KL coefficients according to Eq.~(\ref{eq:6}). The SD
contribution reduces the model mismatch in the error metric (see
Sect.~\ref{sec:mfbd}) and thereby allows the low-order coefficients to
converge to values that take the high-order tails into account.
Instead of averaging the effects of multiple high-order random
wavefronts like \citet{scharmer10high-order}, we here automatically
get a similar effect from the individual wavefronts of the multiple
images in a dataset, provided the set is large enough.

\begin{figure}[!htb]
\centering
\includegraphics[bb=132 436 318 669,width=0.7\linewidth,clip]{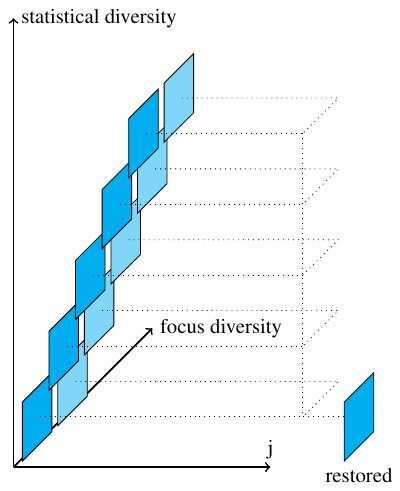}
\caption{Schematic representation of constraints and diversities for
the same data as in Fig.~\ref{fig:constraints_without_tails} but
processed with both FD and SD.}
\label{fig:constraints_with_tails}
\end{figure}

The PD contribution to each wavefront in Eq.~(\ref{eq:pupil}) now
becomes a combination of FD and SD,
\begin{equation}
\theta_{jk} = a_k Z_4 + \tilde{\tau}_{j\Mmfbd}.
\label{eq:13}
\end{equation}
The constraints and diversity scheme is illustrated in
Fig.~\ref{fig:constraints_with_tails}. As in
Fig.~\ref{fig:constraints_without_tails}, all frames include the same
object. Also, frames with the same $j$ index share the same unknown
aberrations and, depending on $k$, may have different FDs. In contrast
to the FD terms, the SDs are independent random realizations for the
different $j$ indices.

\section{Making synthetic data}
\label{sec:synthetic-data}

\begin{figure*}[!thb]
\centering
\begin{minipage}[c]{0.5\linewidth}
\centering
\begin{subfigure}{\imwidth}
\includegraphics[width=\imwidth]{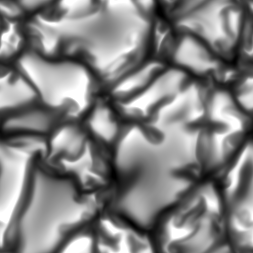}
\caption{}
\label{fig:true_im_538}
\end{subfigure}\quad %
\begin{subfigure}{\imwidth}
\includegraphics[width=\linewidth]{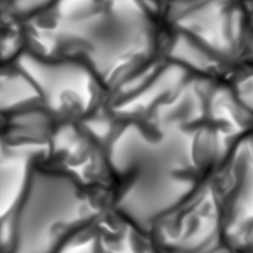}
\caption{}
\label{fig:true_im_630}
\end{subfigure}
\caption{Synthetic ``true'' images. \textbf{(a)} 538~nm.
\textbf{(a)} 630~nm.}
\label{fig:true_im}
\end{minipage}
\qquad
\begin{minipage}[c]{0.45\linewidth}
\centering
\captionof{table}{Image sizes.}
\begin{tabular}[t]{ll}
\hline
\noalign{\vspace{1mm}}
Image type & $N$ \\
\hline
\noalign{\vspace{1mm}}
True image                 & 253 \\
Convolved with PSFs        & 576 \\
MFBD processing            & 384 \\
RMS contrast calculations  & 253 \\
Power spectra              & 254 \\
\hline
\end{tabular}
\tablefoot{Images of size $N\times N$ pixels are used in the
different processing and evaluation steps.}
\label{tab:sizes}
\end{minipage}
\\[7mm]
\end{figure*}

\begin{figure*}[!thb]
\centering
\begin{subfigure}{0.45\linewidth}
\includegraphics[bb=54 46 700 529,width=0.9\linewidth,clip]{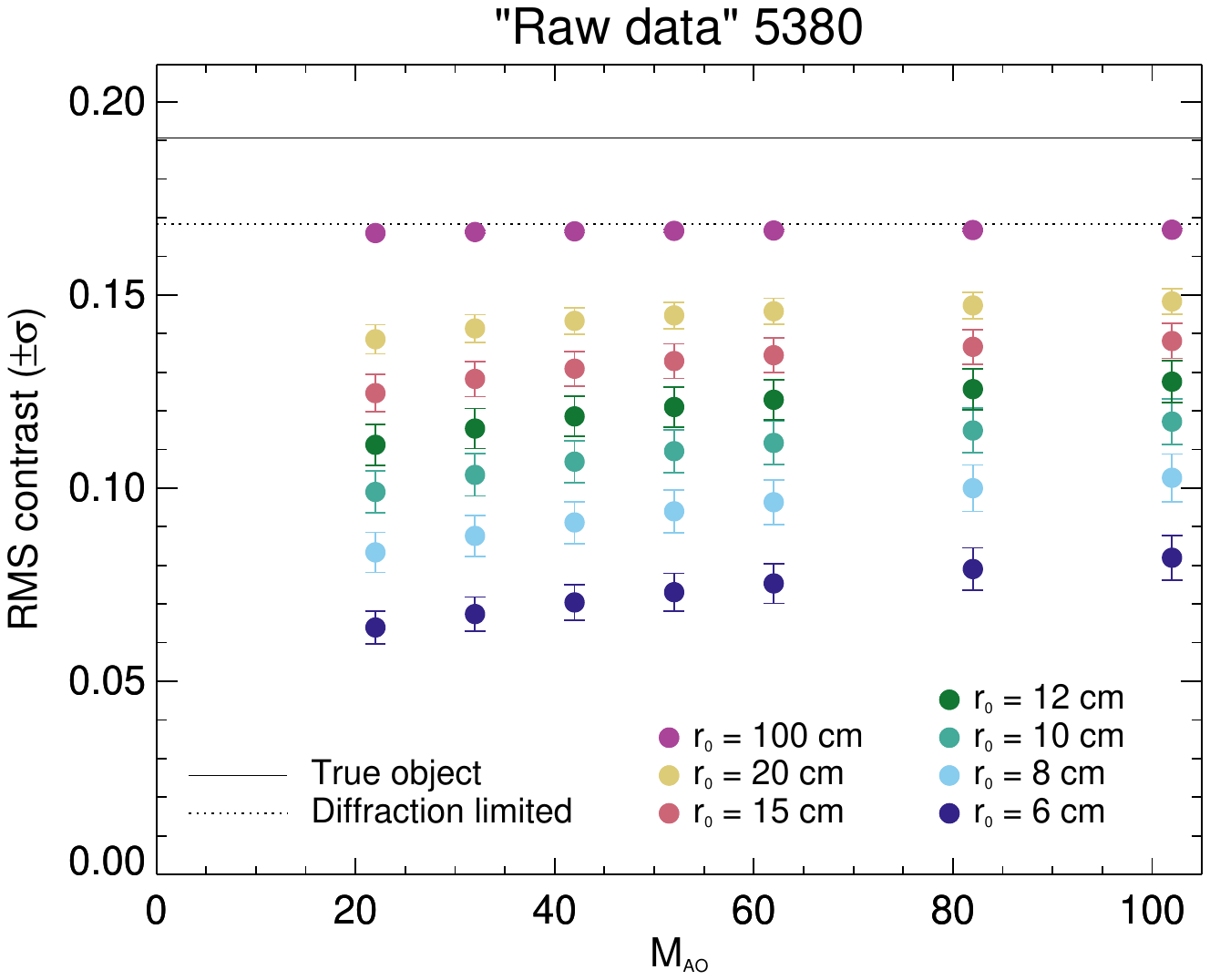}
\caption{\label{fig:contrast_artficial_data_5380_50}}
\end{subfigure}
\hfil
\begin{subfigure}{0.45\linewidth}
\includegraphics[bb=54 46 700 529,width=0.9\linewidth,clip]{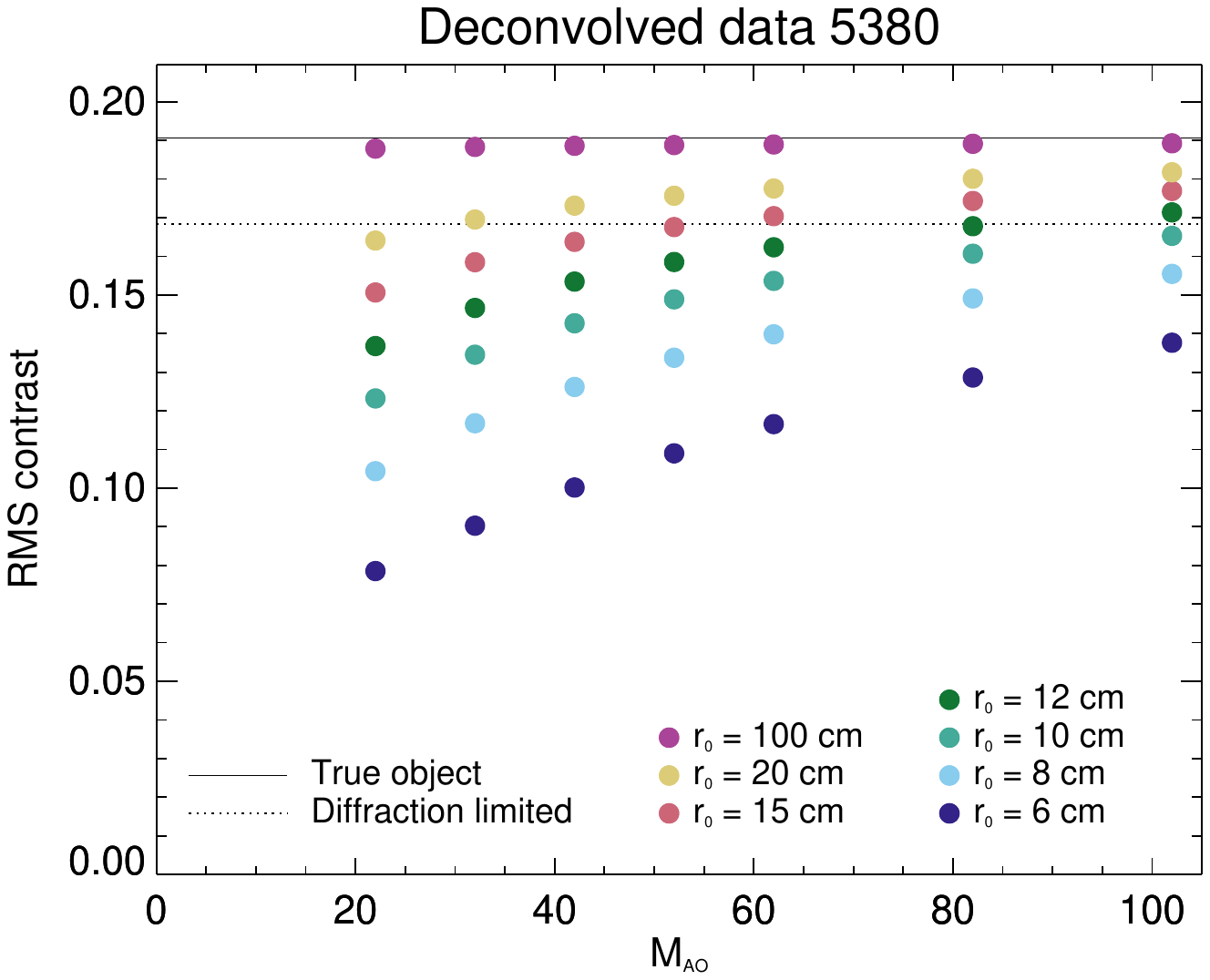}
\caption{\label{fig:contrast_artficial_data_5380_deconvolved}}
\end{subfigure}
\caption{Contrasts for synthetic 538~nm granulation data.
\textbf{(a)} With 50\% AO correction of the $\Mao$ most
significant KL modes. The error bars denote the standard deviation
in each set of images. \textbf{(b)} Multiframe-deconvolved with
the correct values of the coefficients for the $\Mao$ most
significant KL modes, leaving $\tau_{j\Mao}$ uncompensated for.}
\label{fig:contrast_artficial_data_5380}
\end{figure*}

We used IDL\footnote{Interactive Data Language, Harris Geospatial
Solutions, Inc.} code to simulate the image formation of a telescope
with a circular aperture of  97cm diameter with no central obscuration
(like the SST).

\subsection{Wavefronts}
\label{sec:wavefronts}

When making synthetic data for our simulation experiments, we have to
limit the second sum of Eq.~(\ref{eq:5}) at some point, $\Msim$. For
the work presented here, we chose to use 1000 modes with curvature,
which corresponds to the approximation $\Msim = 1002\approx\infty$
with the tip-tilt coefficients set to zero. We note that the truncation
at $\Msim=1002$ modes causes a loss of Strehl ratio (see
Fig.~\ref{fig:wf-and-tails}) but for our purposes it should be
insignificant compared to the Strehls for $M\le102$. With
$\Msim \gg \Mao $, we believe this gives us enough of a tail of
uncorrected modes to reproduce the problem expected with real data.

We simulated 100 wavefronts by drawing 1000 KL
coefficients per wavefront from a normal distribution $\mathcal{N}(0,\sigma_m)$ and
scaling them to different $r_0$ values by multiplication with
$(D/r_0)^{5/6}$ (equivalent to Eq.~(\ref{eq:6})), where $r_0 \in \{6$,
8, 10, 12, 15, 20, $100\}$~cm. The 50\% AO correction was simulated by
multiplying the $\Mao$ first modes with 0.5 for 50\% correction, with
$\Mao \in \{22$, 32, 42, 52, 62, 82, $102\}$. We also made simulated
data with no AO correction.

The pupils and modes were generated with DLMs from the same code base
that generates the modes used in Redux (see Sect.~\ref{sec:software}).
This ensures consistency in making and processing the data.

\subsection{Image data}
\label{sec:image-data}

The ``true'' 253$\times$253-pixel images (see Fig.~\ref{fig:true_im})
correspond to a FOV of 6000~km (or 8\farcs2) squared with an image
scale of 0\farcs0323/pixel. They are 538 and 630~nm snapshots from the
same MHD simulation that \citet{scharmer10high-order} used but sampled
at different time coordinates.

The simulated ``raw'' image data were calculated by convolving the
``true'' images with PSFs corresponding to $\phi_j$ for the different
$\Mao$ and $r_0$ values simulated. The wavefronts were used to make
PSFs at nominal focus as well as for a FD with a  PTV of one~wave.
In addition to  synthetic granulation images, we stored the PSFs to
be used as point-source data. We did not include noise or
anisoplanatism in the simulations.

We wanted oversize images to minimize wrap-around errors with wide
PSF wings, and also to not have the same size for deconvolutions as
for convolutions. The ``true'' images are periodic, and so we repeated the
FOV in both directions and cropped to a size with small prime factors
for efficient FFT operations. We then cropped further before saving
images to FITS files. The MFBD image size was even smaller. The
relevant sizes are summarized in Table~\ref{tab:sizes}. Because the
``true'' FOV is an odd number of pixels, we added one row and column
when calculating power spectra.

\begin{figure*}[!t]
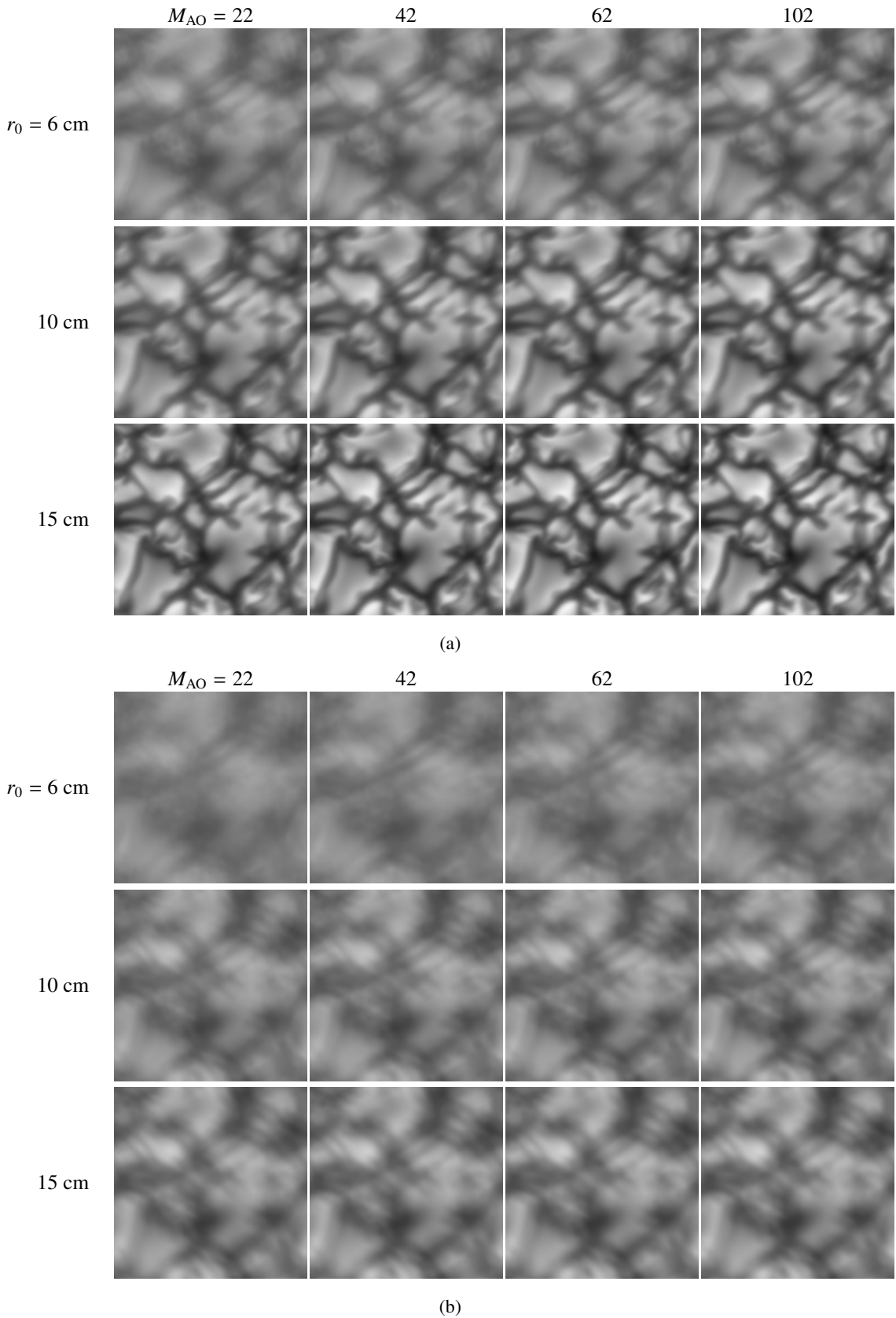

\centering
\begin{subfigure}{\linewidth}
\centering
\def\incgr#1#2{\includegraphics[width=\imwidth]{figs_raw_imf_0000_5380_r0=#1cm_Mao=Mproc=#2_Fao=05}}
\begin{tabular}[t]{rc@{\,}c@{\,}c@{\,}c}
& $\Mao=22$  & 42 & 62  & 102 \\
\raisebox{15mm}{$r_0=$ 6 cm} &\incgr{006}{20}&\incgr{006}{40}&\incgr{006}{60}&\incgr{006}{100} \\
\raisebox{15mm}{10 cm} &\incgr{010}{20}&\incgr{010}{40}&\incgr{010}{60}&\incgr{010}{100} \\
\raisebox{15mm}{15 cm} &\incgr{015}{20}&\incgr{015}{40}&\incgr{015}{60}&\incgr{015}{100} \\
\end{tabular}
\caption{}
\label{fig:raw_im_focus}
\end{subfigure}\\[2mm]
\begin{subfigure}{\linewidth}
\centering
\def\incgr#1#2{\includegraphics[width=\imwidth]{figs_raw_imd_0000_5380_r0=#1cm_Mao=Mproc=#2_Fao=05}}
\begin{tabular}[t]{rc@{\,}c@{\,}c@{\,}c}
& $\Mao=22$ & 42 & 62  & 102 \\
\raisebox{15mm}{$r_0=$ 6 cm} &\incgr{006}{20}&\incgr{006}{40}&\incgr{006}{60}&\incgr{006}{100} \\
\raisebox{15mm}{10 cm} &\incgr{010}{20}&\incgr{010}{40}&\incgr{010}{60}&\incgr{010}{100} \\
\raisebox{15mm}{15 cm} &\incgr{015}{20}&\incgr{015}{40}&\incgr{015}{60}&\incgr{015}{100} \\
\end{tabular}
\caption{}
\label{fig:raw_im_diversity}
\end{subfigure}
\caption{Sample synthetic ``raw'' images with 50\% AO correction up to
mode $\Mao$. Images degraded by first wavefront of a set of 100
random wavefronts. Grayscale as for the ``true'' image in
Fig.~\ref{fig:true_im}. \textbf{(a)} Conventional in-focus images,
\textbf{(b)} FD images.}
\label{fig:raw_im}
\end{figure*}

We processed both 538~nm and 630~nm images for the initial experiment
(Sect.~\ref{sec:constant-r_0}), but as the results were almost
identical, in the following we process only 538~nm data. We show and
discuss only 538~nm data here and in the following sections, except
when explicitly stating otherwise.

The contrasts of these synthetic ``raw''   50\% AO-corrected
conventional focus images are shown in
Fig.~\ref{fig:contrast_artficial_data_5380_50} for varying numbers of
AO-corrected modes. We note the $r_0=100$~cm images approximately reach
the dotted line that represents the contrast of the true image
convolved with the diffraction-limited PSF (the PSF of the theoretical
97~cm aperture without aberrations).

Deconvolution of these data with PSFs based on only the low-order
parts of the wavefronts, $\phi_{j\Mao}$, does not restore the full
contrast except for the $r_0=100$~cm data, as shown in
Fig.~\ref{fig:contrast_artficial_data_5380_deconvolved}. We also made
FD data with a PTV of one wave. Sample images (first wavefront) are
shown in Figs.~\ref{fig:raw_im}. These images illustrate the fact that
the information about the wavefronts manifests itself much more
clearly some distance from focus than in the focal plane.

\section{Processing and results}
\label{sec:processing}

\subsection{Synthetic data with constant $r_0$}
\label{sec:constant-r_0}

\subsubsection{Contrast and error metric}
\label{sec:contr-error-metr}

We MFBD-processed the 50\% AO-corrected granulation and point source
data, with and without FD as well as with and without SD. Adaptive
optics is not capable of completely correcting all the $\Mao$ modes,
and so we need $\Mmfbd\ge \Mao$ to give MFBD the chance to provide
total correction of a known number of modes. This is necessary for the
statistical correction of the higher order modes to work properly. In
general, $\Mmfbd > \Mao$ gave poorer results than $\Mmfbd=\Mao$, and
the granulation contrast grows seemingly without limit above that of
the ``true'' image. We tried $\Mmfbd = 202$, 302, 402, 502, and 602.
This could probably be addressed with a regularization term in the
error metric, penalizing large variation in the wavefront phase.
However, using $\Mmfbd=\Mao$ seems appropriate, and so we base the
remainder of this paper on that.

In the same way that we limit figures to only 538~nm data, we show
results only for subsets of the $r_0$ and $\Mao$ datasets generated
and processed. Showing results for all combinations of $r_0$ and
$\Mao$ leads to figures that are too many or too cluttered, or both.
However, we inspected all the output and the subsets shown are
representative of the entirety of the results.

\begin{figure}[!tbh]
\centering
\includegraphics[bb=54 46 700 529,width=0.8\linewidth,clip]{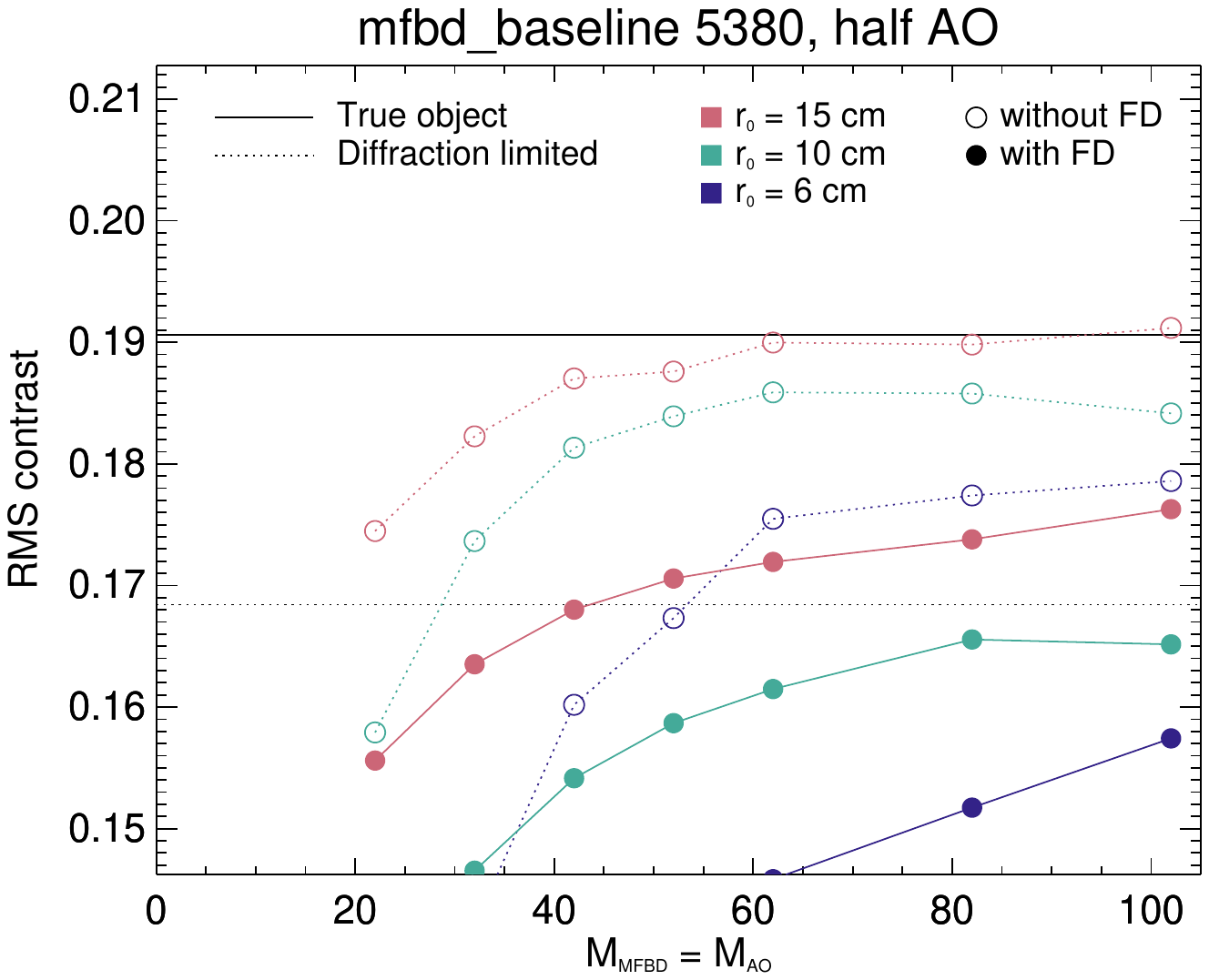}
\caption{MFBD without SD. Restored granulation contrasts.}
\label{fig:contrast_withouttails_Mproc=Mao}
\end{figure}

MFBD processing the synthetic data yields contrasts shown in
Figure~\ref{fig:contrast_withouttails_Mproc=Mao}. The contrasts are
enhanced compared to deconvolution with the correct low-order
coefficients (compare
Fig.~\ref{fig:contrast_artficial_data_5380_deconvolved}), in
particular with larger $r_0$ and with fewer modes corrected by AO and
included in the fit ($\Mao=\Mmfbd$). This enhancement is particularly
strong without FD (open symbols). However, FD (filled symbols), while
constraining the solutions and thereby reducing the contrast
enhancement, does not remove it completely.

\begin{figure}[!tbh]
\centering
\includegraphics[bb=54 46 700 529,width=0.8\linewidth,clip]{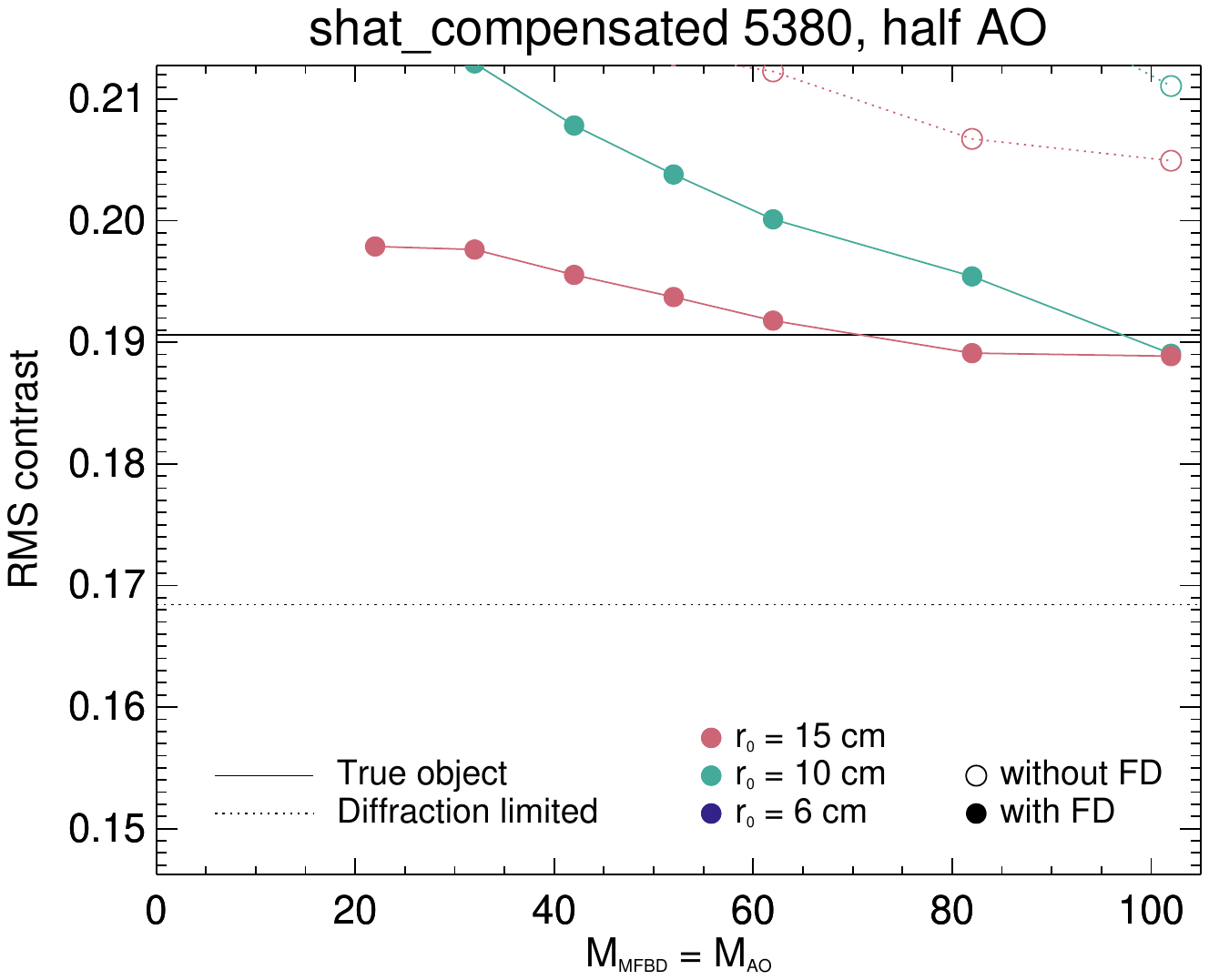}
\caption{MFBD without SD followed by post-restoration compensation
for $\tau_\Mmfbd$. Restored granulation contrasts.}
\label{fig:contrast_Mproc=Mao_shat}
\end{figure}

\citet{scharmer10high-order} did post-restoration compensation for the
uncorrected tail of high-order modes, $\tau_\Mmfbd$, (see also
Sect.~\ref{sec:post-rest-corr}) but only for MFBD without FD. Our
results with that method are shown in
Fig.~\ref{fig:contrast_Mproc=Mao_shat}. Like for
\citet{scharmer10high-order}, our  data compensated in the same way have too much
contrast for MFBD without FD. This is expected due to the
overcompensation relative to the number of included modes mentioned in
the previous paragraph. Only in the best seeing conditions is the
contrast almost correct. The 6cm contrasts are too over-corrected to
even fit in the plot window. Post-correcting data restored with FD
gives better results but there is still mostly too much contrast, in particular
with fewer modes.

\begin{figure}[!tbh]
\centering
\includegraphics[bb=54 46 700 529,width=0.8\linewidth,clip]{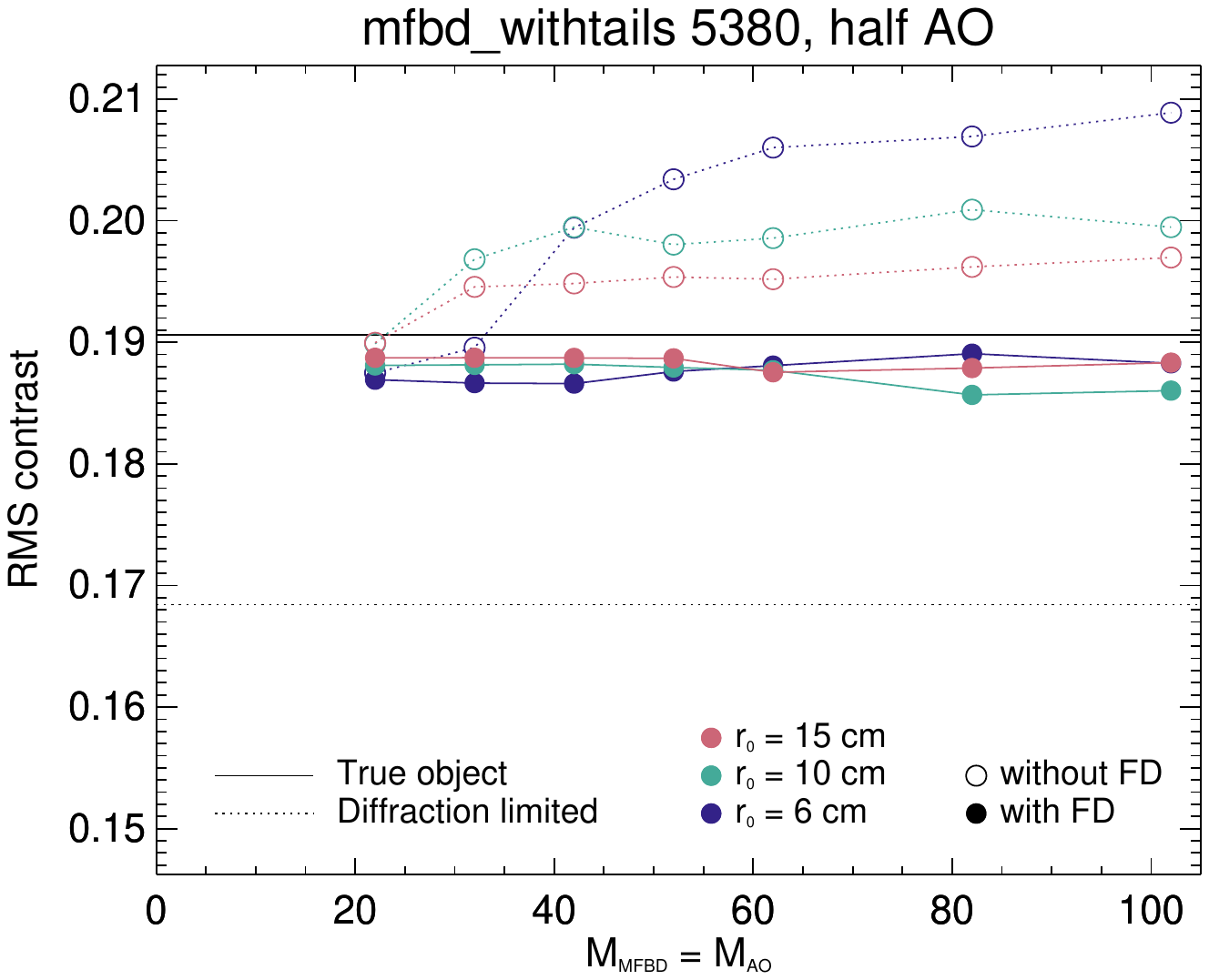}
\caption{MFBD with SD. Restored granulation contrasts.}
\label{fig:contrast_withtails_Mproc=Mao}
\end{figure}

Figure~\ref{fig:contrast_withtails_Mproc=Mao} shows the corresponding
results with SD. We note that for real data we do not know the correct
values of the high-order coefficients. We were therefore careful to use
different random coefficients from those used when making the data.

With both FD and SD (filled symbols), the ``true'' contrast is
recovered very well, and consistently with varying number of modes as
well as varying $r_0$. This is a very encouraging result.

Without FD (open symbols), the contrasts are too high when the number
of modes is large but fairly good when the number of modes is small.
With fewer modes, most of the correction is done with the fixed
$\tilde\tau_{j\Mmfbd}$ contributions and when the number of fitted
parameters is small, it is easier for the MFBD algorithm to converge
to correct values. It is possible that the results for the larger
$\Mmfbd$ can be improved by regularization but this is outside the
scope of this paper. We believe these results clearly demonstrate the need to
collect data with FD if the SD technique is to be used.

\begin{figure*}[!t]
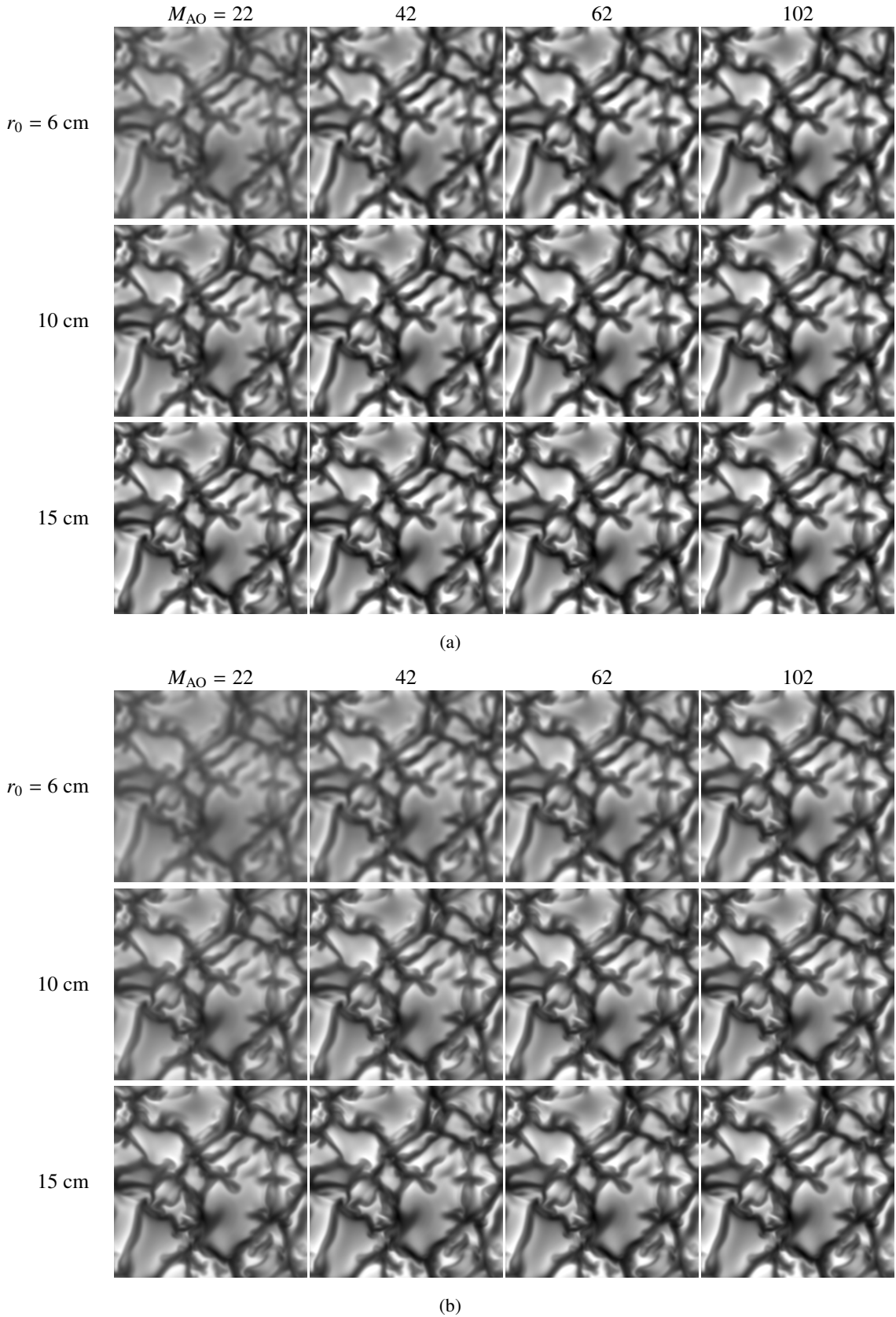

\centering
\begin{subfigure}{\linewidth}
\centering
\def\incgr#1#2{\includegraphics[width=\imwidth]{figs_baseline_im_mfbd_5380_r0=#1cm_Mao=Mproc=#2_Fao=05}}
\begin{tabular}[t]{rc@{\,}c@{\,}c@{\,}c}
& $\Mao=22$ & 42 & 62  & 102 \\
\raisebox{15mm}{$r_0=$ 6 cm} &\incgr{006}{20}&\incgr{006}{40}&\incgr{006}{60}&\incgr{006}{100} \\
\raisebox{15mm}{10 cm} &\incgr{010}{20}&\incgr{010}{40}&\incgr{010}{60}&\incgr{010}{100} \\
\raisebox{15mm}{15 cm} &\incgr{015}{20}&\incgr{015}{40}&\incgr{015}{60}&\incgr{015}{100} \\
\end{tabular}
\caption{}
\label{fig:5380_mfbd_baseline_images_Mproc=Mao}
\end{subfigure}\\[2mm]
\begin{subfigure}{\linewidth}
\centering
\def\incgr#1#2{\includegraphics[width=\imwidth]{figs_baseline_im_jpds_5380_r0=#1cm_Mao=Mproc=#2_Fao=05}}
\begin{tabular}[t]{rc@{\,}c@{\,}c@{\,}c}
& $\Mao=22$ & 42 & 62  & 102 \\
\raisebox{15mm}{$r_0=$ 6 cm} &\incgr{006}{20}&\incgr{006}{40}&\incgr{006}{60}&\incgr{006}{100} \\
\raisebox{15mm}{10 cm} &\incgr{010}{20}&\incgr{010}{40}&\incgr{010}{60}&\incgr{010}{100} \\
\raisebox{15mm}{15 cm} &\incgr{015}{20}&\incgr{015}{40}&\incgr{015}{60}&\incgr{015}{100} \\
\end{tabular}
\caption{}
\label{fig:5380_jpds_baseline_images_Mproc=Mao}
\end{subfigure}
\caption{MFBD without SD, sample restored images. \textbf{(a)}
Without FD. \textbf{(b)} With FD.}
\label{fig:5380_baseline_images_Mproc=Mao}
\end{figure*}

We show MFBD-restored images in
Figs.~\ref{fig:5380_mfbd_baseline_images_Mproc=Mao} (processed without
FD) and \ref{fig:5380_jpds_baseline_images_Mproc=Mao} (with FD). The
images are all scaled to the same grayscale range as the ``true''
image in Fig.~\ref{fig:true_im} and we can see a slight dimming when
smaller numbers of modes were corrected, as well as for smaller $r_0$,
in particular for data processed with FD. The apparent resolution also
suffers slightly in the dimmer images, particularly for the smaller
$r_0$ (see the thinnest features in the intergranular lanes and the
pointy granule in the bottom-left quadrant).

Sample images restored with SD are shown in
Figs.~\ref{fig:5380_mfbd_withtails_images_Mproc=Mao} (without FD) and
\ref{fig:5380_jpds_withtails_images_Mproc=Mao} (with FD). The
improvement in the restoration of the correct contrasts is visible as
less dimming and  more uniform resolution.

\begin{figure*}[!t]
\centering
\begin{subfigure}{\linewidth}
\centering
\def\incgr#1#2{\includegraphics[width=\imwidth]{figs_withtails_im_mfbd_5380_r0=#1cm_Mao=Mproc=#2_Fao=05}}
\begin{tabular}[t]{rc@{\,}c@{\,}c@{\,}c}
& $\Mao=22$ & 42 & 62  & 102 \\
\raisebox{15mm}{$r_0=$ 6 cm} &\incgr{006}{20}&\incgr{006}{40}&\incgr{006}{60}&\incgr{006}{100} \\
\raisebox{15mm}{10 cm} &\incgr{010}{20}&\incgr{010}{40}&\incgr{010}{60}&\incgr{010}{100} \\
\raisebox{15mm}{15 cm} &\incgr{015}{20}&\incgr{015}{40}&\incgr{015}{60}&\incgr{015}{100} \\
\end{tabular}
\caption{}
\label{fig:5380_mfbd_withtails_images_Mproc=Mao}
\end{subfigure}\\[2mm]
\begin{subfigure}{\linewidth}
\centering
\def\incgr#1#2{\includegraphics[width=\imwidth]{figs_withtails_im_jpds_5380_r0=#1cm_Mao=Mproc=#2_Fao=05}}
\begin{tabular}[t]{rc@{\,}c@{\,}c@{\,}c}
& $\Mao=22$ & 42 & 62  & 102 \\
\raisebox{15mm}{$r_0=$ 6 cm} &\incgr{006}{20}&\incgr{006}{40}&\incgr{006}{60}&\incgr{006}{100} \\
\raisebox{15mm}{10 cm} &\incgr{010}{20}&\incgr{010}{40}&\incgr{010}{60}&\incgr{010}{100} \\
\raisebox{15mm}{15 cm} &\incgr{015}{20}&\incgr{015}{40}&\incgr{015}{60}&\incgr{015}{100} \\
\end{tabular}
\caption{}
\label{fig:5380_jpds_withtails_images_Mproc=Mao}
\end{subfigure}
\caption{Sample restored images using MFBD with SD. \textbf{(a)} Without
FD. \textbf{(b)} With FD.}
\label{fig:5380_withtails_images_Mproc=Mao}
\end{figure*}

\begin{figure*}[!t]
\centering
\begin{subfigure}{0.45\linewidth}
\includegraphics[bb=54 46 705 529,width=0.9\linewidth,clip]{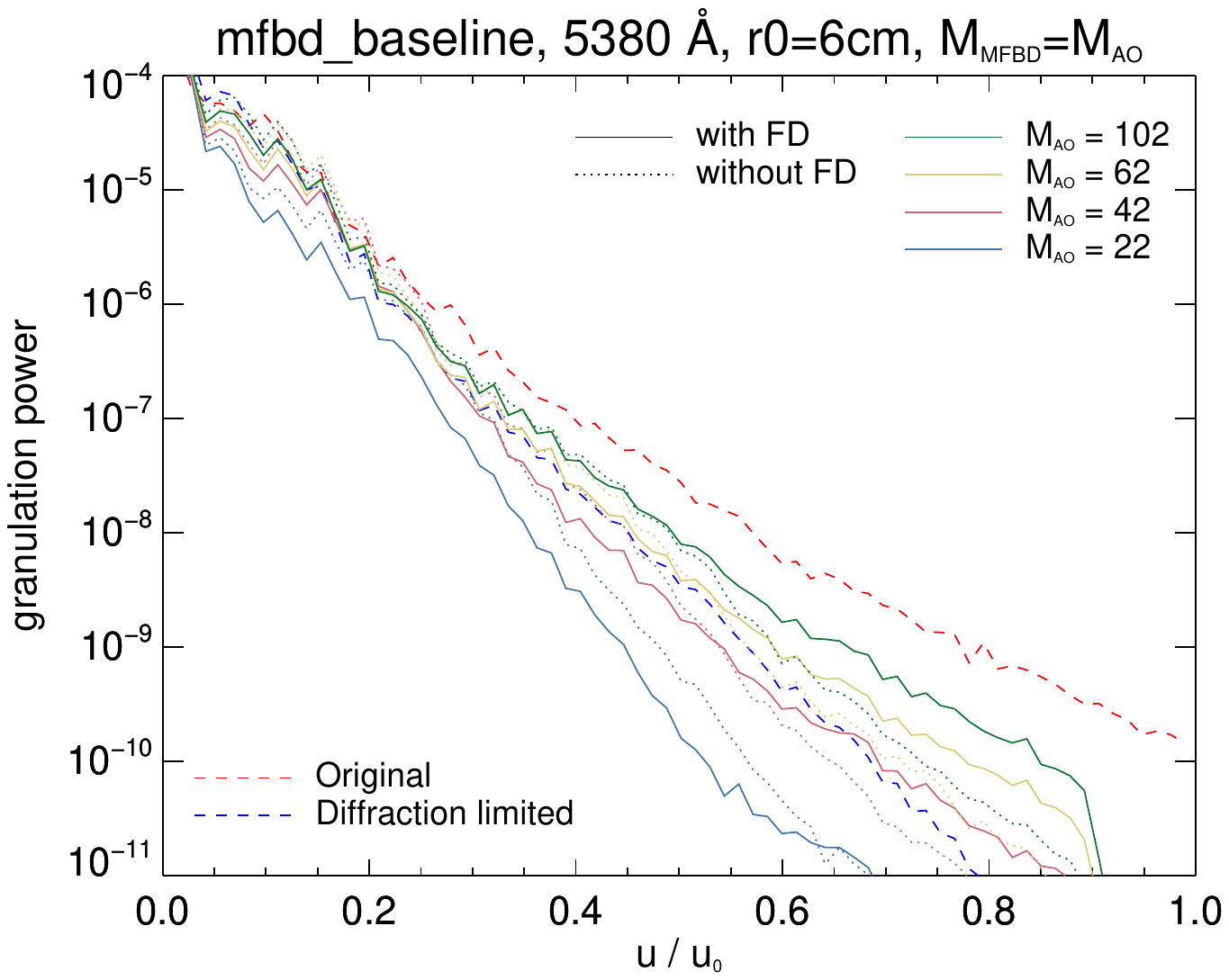}
\caption{}
\end{subfigure}
\hfil
\begin{subfigure}{0.45\linewidth}
\includegraphics[bb=54 46 705 529,width=0.9\linewidth,clip]{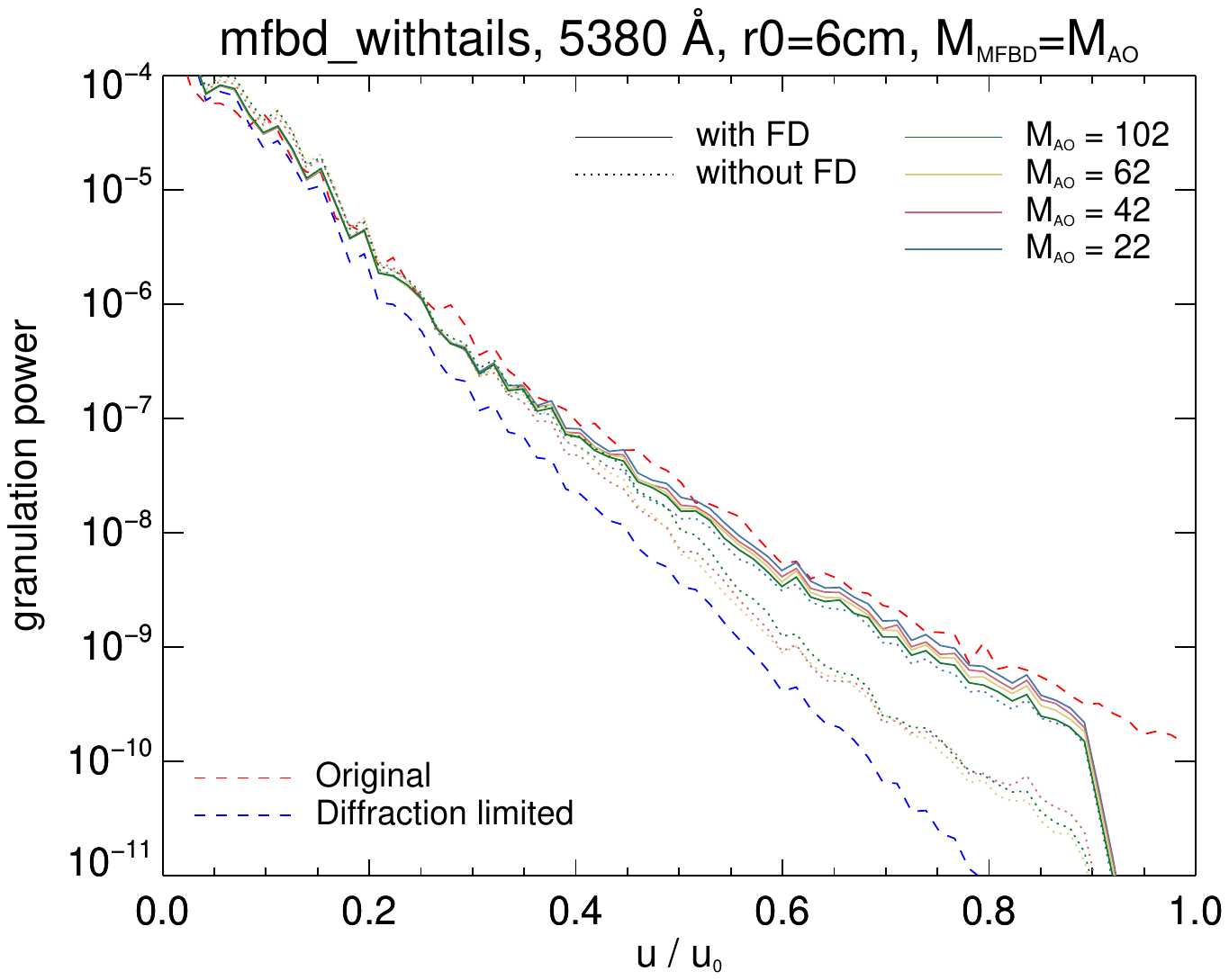}
\caption{}
\end{subfigure}
\caption{Granulation power spectra for $r_0=6$~cm. $\Mmfbd=\Mao$.
\textbf{(a)} Without SD \textbf{(b)} With SD.}
\label{fig:power_5380_6cm_Mproc=Mao}
\vspace{3mm}
\centering
\begin{subfigure}{0.45\linewidth}
\includegraphics[bb=54 46 705 529,width=0.9\linewidth,clip]{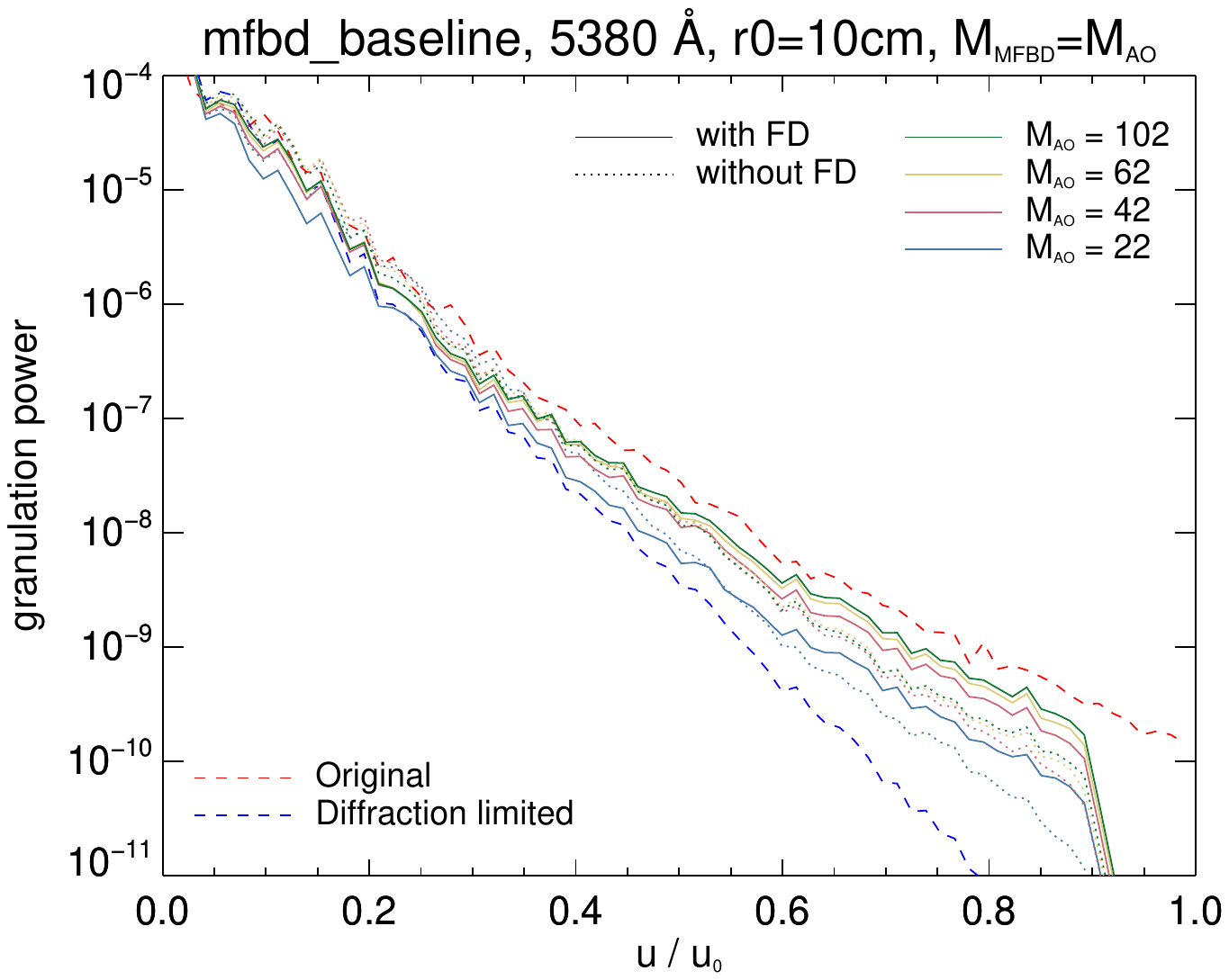}
\caption{}
\end{subfigure}
\hfil
\begin{subfigure}{0.45\linewidth}
\includegraphics[bb=54 46 705 529,width=0.9\linewidth,clip]{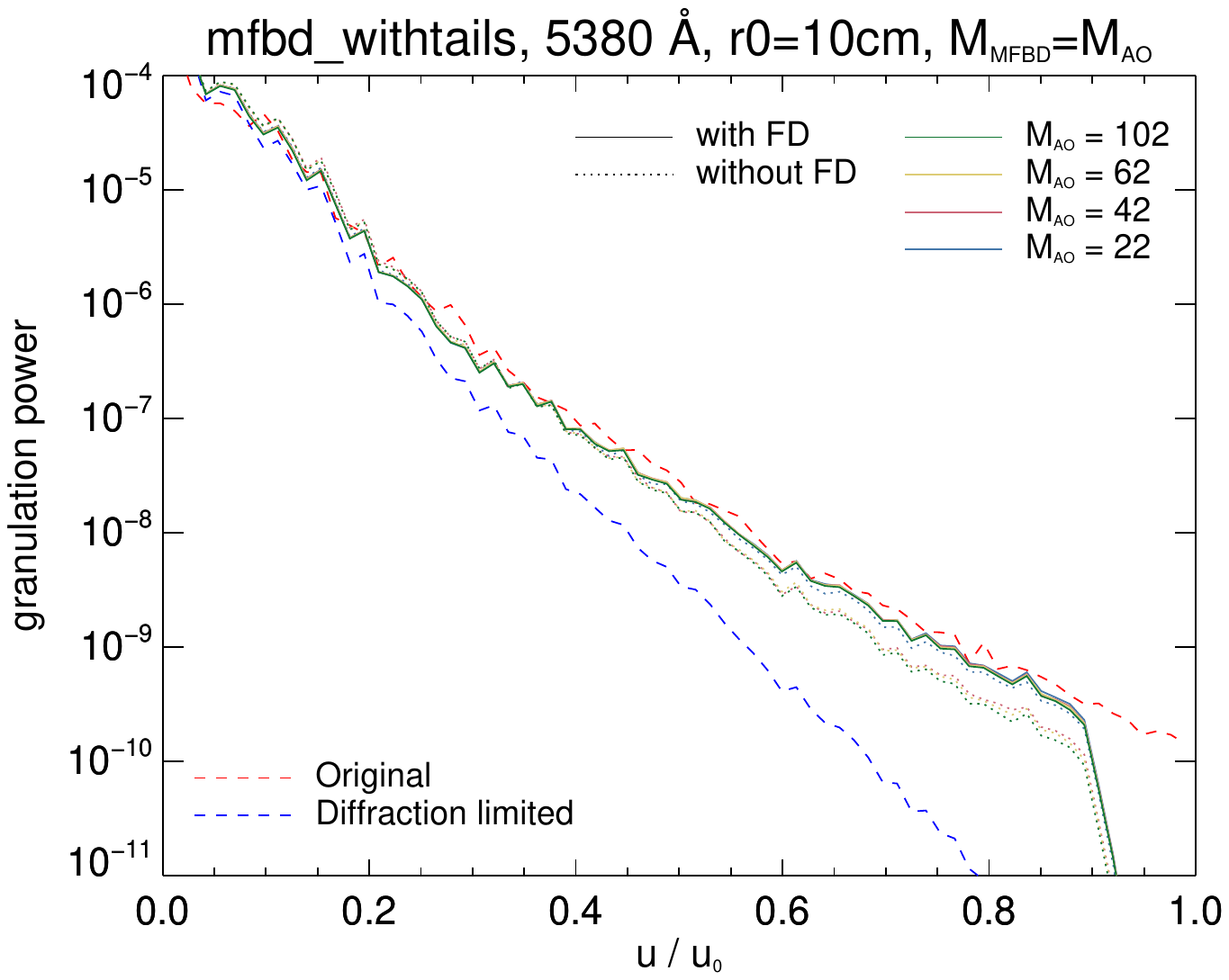}
\caption{}
\end{subfigure}
\caption{Granulation power spectra for $r_0=10$~cm. $\Mmfbd=\Mao$.
\textbf{(a)} Without SD \textbf{(b)} With SD.}
\label{fig:power_5380_10cm_Mproc=Mao}
\vspace{3mm}
\centering
\begin{subfigure}{0.45\linewidth}
\includegraphics[bb=54 46 705 529,width=0.9\linewidth,clip]{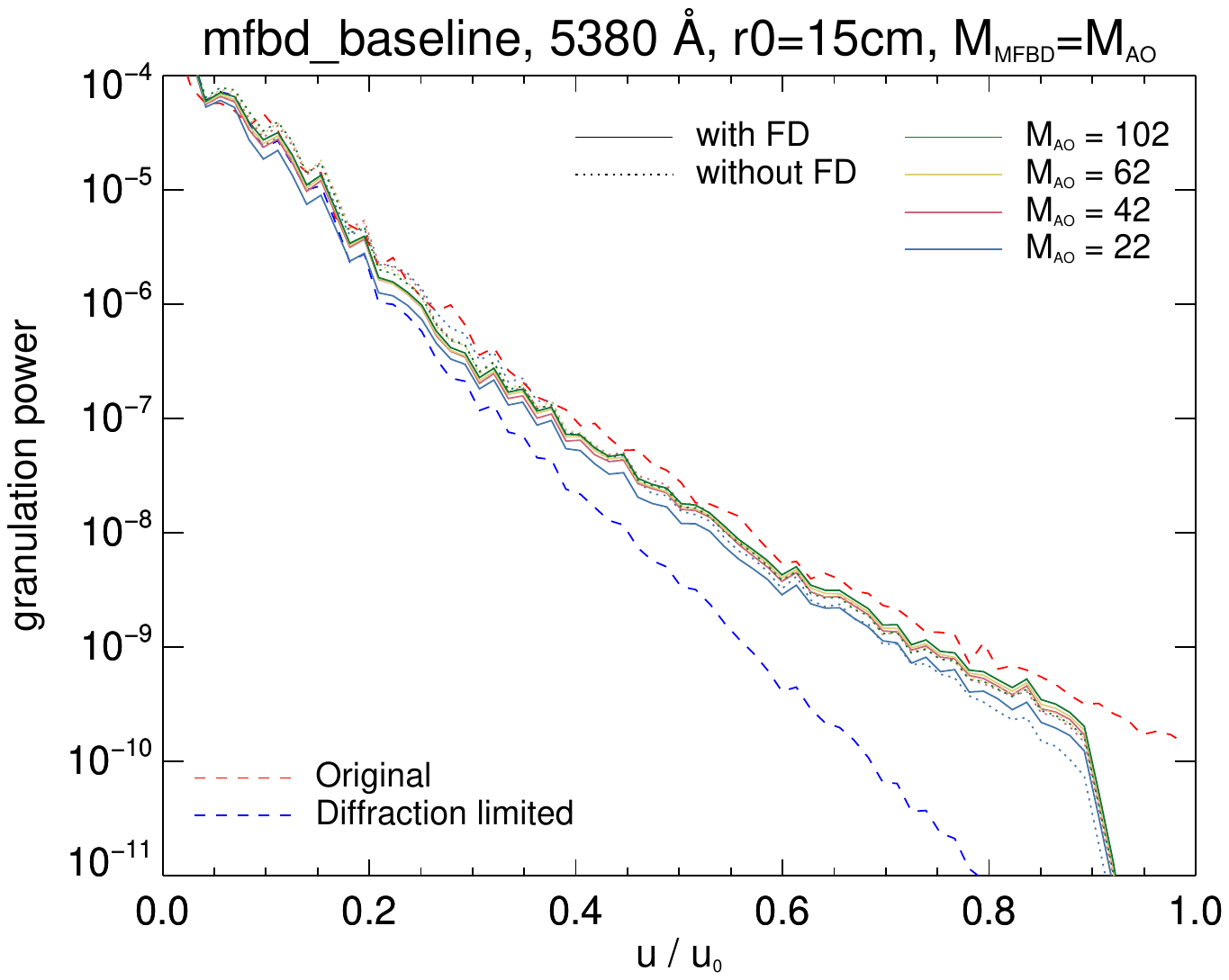}
\caption{}
\end{subfigure}
\hfil
\begin{subfigure}{0.45\linewidth}
\includegraphics[bb=54 46 705 529,width=0.9\linewidth,clip]{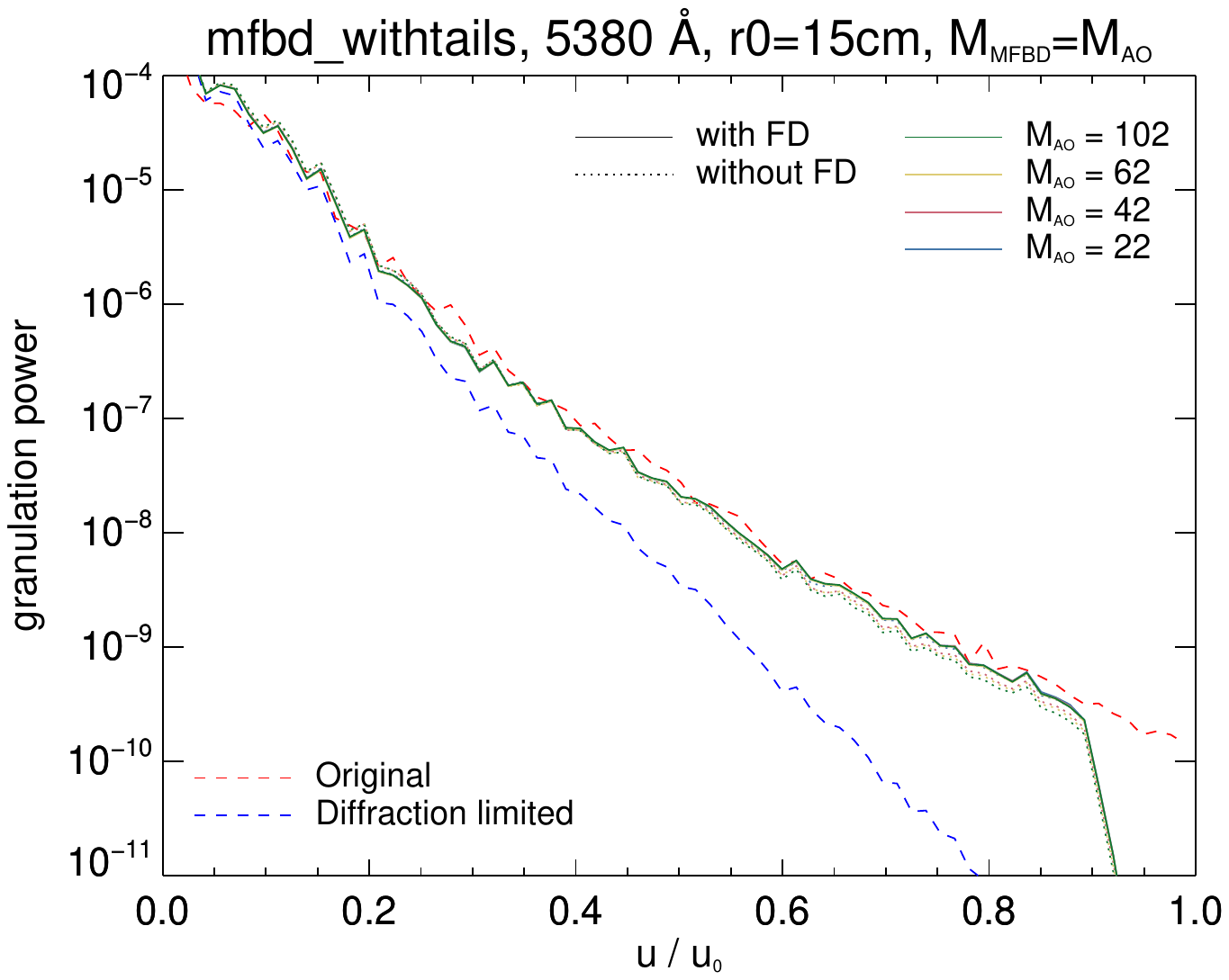}
\caption{}
\end{subfigure}
\caption{Granulation power spectra for $r_0=15$~cm. $\Mmfbd=\Mao$.
\textbf{(a)} Without SD \textbf{(b)} With SD.}
\label{fig:power_5380_15cm_Mproc=Mao}
\end{figure*}

Figures~\ref{fig:power_5380_6cm_Mproc=Mao}--\ref{fig:power_5380_15cm_Mproc=Mao}
show azimuthally averaged power spectra for restored granulation data.
The drop in power at $u/u_0=0.9$ (where $u_0$ is the diffraction
limit) is caused by the low-pass filter $H$ (see Eq.~(\ref{eq:F})),
which is implemented to avoid high-frequency artifacts in the
restoration of real data. With no noise, this filter would extend to
$u_0$ but by default our code limits the filter to
$0.9u_0$.

Without SD (the (a) subfigures), the power spectra are below that of
the true image, and more so with smaller $\Mao$, in particular for the
smaller $r_0$. FD generally helps, in particular in the higher spatial
frequencies, but less for the smaller $\Mao$.

Using SD leads to an overall improvement of the restored power (the (b)
subfigures), in particular together with FD. Combining the two
diversities (solid lines in (b)) brings the power spectra into close
agreement with the true data at all spatial frequencies while SD
without FD does that mostly for the lower spatial frequencies.

The over-correction of contrast without either FD or SD (dotted
lines, (a) subfigures) for the smaller $\Mao$ is caused by an excess
of power at low spatial frequencies. This is where most of the
contrast is formed, as the square of the RMS contrast is proportional
to the integral of the power.

\begin{figure*}[!tbp]
\centering
\begin{subfigure}{0.45\linewidth}
\includegraphics[bb=54 46 700 529,width=0.9\linewidth,clip]{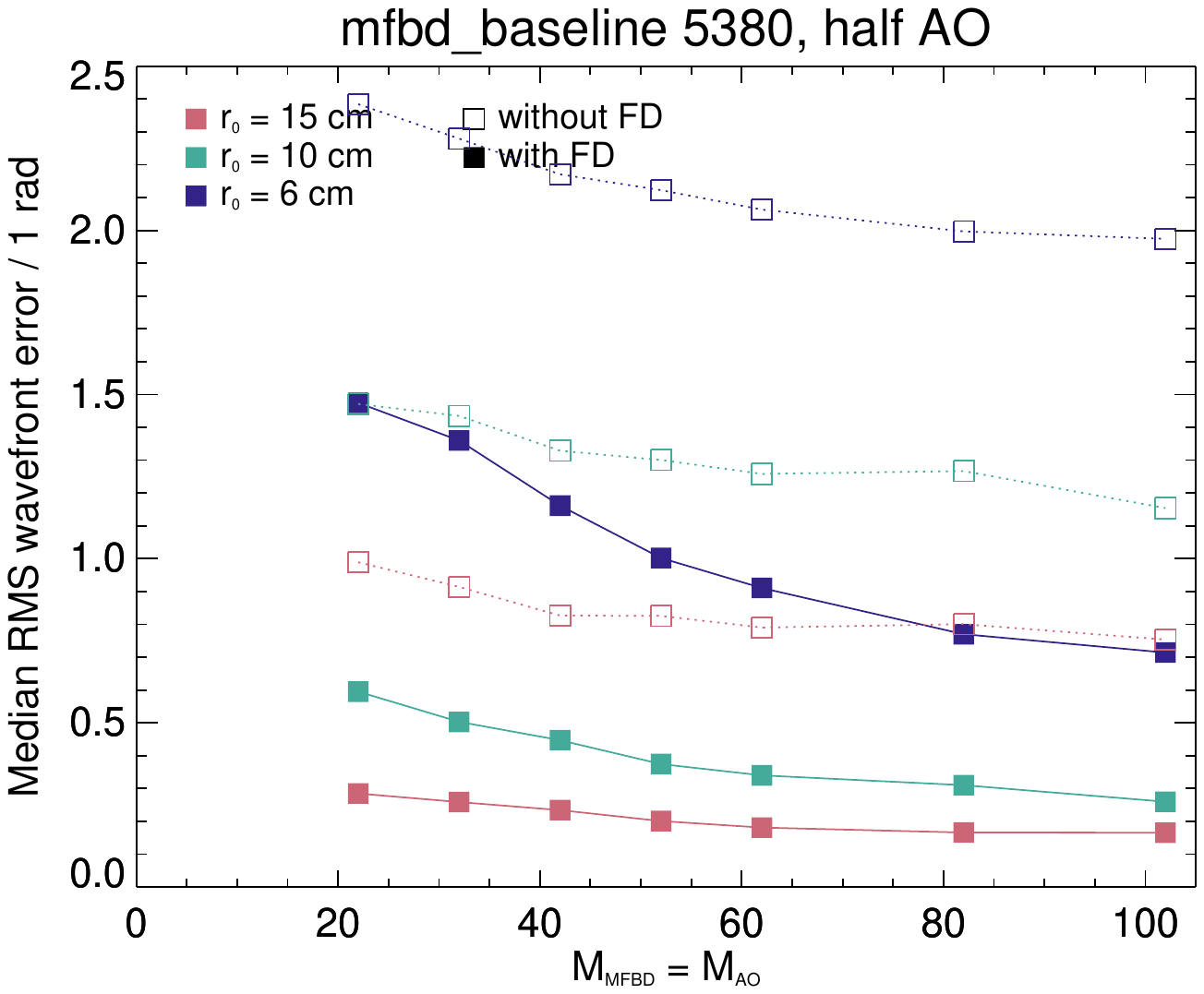}
\caption{}
\label{fig:granulation_wferror_withouttails_Mproc=Mao}
\end{subfigure}
\hfil
\begin{subfigure}{0.45\linewidth}
\includegraphics[bb=54 46 700 529,width=0.9\linewidth,clip]{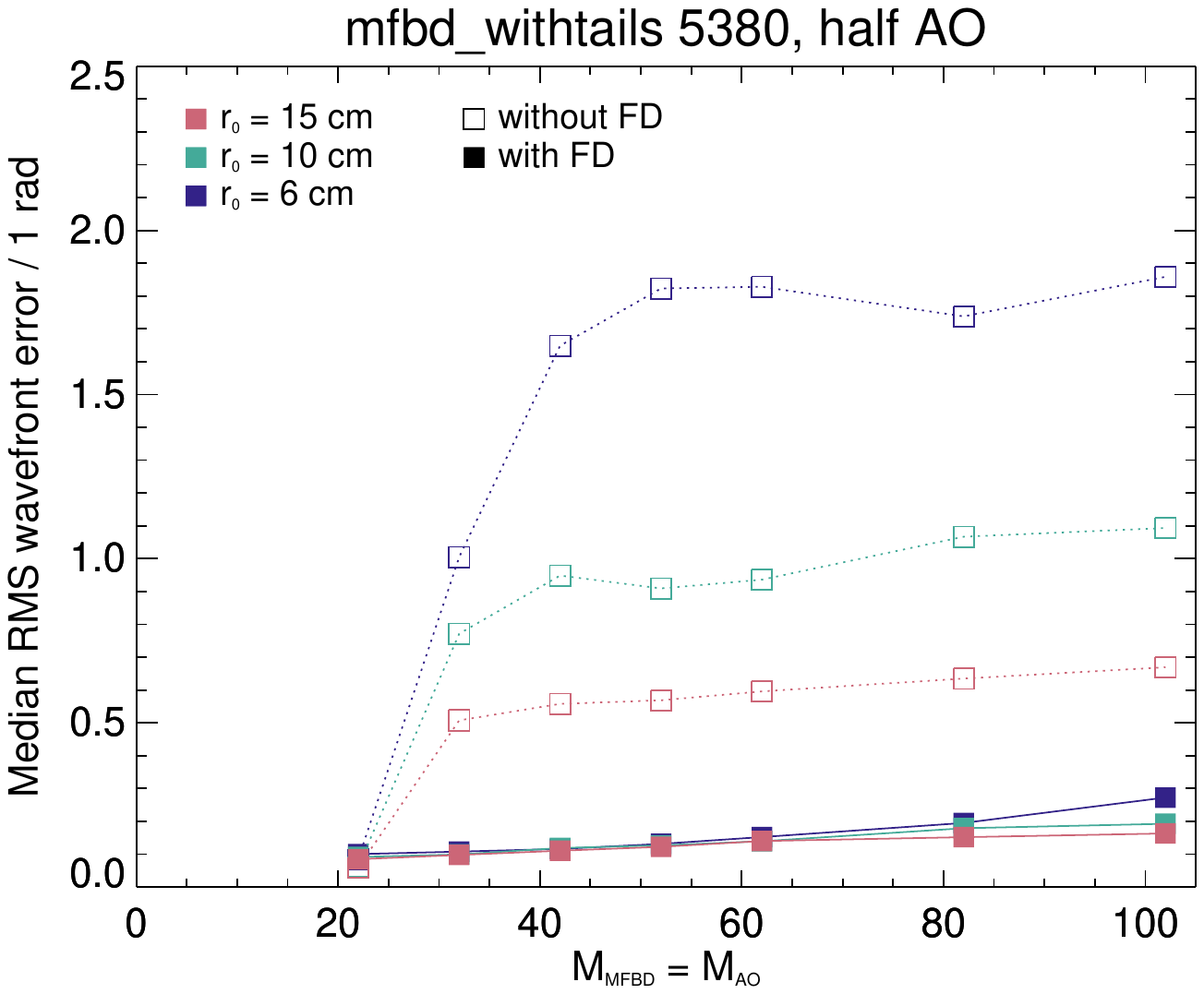}
\caption{}
\label{fig:granulation_wferror_withtails_Mproc=Mao}
\end{subfigure}
\caption{RMS wavefront error not including tip and tilt for
granulation data. \textbf{(a)} Without SD \textbf{(b)} With SD.}
\label{fig:granulation_wferror_Mproc=Mao}
\bigskip
\centering
\begin{subfigure}{0.45\linewidth}
\includegraphics[bb=54 46 700 529,width=0.9\linewidth,clip]{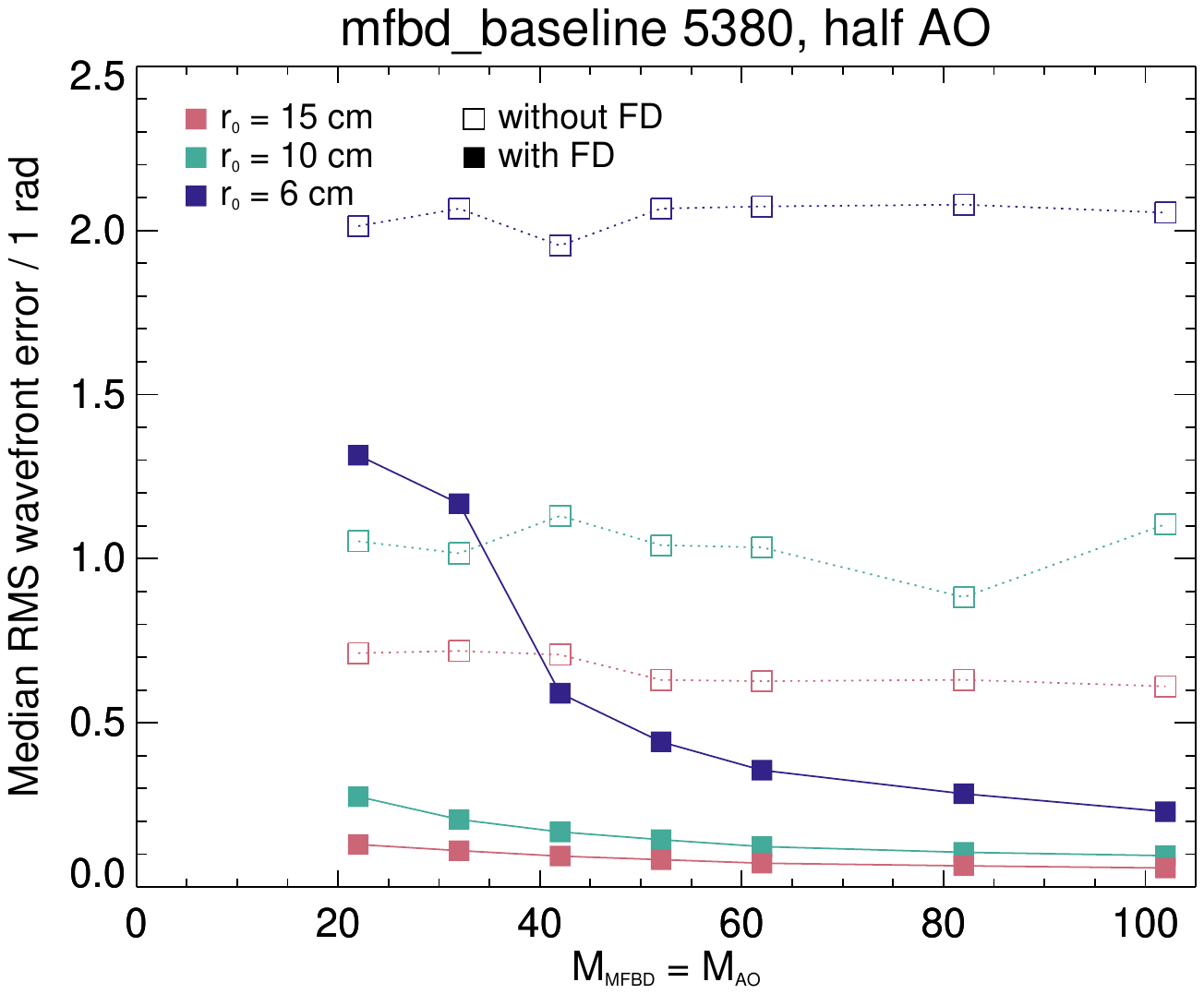}
\caption{}
\label{fig:pointsource_wferror_withouttails_Mproc=Mao}
\end{subfigure}
\hfil
\begin{subfigure}{0.45\linewidth}
\includegraphics[bb=54 46 700 529,width=0.9\linewidth,clip]{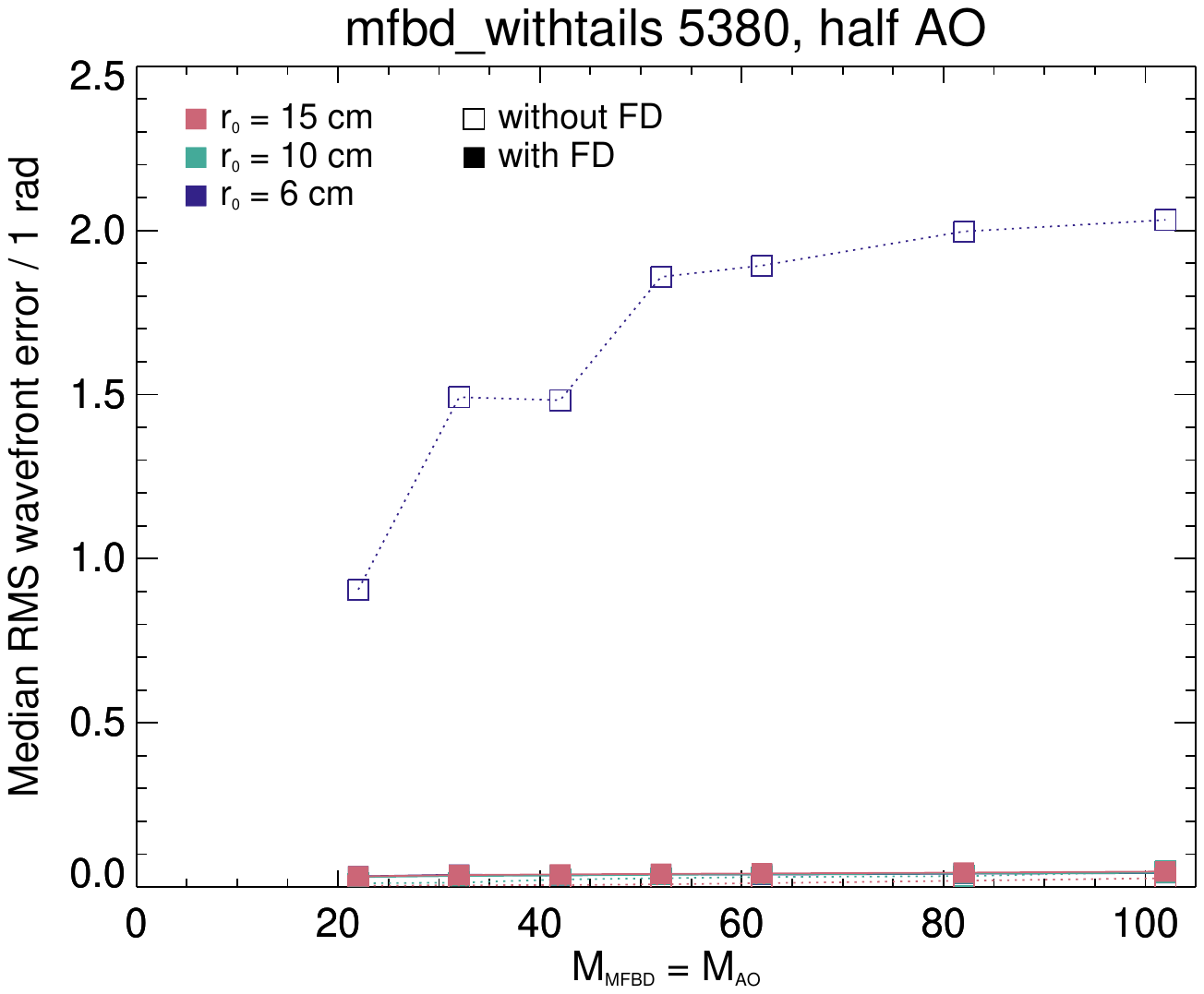}
\caption{}
\label{fig:pointsource_wferror_withtails_Mproc=Mao}
\end{subfigure}
\caption{RMS wavefront error not including tip and tilt for point
source data. \textbf{(a)} Without SD \textbf{(b)} With SD.}
\label{fig:pointsource_wferror_Mproc=Mao}
\bigskip
\bigskip
\end{figure*}

We show that SD improves the restoration of the images but
similar PSFs and OTFs can be made with quite different wavefronts. As
the MFBD model fit is evaluated by the metric $L$ in Eq.~(\ref{eq:2})
involving images and not wavefronts directly, it does not
automatically follow that there is an improvement in the estimated
wavefront coefficients. However, it does seem that SD improves that as
well. We show median wavefront errors in the modes with curvature for
the granulation data in Fig.~\ref{fig:granulation_wferror_Mproc=Mao}
and it is apparent that they are improved significantly when both FD
and SD are used, even for the smallest $r_0$. Without FD, a more
modest improvement can be seen for the smaller numbers of modes.

We also processed point source data sets with the same simulated
aberrations. The wavefront errors of those are shown in
Fig.~\ref{fig:pointsource_wferror_Mproc=Mao}. There are a few things
to note. First, the combination of SD and FD makes granulation a
better WFS than point sources with FD but without SD. With both FD and
SD, the point source wavefront errors are almost zero. And when the
point source data are processed with SD but without FD ((b) open
symbols), for most cases the errors are as small as with FD. We
discuss this further in Sect.~\ref{sec:local-minima}

\begin{figure*}[!t]
\centering
\begin{subfigure}{0.45\linewidth}
\includegraphics[bb=54 46 700 529,width=0.9\linewidth,clip]{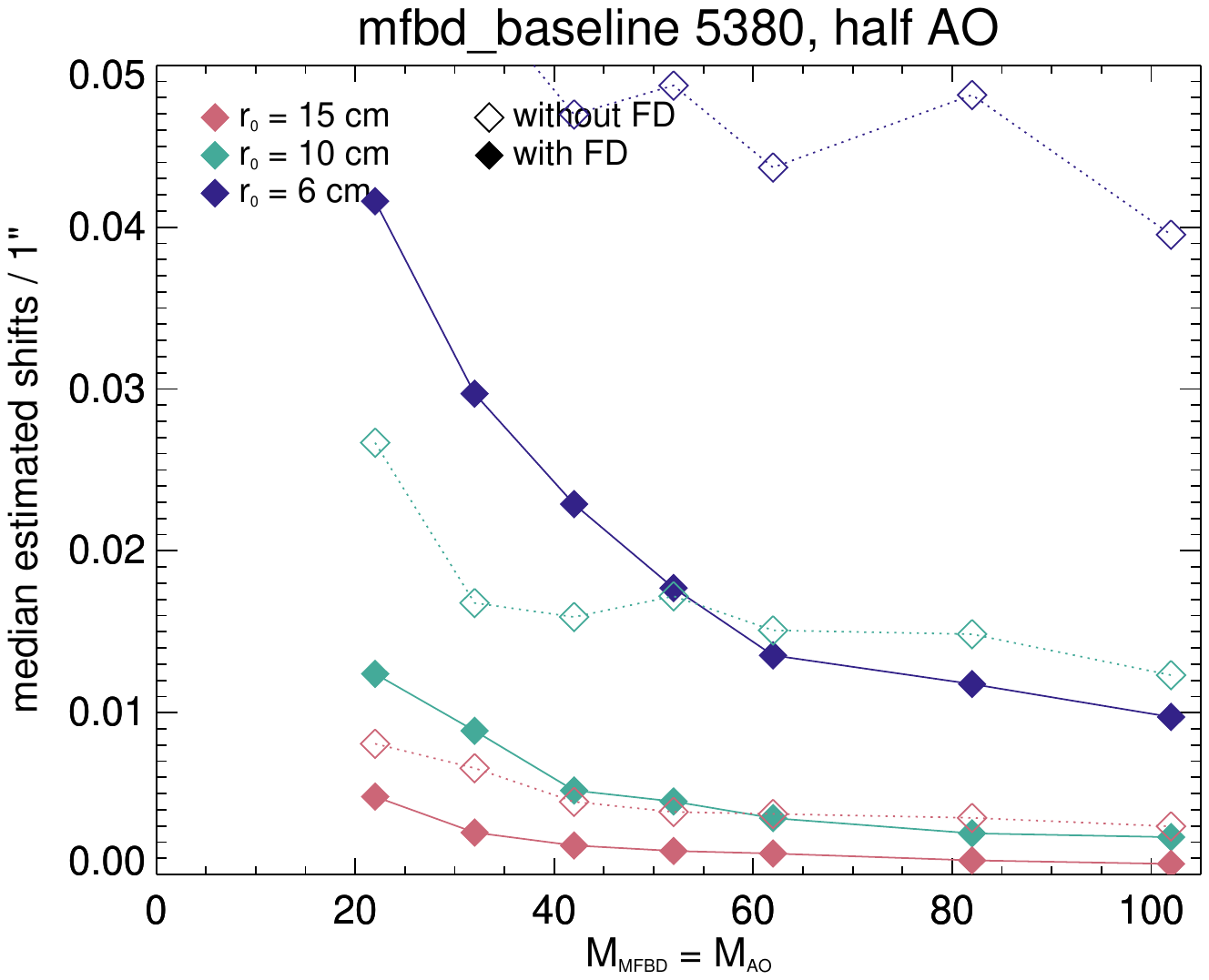}
\caption{}
\label{fig:granulation_tlterror_withouttails_Mproc=Mao}
\end{subfigure}
\hfil
\begin{subfigure}{0.45\linewidth}
\includegraphics[bb=54 46 700 529,width=0.9\linewidth,clip]{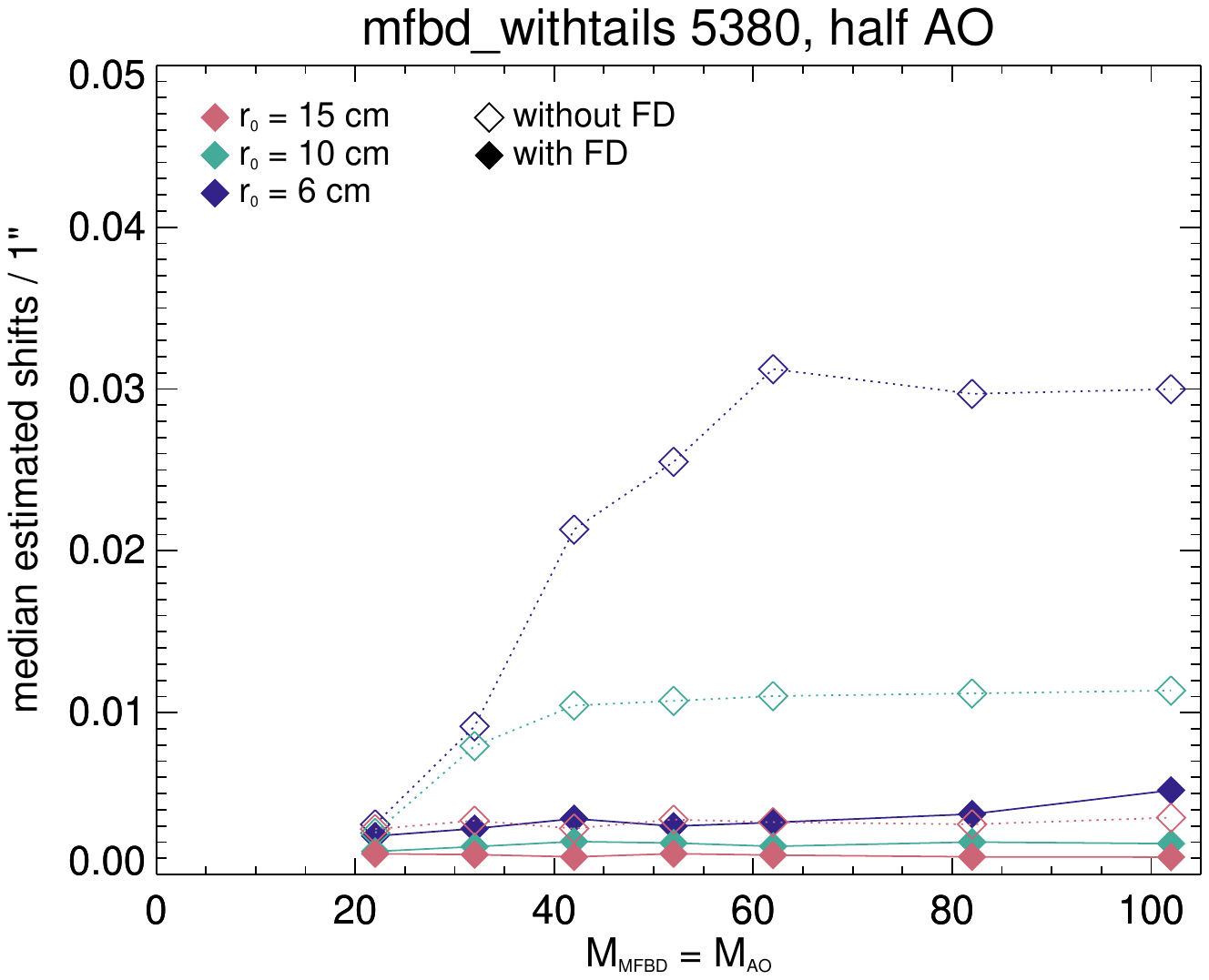}
\caption{}
\label{fig:granulation_tlterror_withtails_Mproc=Mao}
\end{subfigure}
\caption{Wavefront tilt errors as image shifts for granulation data.
\textbf{(a)} Without SD \textbf{(b)} With SD.}
\label{fig:granulation_tlterror_Mproc=Mao}
\bigskip
\centering
\begin{subfigure}{0.45\linewidth}
\includegraphics[bb=54 46 700 529,width=0.9\linewidth,clip]{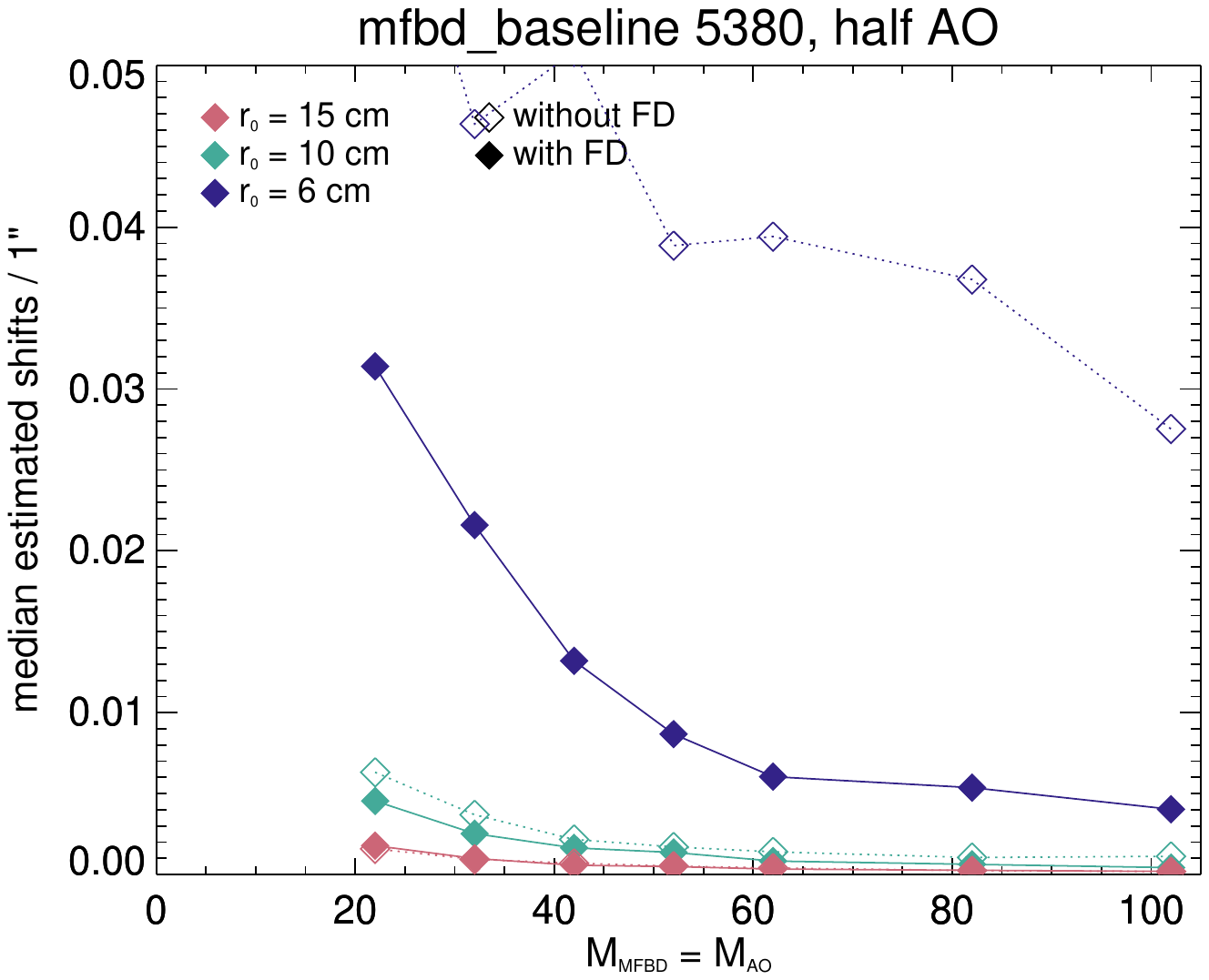}
\caption{}
\label{fig:pointsource_tlterror_withouttails_Mproc=Mao}
\end{subfigure}
\hfil
\begin{subfigure}{0.45\linewidth}
\includegraphics[bb=54 46 700 529,width=0.9\linewidth,clip]{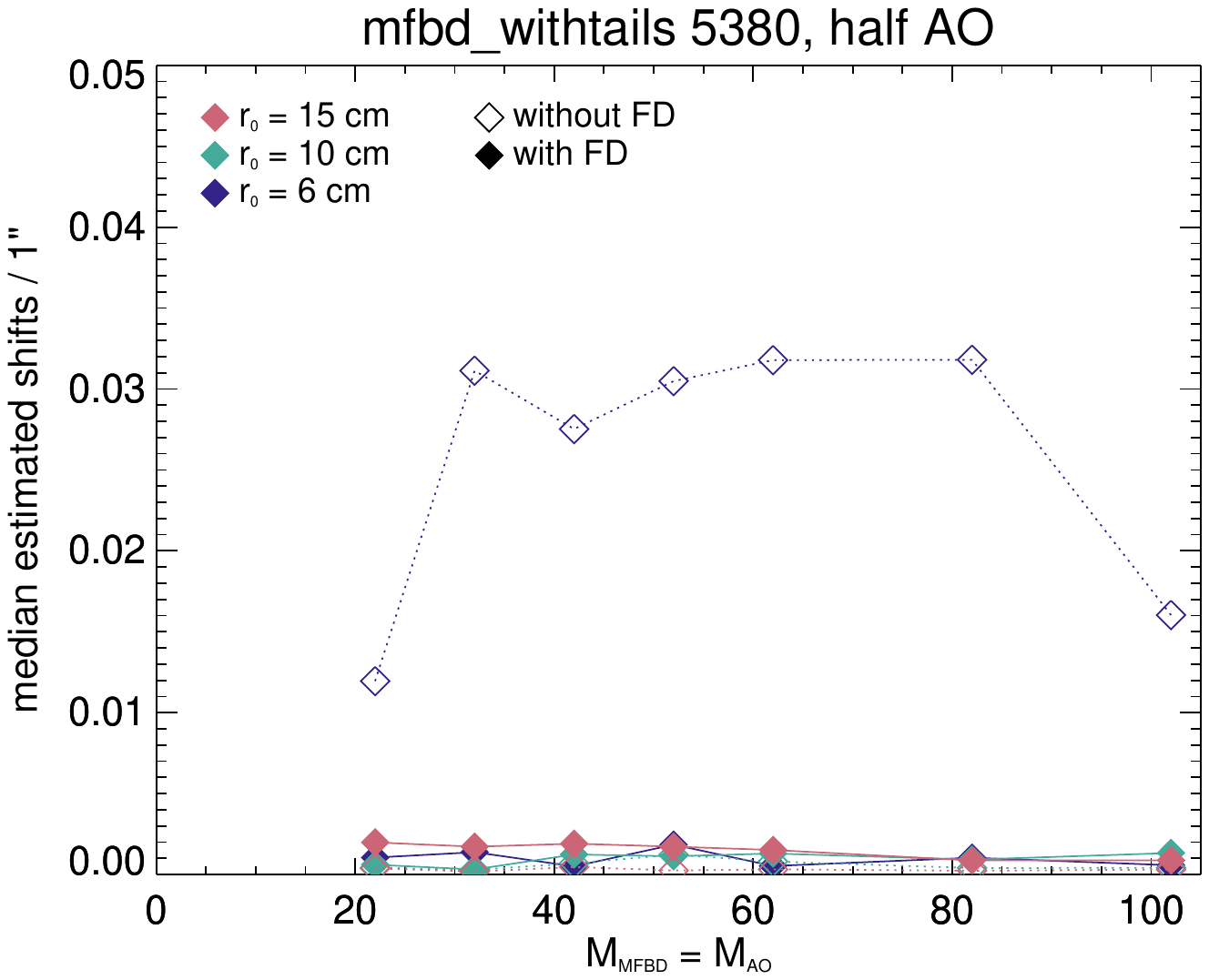}
\caption{}
\label{fig:pointsource_tlterror_withtails_Mproc=Mao}
\end{subfigure}
\caption{Wavefront tilt errors as image shifts for point source
data. \textbf{(a)} Without SD \textbf{(b)} With SD.}
\label{fig:pointsource_tlterror_Mproc=Mao}
\end{figure*}

The noncurvature wavefront errors, that is, the tip-tilt components,
require separate evaluation. These modes are not included when
making the synthetic data; however, they are included in the
parameterization of the wavefronts used in the MFBD-processing, in the
same way as when we process real data. Their coefficients are allowed
to change from their initial zeros, and so the estimated tip-tilt
contributions are wavefront errors that correspond to shifts in the
relative positions of the 100 exposures included in the data sets.

Figures~\ref{fig:granulation_tlterror_Mproc=Mao} and
\ref{fig:pointsource_tlterror_Mproc=Mao} show the median lengths of
the image shift vectors, which demonstrate that, for granulation, SD
in fact improves the estimated tilts, both with and without FD. With
FD, the improvements are significant, in particular for low $\Mmfbd$
and $r_0$. The tilt errors with both SD and FD are consistently small,
just a few (single digits) milliarcsecs regardless of $r_0$. The
comparatively poor results without SD for low $\Mmfbd$ and $r_0$
appear to correlate with the results for the wavefront errors in
Figs.~\ref{fig:granulation_wferror_Mproc=Mao} and
\ref{fig:pointsource_wferror_Mproc=Mao}, which is expected because we
have seen before that errors in tip-tilt couple with errors in
antisymmetric modes like coma, which correspond to asymmetries in the
PSFs.
For point sources, the results are similar, except that at larger
$\Mmfbd$ the shifts estimated with SD and FD are slightly larger than
for SD without FD. We note that in both cases the errors are less than
2 milliarcsec.

\subsubsection{Processing time and convergence}
\label{sec:proc-time-conv}

\begin{figure*}[!htbp]
\centering
\begin{subfigure}{0.45\linewidth}
\includegraphics[bb=50 46 710 529,width=0.95\linewidth,clip]{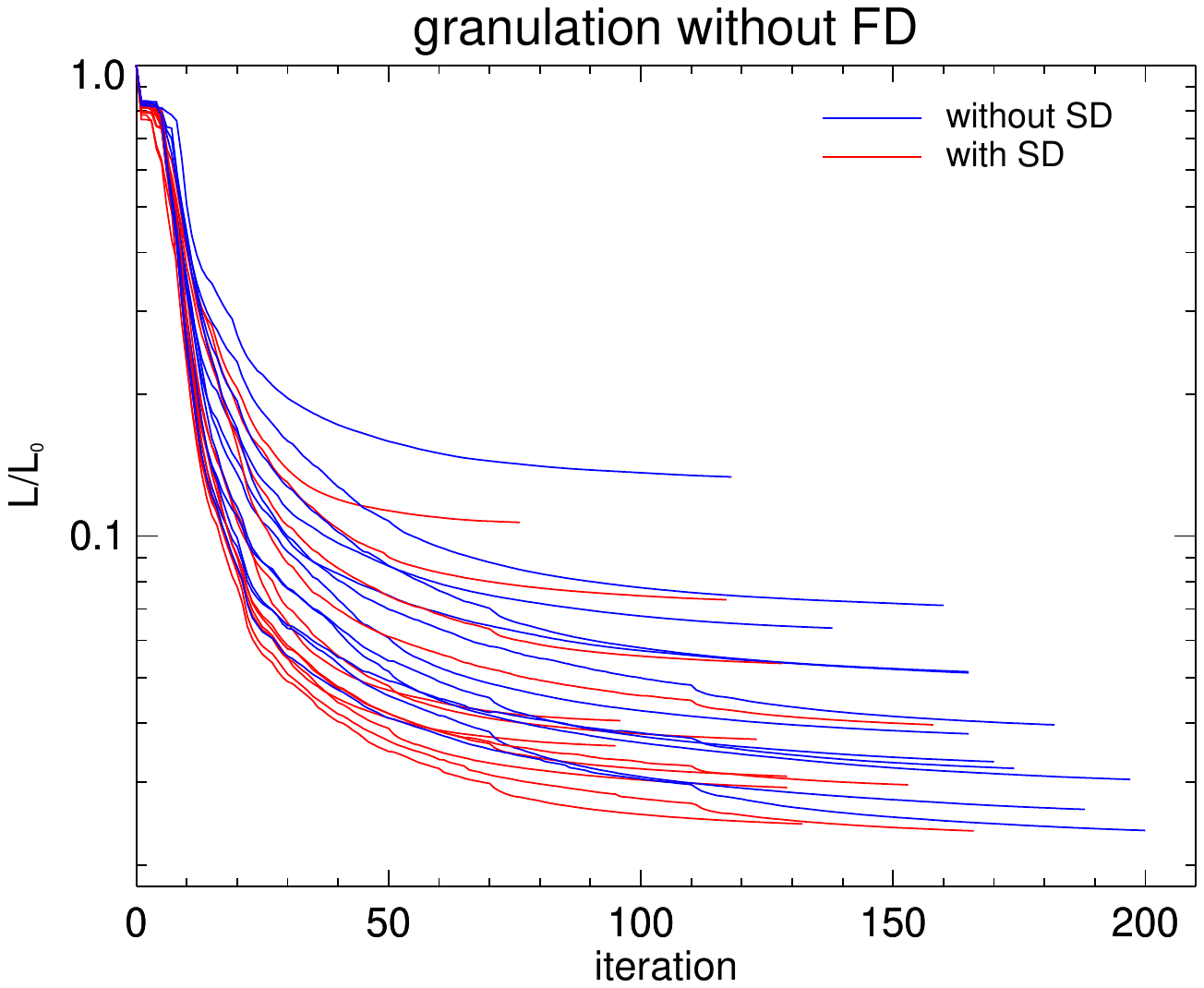}
\caption{}
\label{fig:convergence_granulation_withoutFD}
\end{subfigure}
\hfil
\begin{subfigure}{0.45\linewidth}
\includegraphics[bb=50 46 710 529,width=0.95\linewidth,clip]{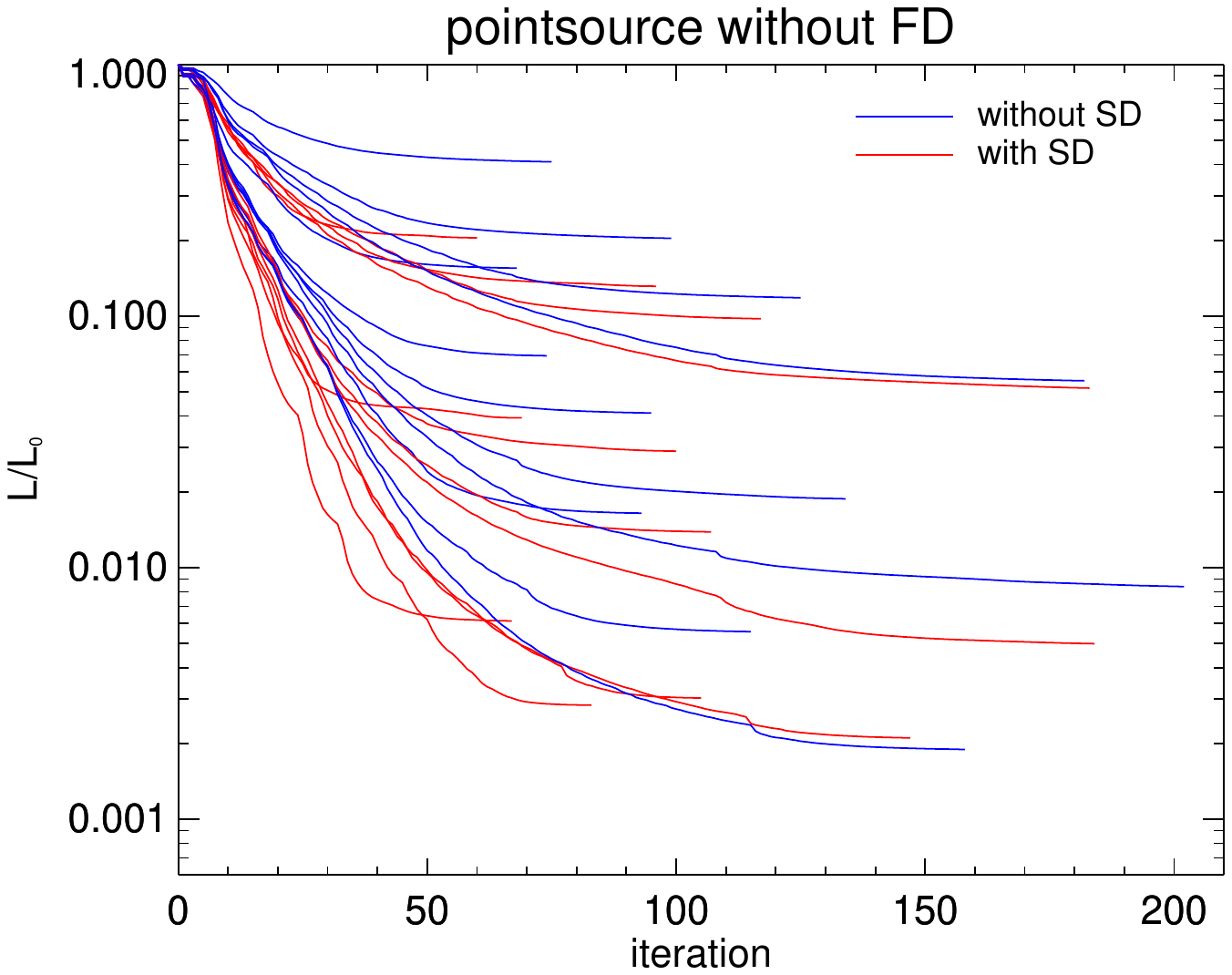}
\caption{}
\label{fig:convergence_pointsource_withoutFD}
\end{subfigure}
\\[1mm]
\begin{subfigure}{0.45\linewidth}
\includegraphics[bb=50 46 710 529,width=0.95\linewidth,clip]{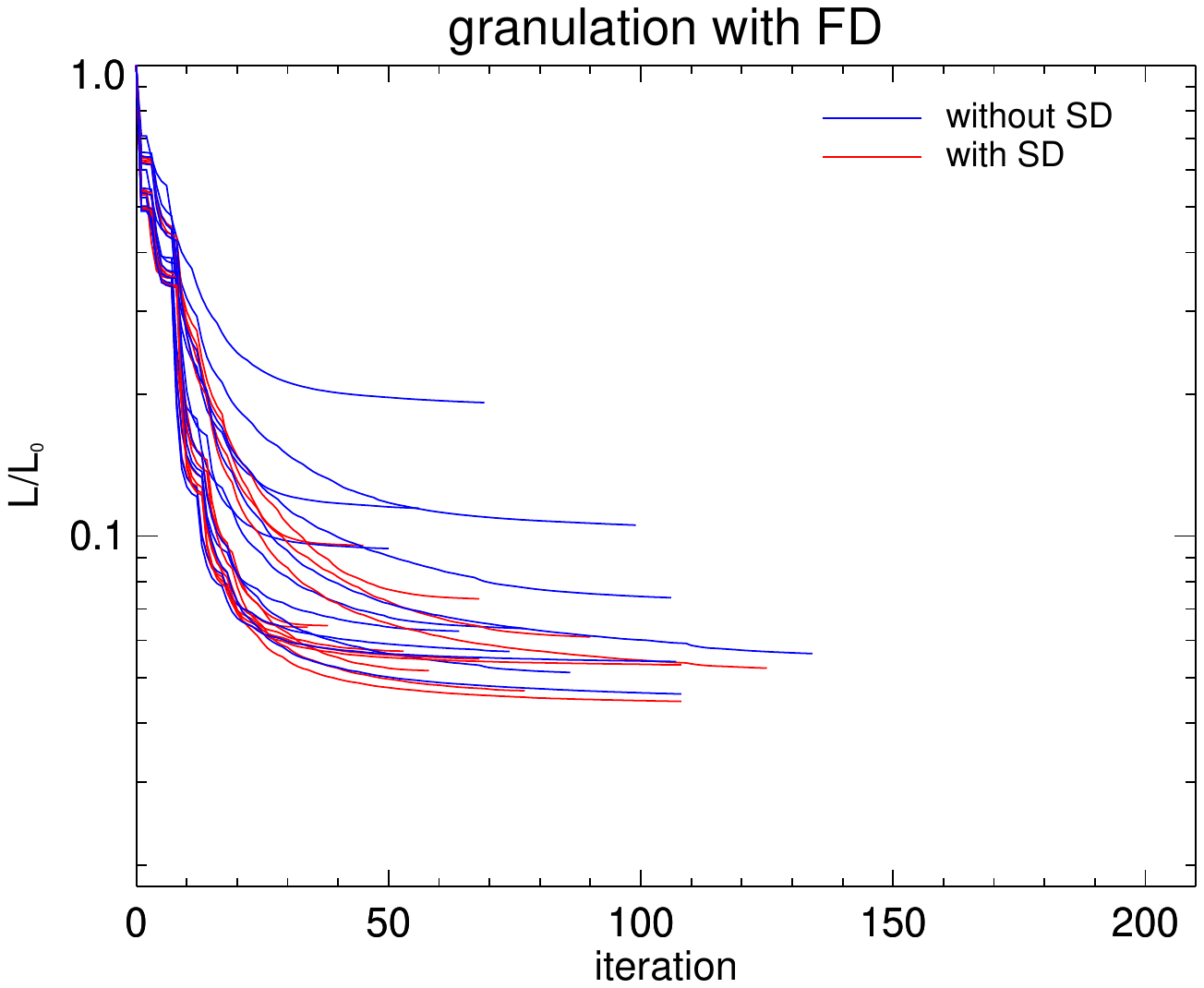}
\caption{}
\label{fig:convergence_granulation_withFD}
\end{subfigure}
\hfil
\begin{subfigure}{0.45\linewidth}
\includegraphics[bb=50 46 710 529,width=0.95\linewidth,clip]{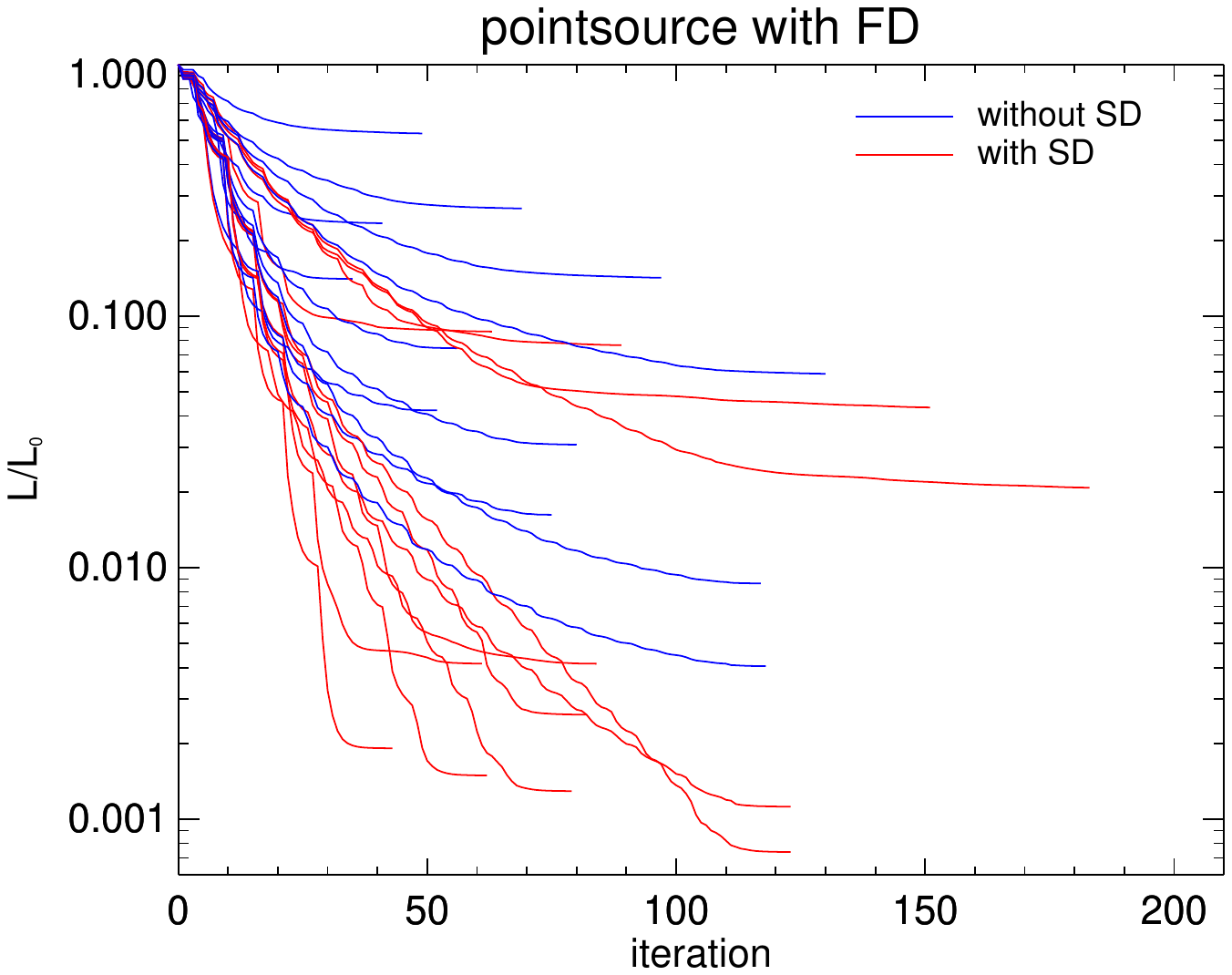}
\caption{}
\label{fig:convergence_pointsource_withFD}
\end{subfigure}
\caption{Convergence for processing of datasets with
$r_0\in\{6, 10, 15\}$~cm and
$\Mmfbd=\Mao \in \{22, 42, 62, 102\}$. $L/L_0$ is the error metric
normalized with its initial value. \textbf{(a)} Granulation
without FD. \textbf{(b)} Point source without FD. \textbf{(c)}
Granulation with FD. \textbf{(d)} Point source with FD.}
\label{fig:convergence}
\end{figure*}

Whether or not SD comes with a penalty in processing time is a
legitimate question. The number of unknowns does not increase when we
add the SD, and so each iteration takes the same amount of processing
(dominated by computing gradients) with or without it. The convergence
depends on the optimization method used and the exact number of
iterations depends on the convergence criterion used. The MOMFBD code
implements multiple methods but we tend to use conjugate gradients.

Figure~\ref{fig:convergence} shows the convergence of the error metric
for a subset of the $\Mmfbd$ and $r_0$ datasets. We show all
combinations of using or not using FD and/or\ SD for both
granulation data and point sources. These plots, with separate curves
for many datasets, do not allow detailed comparisons, but it is
obvious that using SD is generally associated with faster reduction in
metric than not using SD. This holds regardless of whether FD is used
(panels (b) and (d)) or not (panels (a) and (c)). We note that the metric is
normalized to the initial value, corresponding to zero KL
coefficients. As the initial metric for the SD processing is already
compensated for the high-order tails, the final differences between SD
and no SD are slightly larger than shown in the figure but the effect
is not dramatic.

The number of iterations seems to be smaller with SD than without,
except for point sources and FD (d), but this is also where the
improvement in converged error metric is the greatest. In the other
panels of Fig. 19, the improvement in error metric can be seen to be modest in many cases,
which is consistent with very shallow minima in the error metric.
The wavy pattern superimposed on the metric curves is caused by a
strategy for reducing the risk for convergence to local minima, which is
discussed in Sect.~\ref{sec:local-minima}.

\subsubsection{Local minima}
\label{sec:local-minima}

For point source data, with SD but without FD, the wavefront errors
are almost zero except for $r_0=6$~cm; see
Fig.~\ref{fig:pointsource_wferror_withtails_Mproc=Mao}. For 6~cm, the
errors are about as large as with granulation data. This suggests that
the MFBD model fits converged to local minima and that there are
global minima at the correct solutions -- perhaps for both point
sources and granulation.

In the MOMFBD code (see Sect.~\ref{sec:software}), the following
strategy for avoiding local minima is implemented. The fitting starts
with a small subset of the modes, by default the first $M_1=5$ modes;
it lets those converge for a few iterations, continues with $M_2=10$
modes, lets them converge, and so on, until all the $\Mmfbd$ modes are
included in the fit. In other words, we are fitting first
$\phi_{jM_1}$, then $\phi_{jM_2},\phi_{jM_3}$, and so on, and finally
$\phi_{j\Mmfbd}$. By allowing the more significant low-order modes to
get started in the right direction, this scheme improves the chances
that the fit converges to the global minimum. The wavy pattern in
Fig.~\ref{fig:convergence} and mentioned above is caused by these
repeated increases in the number of fitted coefficients.

However, as SD also improves the convergence, and reduces the
risk for convergence to local minima, we may want to modify this
scheme. SD with $\tilde\tau_{j\Mmfbd}$ means there is still a model
mismatch during this process of increasing the number of fitted modes.
When $M_i$ low-order modes are fitted, and $\tilde\tau_{j\Mmfbd}$
statistically represents the most high-order modes, there is still a
mismatch in that modes $M_i+1$ through $\Mmfbd$ are not accounted for.
One way to further improve the chances of global minimization might be
to include the appropriate SDs while stepping up the number of fitted
modes: when $\phi_{jM_i}$ is fitted, use $\tilde\tau_{j{M_i}}$ as the
SD.

For data that are not AO-corrected, this is straight forward. For
AO-corrected data, we have to address the fact that a subset of the
unfitted modes are partially AO corrected, and so we should take this into
account when calculating $\tilde\tau_{j{M_i}}$. At this point, one might question whether or not we have to
calibrate the efficiencies with which the AO corrects the modes, as is the case
for Speckle or the method of \citetads{2021A&A...653A..17S}, and take
decorrelation of modes with distance from the WFS lock point into
account. We believe that this would not have to be done exactly, as it
is not involved in the final convergence with $\Mmfbd$ fitted modes
and $\tilde\tau_{j\Mao}$ as the SDs. It should be possible to use very
approximate efficiencies. This is something to keep in mind when
actually implementing the improvement of this strategy.

\subsection{Synthetic data with varying $r_0$}
\label{sec:varying-r_0}

For real observations, $r_0$ is not constant over the $\sim$10~s
period it takes to collect a dataset with a tuning
spectro(polari)meter like SST/CRISP or SST/CHROMIS, and so for a more
realistic simulation we need to allow for variations in $r_0$. This is
no problem for the SD concept, as each exposure $j$ can have its own
diversity coefficient, scaled to the $r_0$ at the time of that
exposure.

\citet{scharmer10high-order} demonstrated that variations in $r_0$
measured with the wide-field WFS of \citet{scharmer10s-dimm+} with a
temporal resolution of 2~s correlated with the RMS contrast in raw
granulation data. \citetads{2019A&A...626A..55S} showed 2~s $r_0$
measurements from the SST AO WFS (upgraded in 2013) that also
correlate with raw granulation data. These correlations suggest that
these WFSs measure the total wavefronts from the whole height
distribution of turbulence along the line of sight. In their temporal
$r_0$ plots, we can see that these meaningful $r_0$ measurements can
easily vary from 5 and 20~cm in a few seconds. We therefore
constructed synthetic data from sets of 100 atmospheric wavefronts
with $r_0$ values spaced evenly in the ranges 10--14~cm, 8--16~cm, and
6--18~cm, respectively. (There are no moving phase sheets involved in
the simulations, and so the $j$ index is not a real time coordinate.
Therefore it is of little importance that the $r_0$ values are
``sorted'' this way.)

\begin{figure*}[!tbhp]
\centering
\begin{subfigure}{0.45\linewidth}
\includegraphics[bb=54 46 700 529,width=0.9\linewidth,clip]{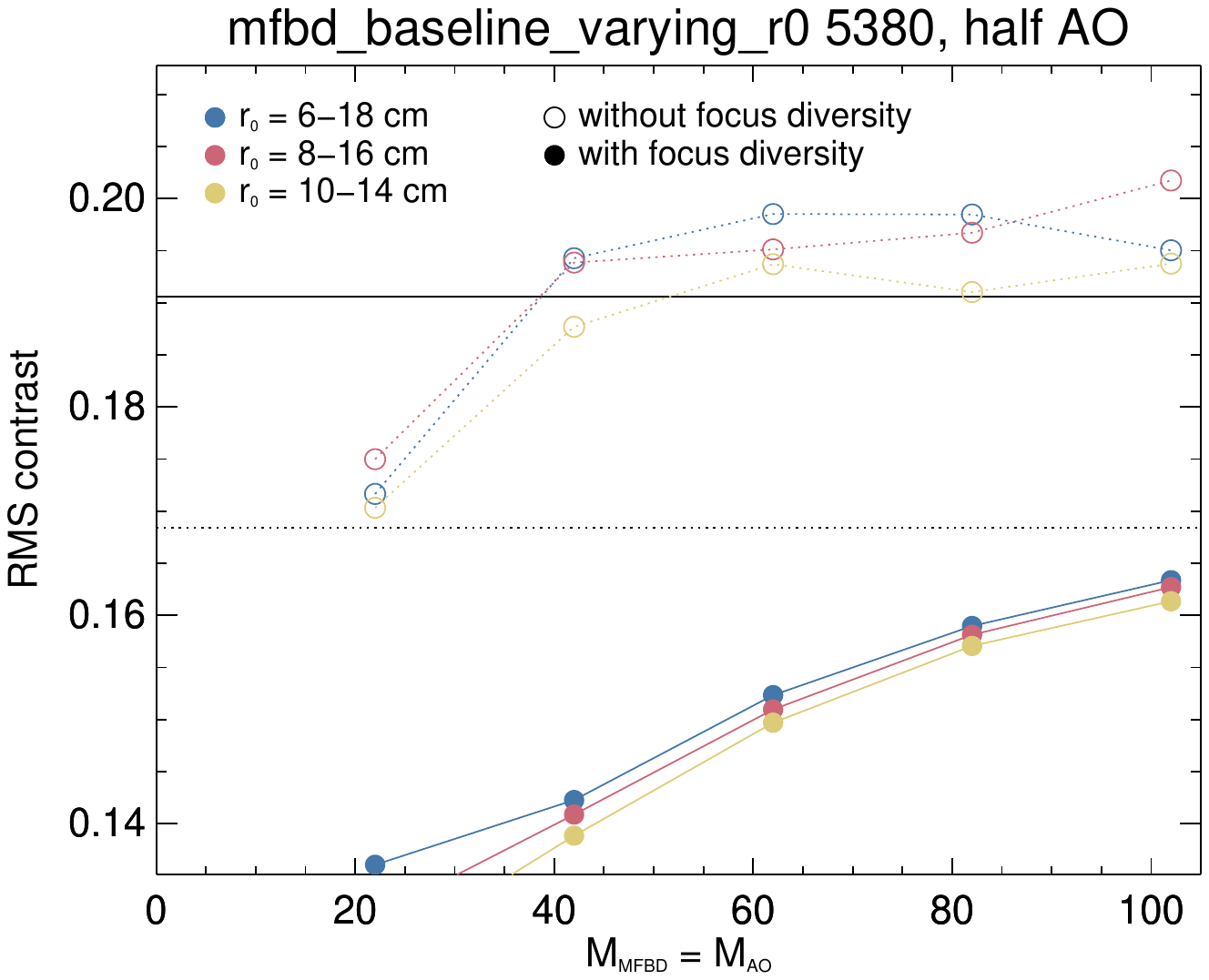}
\caption{}
\end{subfigure}
\hfil
\begin{subfigure}{0.45\linewidth}
\includegraphics[bb=54 46 700 529,width=0.9\linewidth,clip]{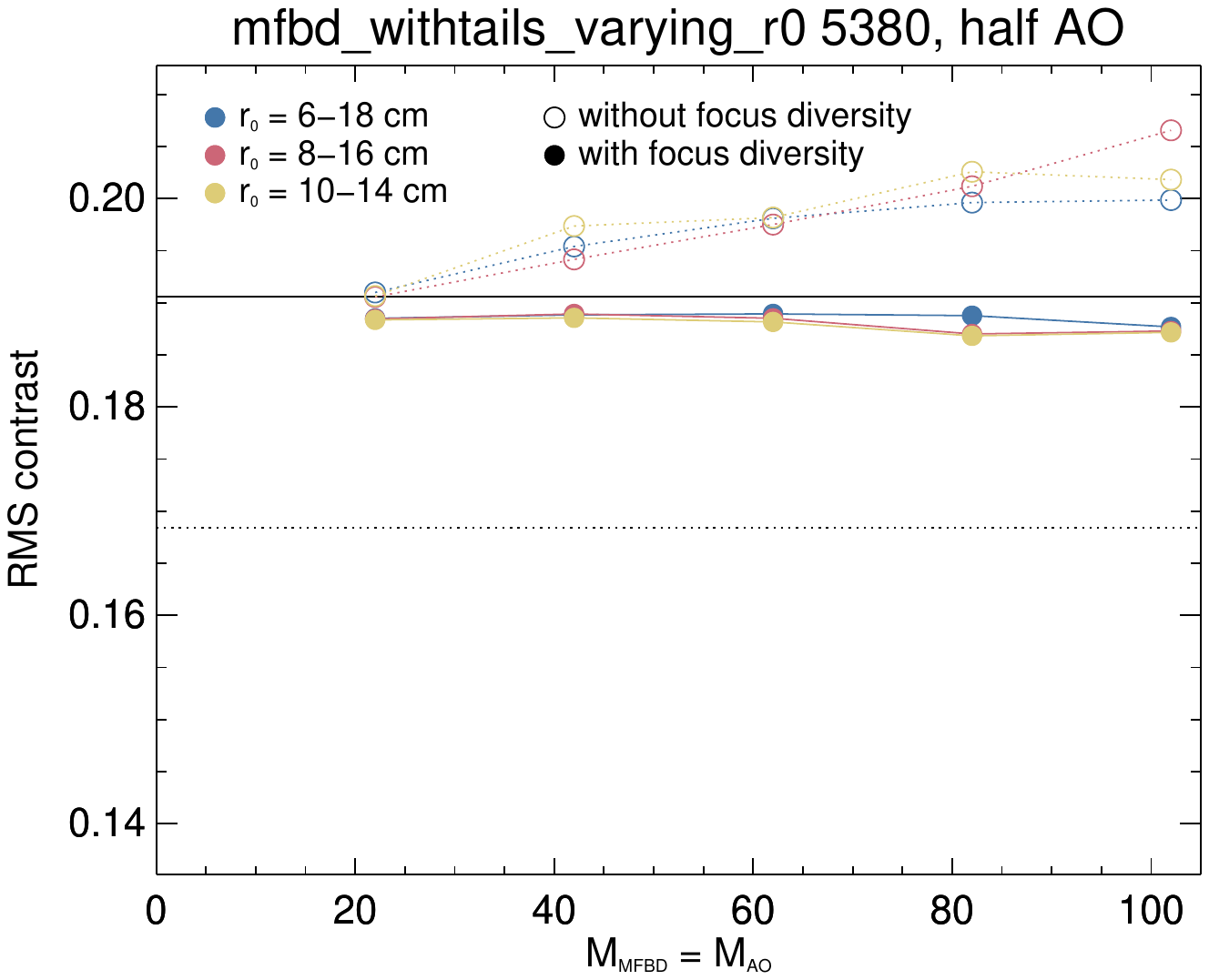}
\caption{}
\end{subfigure}
\caption{Restored granulation contrast for varying $r_0$ and
$\Mmfbd=\Mao$. \textbf{(a)} Without SD. \textbf{(b)} With SD.}
\label{fig:contrast_Mproc=Mao_var_r0}
\end{figure*}

We then MFBD-processed these data as before, with and without FD and with
and without SD, and the $\tilde\tau_{j\Mmfbd}$ SD contributions scaled
to the individual $r_0$ values for each simulated wavefront. We
processed only a subset of the different $\Mao$ values. The MFBD
inversions were initialized with the wavefront coefficients at their
true values, because we wanted to evaluate the optimum fits rather
than the risk of convergence to local minima.

The restored contrasts, shown in
Fig.~\ref{fig:contrast_Mproc=Mao_var_r0}, are remarkably similar for
the different $r_0$ distributions. Comparing the results with constant
$r_0$ in Figs.~\ref{fig:contrast_withouttails_Mproc=Mao}
and~\ref{fig:contrast_withtails_Mproc=Mao}, the FD contrasts without
SD for varying $r_0$ are similar to those of the constant $r_0=6$, 8, or
10~cm datasets, while with SD the contrasts are  consistently correct for
constant and varying $r_0$. Without FD, the contrasts without SD are
overestimated for varying $r_0$ while they are not for constant
$r_0$. With SD, processing the varying $r_0$ data without FD gives
similar results to the constant $r_0=12$, 15, and 20~cm data.
Varying $r_0$ does not seem to be a problem for the SD methods.

\subsection{Errors in $r_0$}
\label{sec:errors-r_0}

The SD component $\tilde\tau_{j\Mmfbd}$ depends on $r_0$, which begs
the questions of how sensitive the proposed method is to errors in
$r_0$ and of whether or not it would be possible to include $r_0$ as
an unknown parameter in the MFBD model fitting. We can examine some
aspects of these questions without actually implementing the
functionality in the redux MOMFBD code. We can investigate the effects
of errors in the $r_0$ values used to calculate the SD
$\tilde\tau_{j\Mmfbd}$ components.

The errors to consider may be guided by the 2~s resolution $r_0$
measurements mentioned in Sect.~\ref{sec:varying-r_0}. There may be
smoothed-out variations on even shorter timescales, although they can
be expected to be much smaller than the variations in the measurements
themselves. This would apply to both negative and positive peaks. It
seems prudent to test with random errors of at least a few percent.
Below, we also examine the effect of systematic errors of similar
magnitudes in the measured~$r_0$.

For testing the sensitivity to such errors, we used the varying-$r_0$
data from Sect.~\ref{sec:varying-r_0}. We scaled the
$\tilde\tau_{j\Mmfbd}$ components to $r_0$ values with
normal-distributed random errors added to the true values, as well as
systematic errors. We processed with $\Mmfbd = \Mao$ and initialized
to the correct $\Mmfbd$ wavefront coefficients as above.

\begin{figure*}[!tbhp]
\centering
\begin{subfigure}{0.45\linewidth}
\includegraphics[bb=34 46 700 529,width=0.9\linewidth,clip]{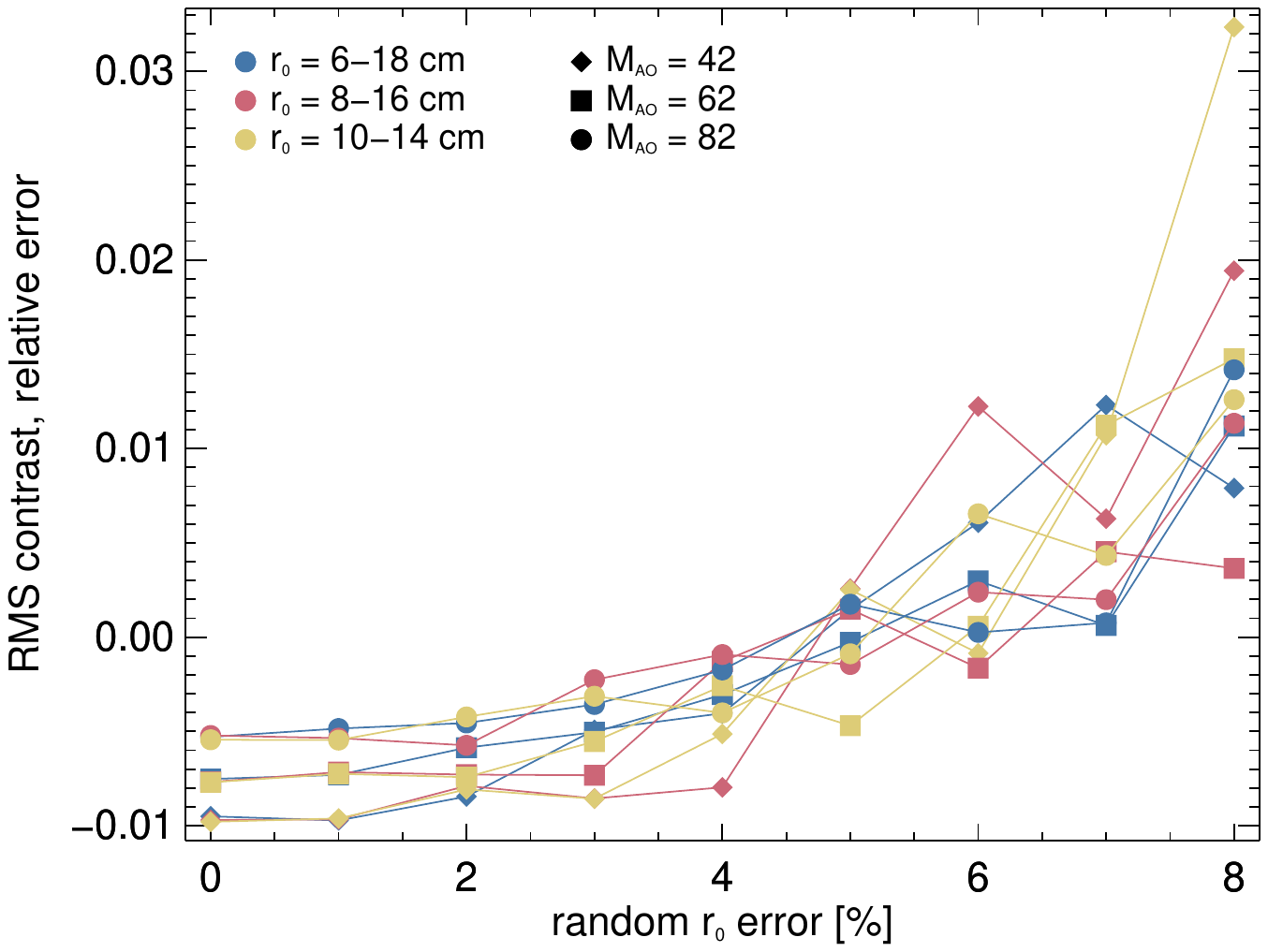}
\caption{}
\end{subfigure}
\hfil
\begin{subfigure}{0.45\linewidth}
\includegraphics[bb=34 46 700 529,width=0.9\linewidth,clip]{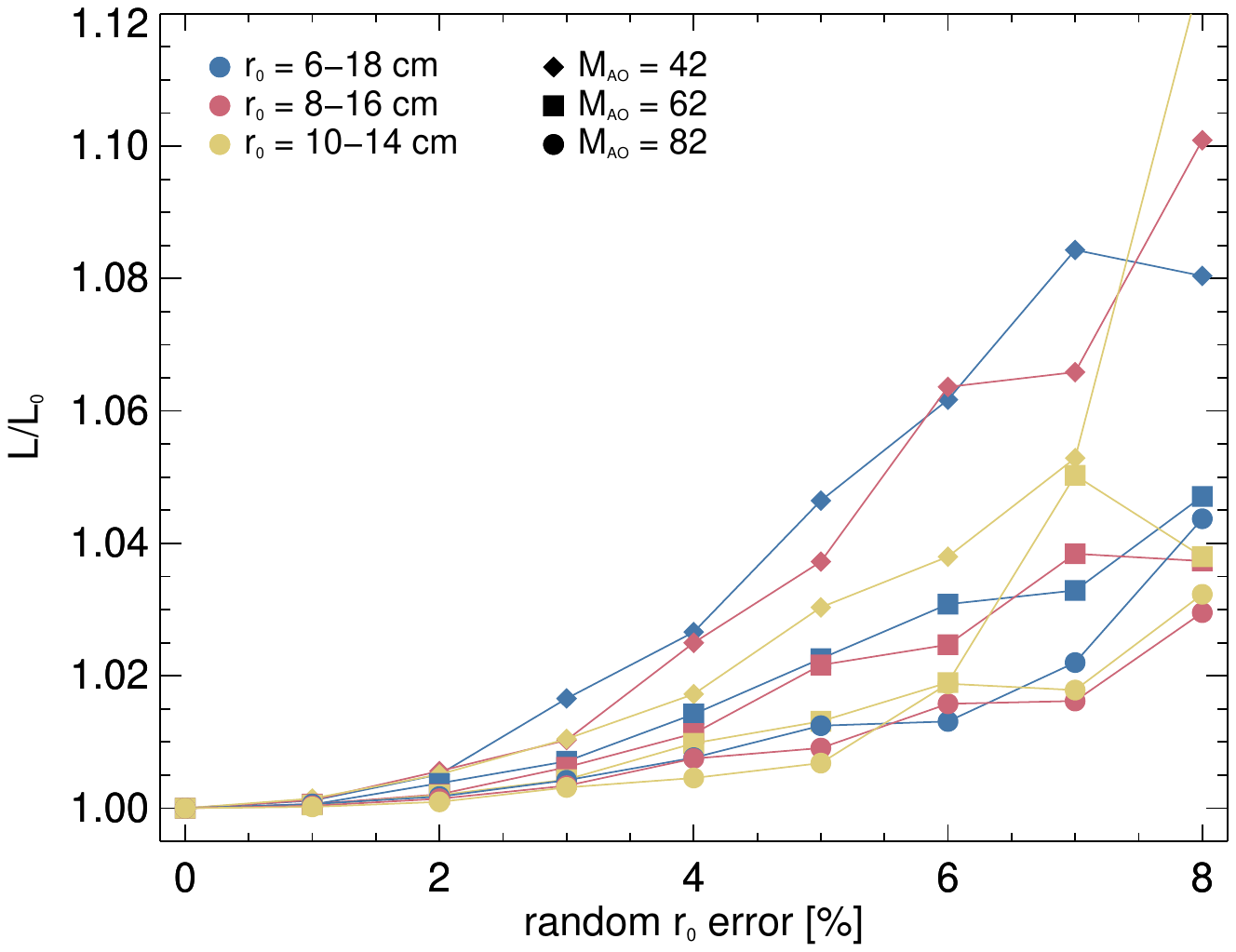}
\caption{}
\end{subfigure}
\caption{Results from using MFBD with FD and SD on the same data as
in Fig.~\ref{fig:contrast_Mproc=Mao_var_r0} but with random errors
in the assumed $r_0$. \textbf{(a)}~RMS contrast, relative error.
\textbf{(b)}~Converged error metric $L$, normalized to zero error
in $r_0$.}
\label{fig:jpds_granulation_with_randerrors}
\end{figure*}

Figure~\ref{fig:jpds_granulation_with_randerrors} shows the results
with random errors in $r_0$. The errors were drawn from a normal
distribution with standard deviations in the range 0\%--8\%. We
evaluated the error in the restored RMS contrast (the difference
between the true image contrast and the estimated contrast) as well as
the error metric $L$; see Eq.~(\ref{eq:2}). The contrast errors are
small, less than 1\% for random errors $\la6$\%. The normalized error
metric (unity for zero $r_0$ error) is clearly at a minimum near zero
error, and so it might be possible to converge to correct $r_0$ values by
including the SDs in the model fitting. In spite of initializing with
the correct wavefront coefficients, the error metrics show signs of
convergence to local minima for random errors larger than $\sim$6\%.

\begin{figure*}[!tbhp]
\centering
\begin{subfigure}{0.45\linewidth}
\includegraphics[bb=34 46 700 529,width=0.9\linewidth,clip]{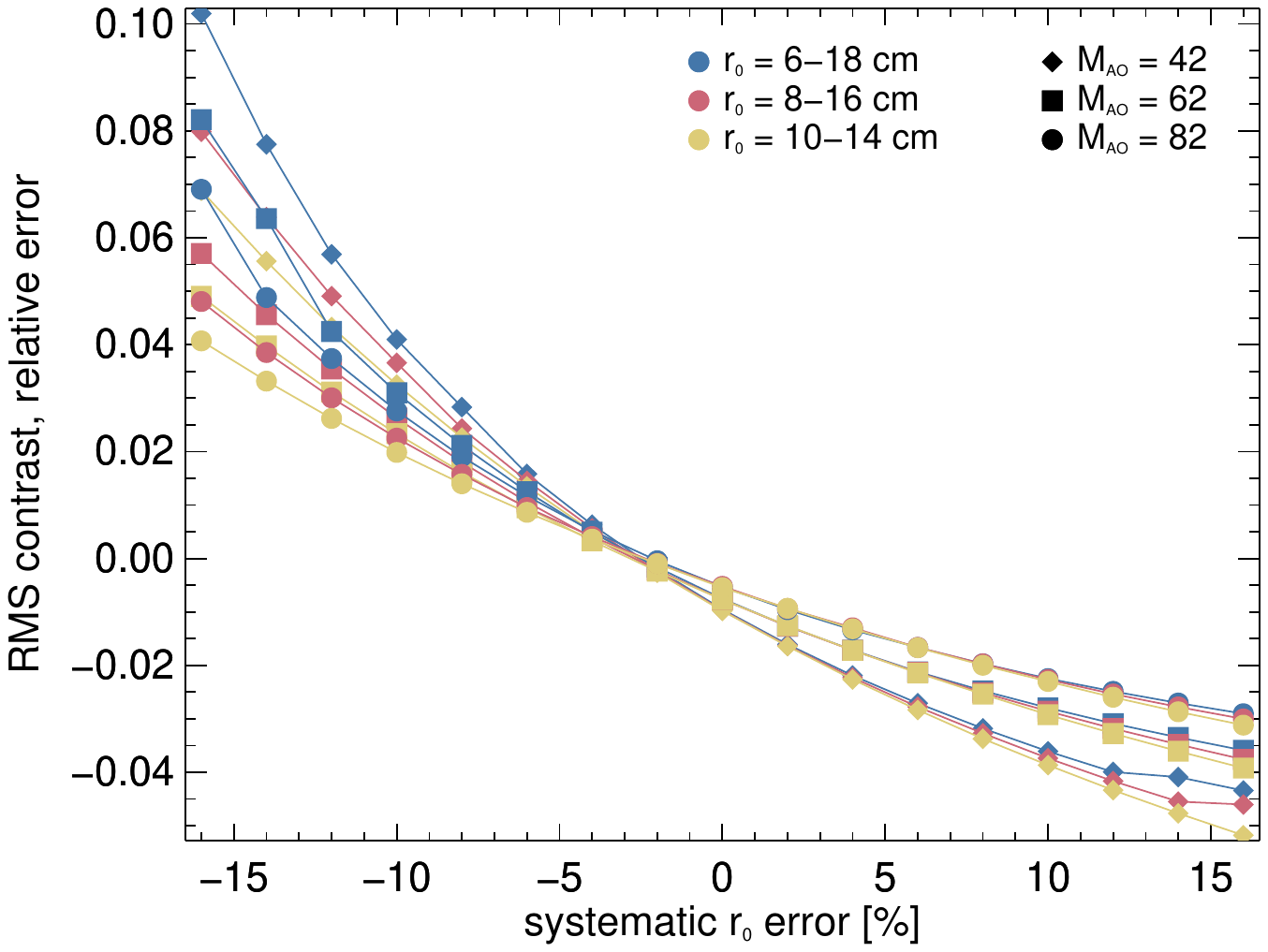}
\caption{}
\end{subfigure}
\hfil
\begin{subfigure}{0.45\linewidth}
\includegraphics[bb=34 46 700 529,width=0.9\linewidth,clip]{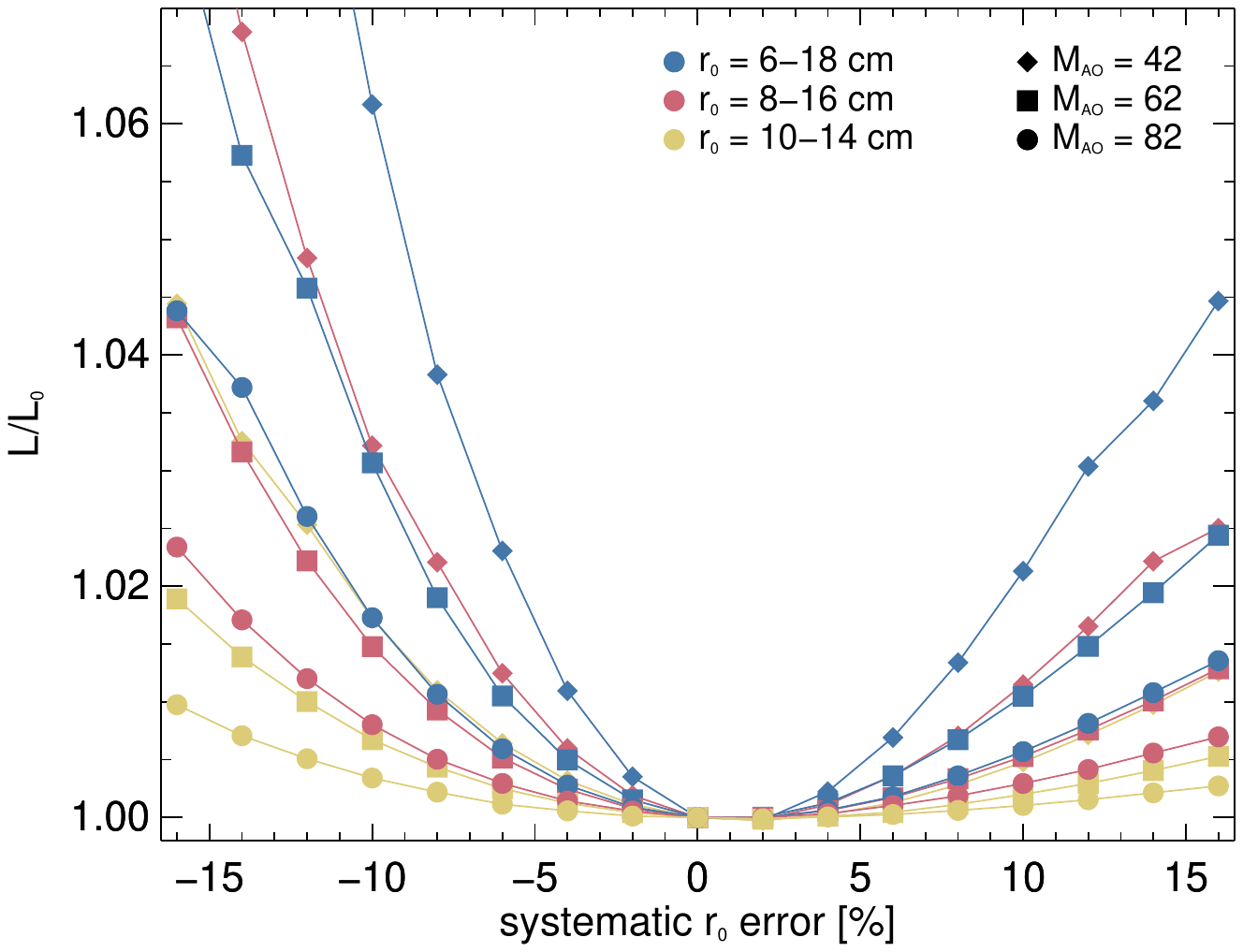}
\caption{}
\end{subfigure}
\caption{Results from using MFBD with FD and SD on the same data as
in Fig.~\ref{fig:contrast_Mproc=Mao_var_r0} but with systematic
errors in the assumed $r_0$. \textbf{(a)}~RMS contrast, relative
error. \textbf{(b)}~Converged error metric $L$, normalized to zero
error in $r_0$.}
\label{fig:jpds_granulation_with_systerrors}
\end{figure*}

Figure~\ref{fig:jpds_granulation_with_systerrors} shows the results
from adding systematic errors in the range $-16\%$ to 16\% to $r_0$. The
contrast errors are larger for the data sets with a larger range of
$r_0$ values, in particular with less AO correction. The contrast and
metric curves are smoother than for the random errors. There are clear
minima in the metric near 0\% error for all the data sets, which
suggests it might be possible to fit for these errors as well.

\subsection{Pretending varying $r_0$ are constant}
\label{sec:assum-r_0-const}

\begin{figure*}[!tbh]
\centering
\begin{subfigure}{0.45\linewidth}
\includegraphics[bb=34 46 700 529,width=0.9\linewidth,clip]{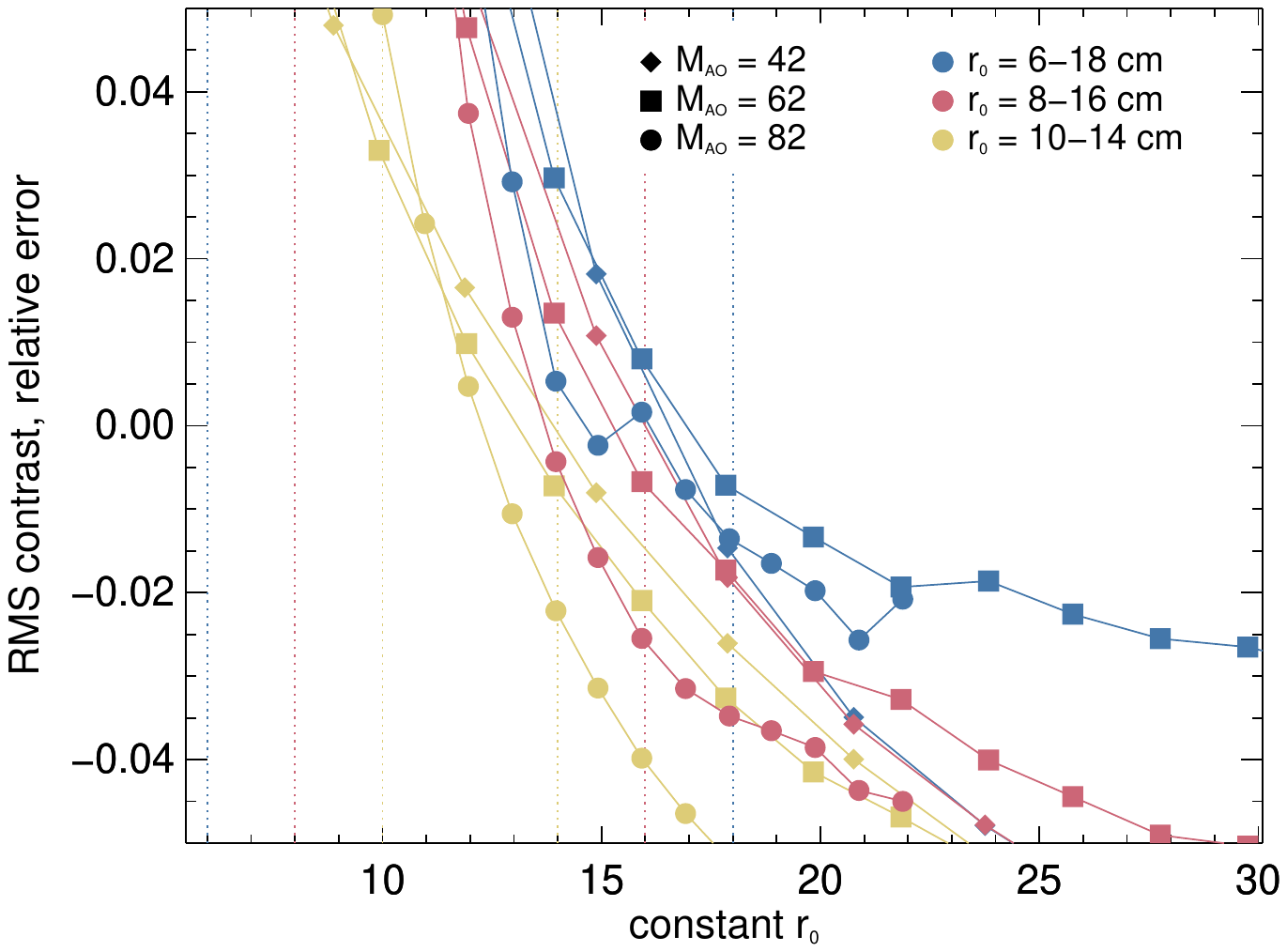}
\caption{}
\end{subfigure}
\hfil
\begin{subfigure}{0.45\linewidth}
\includegraphics[bb=34 46 700 529,width=0.9\linewidth,clip]{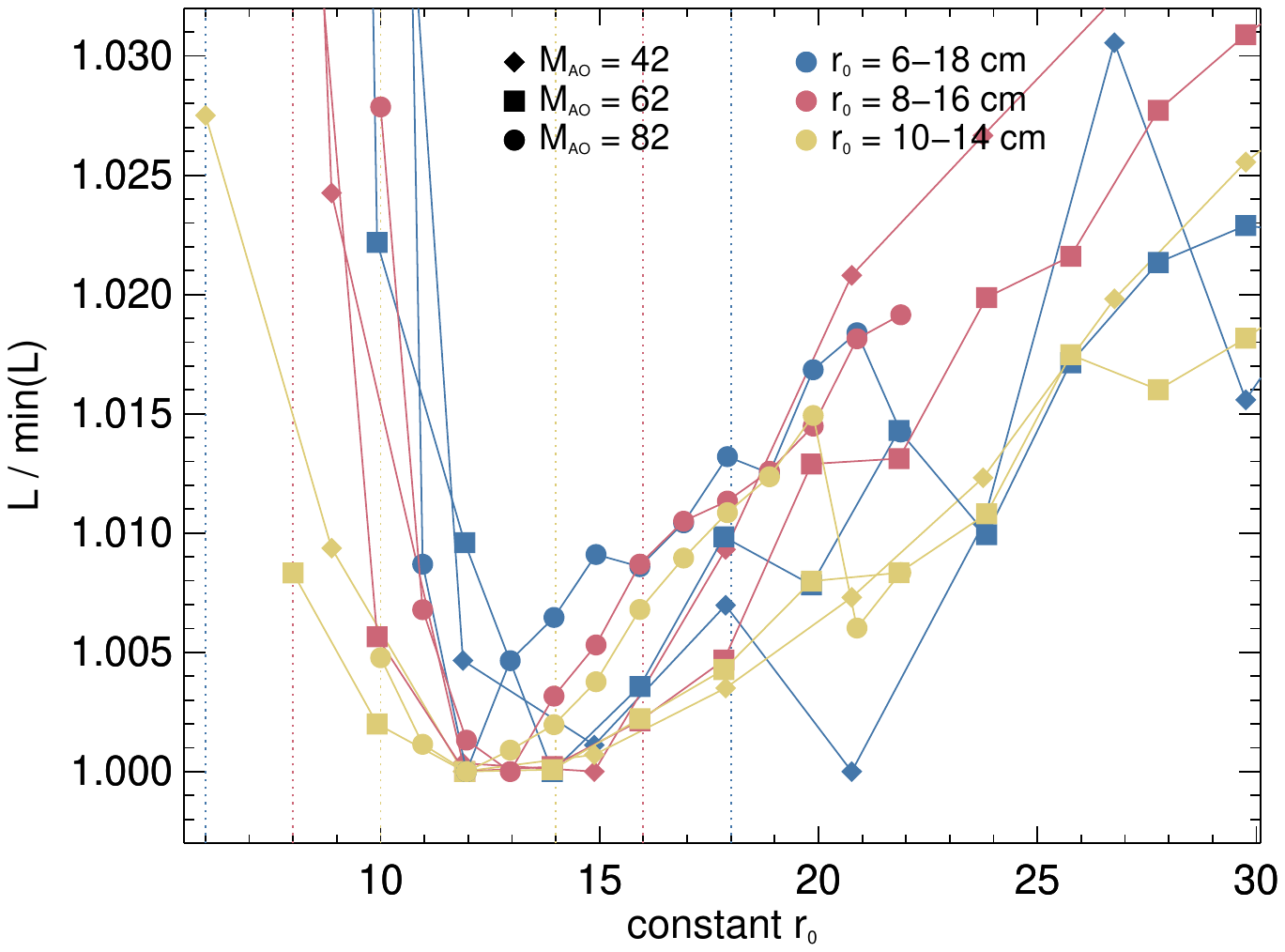}
\caption{}
\end{subfigure}
\caption{Results from using MFBD with FD and SD on the same data as
in Fig.~\ref{fig:contrast_Mproc=Mao_var_r0} but with a constant
assumed $r_0$. The vertical dotted lines represent the limits of
the $r_0$ intervals. \textbf{(a)}~RMS contrast, relative error.
\textbf{(b)}~Converged error metric $L$, normalized to its
minimum.}
\label{fig:jpds_granulation_pretend}
\end{figure*}

In the spirit of Speckle, we might also want to process data with
varying $r_0$ as if $r_0$ were constant during the collection
interval.
Figure~\ref{fig:jpds_granulation_pretend} shows the impact of this
hypothetical situation on the restored RMS contrast as a function of
the constant $r_0$ used when calculating the SD coefficients for the
three datasets with varying $r_0$, as well as the converged error
metric. As when processing the random and systematic errors in $r_0$,
we initialized with the correct $\Mmfbd$ coefficients.

The contrast errors are fairly large, even with $r_0$ within the range
of the values included in the input. There is no clear indication for
how the constant $r_0$ should be calculated. The arithmetic average is
12~cm for all three datasets, while the optimum assumed $r_0$ seem to
vary within the upper half of the individual ranges based on where the
contrasts cross the correct value. The error metrics seem to have
minima around the crossing points, but these are too shallow (and, in
the 6--18cm case, show too much variability, probably because of local
minima) to give any confidence that including a constant $r_0$ would
converge to a useful value if included as an unknown in the MFBD
processing. Assuming a constant $r_0$ value for the SD processing
would probably require figuring out how to properly average $r_0$
values measured within the collection interval. Furthermore, if the
individual $r_0$ values are available, there is no reason not to use
them.

\subsection{Software}
\label{sec:software}

For the processing, we used the Redux fork
\citepads{2021A&A...653A..68L} of the MOMFBD code by
\citetads{2005SoPh..228..191V} based on the linear equality
constraint (LEC) formulation of MFBD and PD by
\citetads{2002SPIE.4792..146L}. Redux is maintained by one of the authors of the  present paper
(T.H.).

In the Redux code, KL modes are by default implemented as sums of
Zernike polynomials 2--2000 following \cite{roddier90atmospheric}. An
exception is that we use pure Zernike tip and tilt modes instead of
the KL counterparts. This is in order to make it easier to realize
(parts of) large tip and tilt coefficients as a change in subfield
position. This makes the tip/tilt modes not quite orthogonal to the
first few of the following modes and might change the covariance
slightly, but should not cause any problem for the higher order modes
that make up $\tau_M$.

We note that the LEC formulation of MFBD has no problems with
representing the SD mode of processing, which was completely unanticipated when
it was devised 20 years ago. Only relatively minor changes to the code
were required: mainly the reading of diversity modes from a file
instead of internally generating the Zernike focus mode as for FD
only. There were also a few changes related to preprocessing of raw
data, as the code is by default set up for the processing of real
solar data with an extended object that is larger than the FOV. The
code can now optionally skip certain steps related to intensity
normalization and windowing and leave it to the user to provide the
data in the appropriate state.

\section{Discussion}
\label{sec:discussion}

Our results are obtained with very idealized data (no noise, no
anisoplanatism, simple AO correction with the same modes as used in
the MFBD processing) and do not necessarily indicate that the
performance would be the same with real data. However, comparisons
with the same data, processed both with and without SD, should show
the advantages of using SD, but the extent of these advantages remains to be
evaluated with real data for particular instruments and is outside the
scope of this paper.

The SD representation of the high-order mode contribution combines the
advantages of MFBD-based methods with some of those of Speckle. As is the case
for Speckle, the contrast and power spectrum is corrected for the full
range of wavefront modes. But an important advantage with MFBD over
Speckle is retained, namely that the low-order modes are corrected without
calibration of the performance of the AO in the presence of
anisoplanatism. Moreover, the MFBD approach blindly compensates for
all low-order aberrations irrespective of their origin being from the
optics or from the atmospheric turbulence, and regardless of whether
these aberrations obey Kolmogorov statistics. The high-order modes
that are not AO corrected are also not subject to varying correction
over the FOV. Although the wavefronts vary spatially due to
anisoplanatism, we can still expect $r_0$ to be the same over the FOV.

The procedure for calculating $\tilde\tau_{j\Mmfbd}$ used in this work
still uses finite sums of modes, although to $\Msim \gg \Mmfbd$. In
hindsight, it would have been better to make nonparameterized
simulated wavefronts \citepads[e.g.,]{2012ApOpt..51.7953K} and instead
subtract the best-fit $\phi_\Mmfbd$ wavefront, which is similar to the
procedure used by \citetads{2021A&A...653A..17S}. We plan to switch to
this approach when continuing the development of this method and adapting it to
real data.

The results seem fairly robust to errors in $r_0$. In particular,
random errors around $r_0$ measured with a temporal resolution of a
few seconds should not be a problem. The method is formulated so that
if $r_0$ measurements with temporal resolution are known, they are a
natural part of the problem formulation. Some of our results indicate
it might be possible to estimate $r_0$ as an additional fit parameter
along with the low-order wavefront mode parameters.

It seems possible to process data with varying $r_0$ as if $r_0$ were
constant during the collection interval. However, as the SD method
fully supports using varying $r_0$ we do not see any advantage of using
this option, in particular as it is not obvious how to average the
$r_0$ values.

Normally, the better AO corrections are considered to be those that
remove more of the wavefront, because uncorrected aberrations
typically reduce the contrast and hide signal in the noise. On the
other hand, MFBD converges to better solutions when it has fewer
parameters to fit. With SD, we want the number of fitted modes to be
at least as large as the number of AO-corrected modes. If the
observations are not noise limited, this potentially creates a
trade-off between AO correction with many modes and using a smaller
number of well-defined modes. The lower limit for the number of modes
to use is then set by the ability of the image-formation model to
represent the resolution-critical cores of the PSFs.

We have shown the results of simulation experiments with synthetic
data mimicking AO-corrected observational data in that the
coefficients of the $\Mao$ lowest order modes were reduced by half
from the random values following Kolmogorov statistics. While these
data are the most interesting for our own processing needs, we did
also process synthetic data without such simulated AO correction. We
do not show those results here and we do not discuss them in detail.
However, we do note that the SD extension of MFBD image restoration
works for such data as well. The data are more degraded for the same
$r_0$ than the AO-corrected data, and so the results are worse, as expected.
However, as with AO correction, SD does improve the wavefront sensing
and restored contrasts. Without AO correction, SD is easier to use in
the sense that there is no $\Mao$ that limits the choice of $\Mmfbd$
and there is no partial correction to take into account when
increasing the number of fitted modes in steps from 5 to $\Mmfbd$.
We believe it would be beneficial to include SD in all MFBD problems
where the wavefront aberrations follow Kolmogorov statistics. Or
rather, for all problems where the statistics are known.

In the formulation of \cite{scharmer10high-order}, the corrective
transfer function is based on an ensemble average of many OTFs based
on independent random wavefronts with the correct $r_0$. We can use a
single random wavefront per exposure in the SD method presented here
because our data sets consist of many exposures, and so the SD
correction does include an ensemble of independent random wavefronts
because of that. However, MOMFBD processing of line scans from imaging
spectro(polari)meters such as CRISP or CHROMIS \citep[see,
e.g.,][]{2021A&A...653A..68L} may require more consideration. While
such datasets usually have approximately 100 or more WB images
collected during the scan, for each NB tuning (and polarimetric state)
only about 10 exposures are usually collected. This may be too few for
a single random wavefront per state. If this is the case, the optimum
processing strategy may be to use the post-restoration method of
\cite{scharmer10high-order} on the NB data based on the low-order
wavefronts estimated with SD. This is an issue that should be
addressed when we adapt SD for use with real SST data.

The NN MFBD method of \citetads{2021A&A...646A.100A} shares the finite
wavefront parameterization with traditional MFBD and should have the
same problem with restoring to correct power and contrast. We expect
that such NN methods would benefit from including SD in their image
formation model, just like traditional MFBD.

Speckle for AO-corrected data no longer needs $r_0$ measured from the
spectral ratio, as $r_0$ can be obtained from the AO WFS.
However, the spectral ratio is instead used for measuring the
efficiencies of the AO correction for the corrected modes, and for monitoring
how those efficiencies degrade with distance from the WFS lock point.
Interestingly, \citetads{2017A&A...607A..83P} added random noise to
the efficiencies and found that this did not change the photometry of
the restored data significantly. This is similar to our experience; we find that the details in the high-order contributions to the wavefronts are of little importance, as long as the statistics are correct.

\balance

\section{Conclusions}
\label{sec:conclusions}

We present a novel extension to the family of versatile
wavefront-sensing and image-restoration methods that are based on
MFBD, including PD, which we refer to as statistical diversity (SD). This is used to
statistically represent high-order tail contributions to the unknown
wavefronts so that the high-order mode coefficients do not have to be
included as unknowns in the fitted model parameters. We tested SD with
ideal, synthetic data. When combined with traditional PD, or FD as we
refer to it in this paper, it comes with some important improvements:
(1) The PSFs include the wide wing ``halos'' that are responsible for a
major contribution to stray light in the restored images, and the results
are more consistent contrasts and power spectra in the restored
images. (2) The accuracy of the parameter estimates that are
included in the fitted model is improved significantly.

Statistical diversity does not increase the number of free model
parameters, and does not require calibration of the AO correction,
either at the WFS lock point or as it degrades with distance from it.
SD combines the system-diagnostic ability of MFBD-based methods
(measures fixed aberrations and partially AO-corrected aberrations of
low order, which affect the core of the PSFs necessary for restoring
diffraction-limited resolution) with the known statistics of
atmospheric turbulence exploited by Speckle methods (compensating for
high-order modes that mainly affect the power spectrum and thereby the
contrast). The method also has the potential to improve the
convergence, in that the low-order parameters seem to reach their
final values more easily when the total effect of multiple high-order
modes is taken into account. The convergence could be improved further
by adapting the SD contribution to the existing scheme of starting
with only a few modes and then increasing the number of fitted modes
step by step until the full $\Mmfbd$ is reached.

\begin{acknowledgements}
MGL is grateful to Göran Scharmer for carefully reading the
manuscript and discussing our results, to Friedrich Wöger for
interesting discussions and for patiently answering questions about
Speckle methods, and to Michiel van Noort for explaining details
about his recent work. This work was carried out as a part of the
SOLARNET project, supported by the European Union’s Horizon 2020
research and innovation programme under grant agreement No. 824135.
The Institute for Solar Physics is supported by a grant for research
infrastructures of national importance from the Swedish Research
Council (registration number 2017-00625).
\end{acknowledgements}

\end{document}